\documentclass[aps,prl,showpacs,showkeys,amsmath,amssymb,
onecolumn,
floatfix,
superscriptaddress
]{revtex4}

\usepackage{amsmath}
\usepackage{graphicx}
\usepackage{multirow}
\usepackage{longtable}
\usepackage{rotating}

\usepackage{amsmath,amstext,amsfonts,amsbsy,amssymb,amscd,bbm,epsfig,subfigure}

\usepackage{color}

\def\be{\begin{equation}}
\def\ee{\end{equation}}
\def\bc{\begin{center}}
\def\ec{\end{center}}
\def\bea{\begin{eqnarray}}
\def\eea{\end{eqnarray}}
\newcommand{\bi}{\begin{itemize}}
\newcommand{\ei}{\end{itemize}}                 

\newcommand{\ba}{\begin{array}{c}}
\newcommand{\bad}{\begin{array}{ccc}}
\newcommand{\ea}{\end{array}}

\newcommand{\ov}{\overline}

\newcommand{\bra}[1]{\ensuremath{\langle #1 |}}   
\newcommand{\ket}[1]{\ensuremath{| #1 \rangle}}   
\newcommand{\eps}{\varepsilon}
\newcommand{\ldm}{\ensuremath{{\Delta m_{31}^2}}}          
\newcommand{\sdm}{\ensuremath{{\Delta m_{21}^2}}}

\allowdisplaybreaks[4] 

\begin{document}

\vspace*{-3.cm}
\begin{flushright}
RM3-TH/14-17\\
SISSA 62/2014/FISI
\end{flushright}
\vspace*{0.7cm}

\vspace*{0.2cm}

\title{ \Large Probing new physics scenarios in accelerator and reactor neutrino experiments}

\author{A.~Di~Iura}
\affiliation{Dipartimento di Matematica e Fisica, 
Universit\`{a} di Roma Tre, Via della Vasca Navale 84, I-00146 Rome}
\author{I.~Girardi}
\affiliation{SISSA/INFN, Via Bonomea 265, I-34136 Trieste, Italy }
\author{D.~Meloni}
\affiliation{Dipartimento di Matematica e Fisica, 
Universit\`{a} di Roma Tre, Via della Vasca Navale 84, I-00146 Rome}

\date{\today}

\begin{abstract}

We perform a detailed combined fit to the $\ov \nu_e \rightarrow \ov \nu_e$ disappearence data
of the Daya Bay experiment and the appearance $\nu_{\mu} \rightarrow \nu_{e}$ and disappearance
$\nu_{\mu} \rightarrow \nu_{\mu}$ data of the Tokai to Kamioka (T2K) one in the presence of two models of new physics affecting 
neutrino oscillations, namely a model where sterile neutrinos can propagate in a 
large compactified extra dimension and a model where non-standard interactions (NSI) affect the neutrino production and 
detection. We find that the Daya Bay 
$\oplus$ T2K data combination 
constrains the largest radius of the compactified extra dimensions to be $R\lesssim 0.17$ $\mu {\rm m}$ at 2$\sigma$ C.L. 
(for the inverted ordering of the neutrino mass spectrum) and the relevant NSI parameters in the range ${\mathcal O}(10^{-3})-{\mathcal O}(10^{-2})$,
for particular choices of the charged parity violating phases.
\end{abstract}

\maketitle

\section{Introduction}
After the recent discovery of the reactor angle $\theta_{13}$ and its measurement in the Daya Bay \cite{An:2013uza} and RENO 
\cite{Ahn:2012nd} reactor 
experiments,
experimental efforts in the neutrino sector are now devoted to establishing the presence of charged parity (CP) violation
in the lepton sector, the neutrino mass ordering and the absolute neutrino mass scale.
In fact, the relatively large value of $\theta_{13}\sim 9^o$ opens up the
possibility of searching for possible non-vanishing CP violating phase $\delta$ in the Long Baseline (LBL)
neutrino experiments, Tokai to Kamioka (T2K) \cite{Abe:2013hdq} and NO$\nu$A \cite{Ayres:2004js}
and in future LBL experiments such as Hyper-Kamiokande \cite{Abe:2011ts} and the 
Long-Baseline Neutrino Experiment (LBNE) \cite{Akiri:2011dv}.
The recent observation of 28 electron neutrino events in T2K \cite{Abe:2013hdq}
confirmed the $\nu_{\mu} \rightarrow \nu_e$ transition at more than 7$\sigma$
and provided a first (although weak) indication for the value of $\delta$. 
Indeed a preliminary combined joint analysis \cite{Giganti:NOW2014} of the appearance and disappearance channels in 
T2K, which also includes the reactor constraints on $\theta_{13}$, 
disfavors $\delta/\pi \in [0.1,0.8]$ at more than 90\% C.L., with a best fit point around $\delta = 3 \pi/2$. 
This shows the large increase of sensitivity in the determination of $\delta$ when
performing a combined analysis of reactor and super-beam data \cite{Capozzi:2013csa,Forero:2014bxa,GonzalezGarcia:2012sz,Abe:2014tzr}.
\\
The strength of such a procedure can also be used to test the presence of 
physics beyond the Standard Model (SM) in the neutrino sector (affecting 
neutrino oscillation probabilities)
and, to some extent,  to analyze its
impact on the determination of the standard oscillation parameters.
In this paper, we consider two possible such scenarios:
the so called non-standard neutrino interactions (NSI), where 
the neutrino interactions with ordinary matter are parametrized at low energy in terms of effective flavour-dependent
couplings $\varepsilon_{\alpha \beta}$ \cite{Grossman:1995wx},
and the large extra dimensions
(LED) model, where sterile neutrinos can propagate in a larger than three dimensional space
whereas the SM left-handed neutrinos are confined to a four dimensional (4-D) space-time brane \cite{Barbieri,Mohapatra,Davoudiasl:2002fq}.
The effects of sterile neutrinos have been also recently studied in the context of short-baseline oscillation experiments and in the beta spectrum as measurable by KATRIN-like experiments (e.g. see \cite{BastoGonzalez:2012me}).

The aim of this paper is to take full advantage of whole T2K 
($\nu_\mu \to \nu_e$ appearance \cite{Abe:2013hdq} and $\nu_\mu \to \nu_\mu$ disappearance \cite{Abe:2014ugx}) 
and $\bar \nu_e \to \bar \nu_e$ Daya Bay \cite{An:2013zwz} data in order to:
\begin{itemize}
\item study in details 
how the correlation between the standard oscillation parameters
are modified by the presence of new physics 
\cite{Girardi:2014kca,Girardi:2014gna};
\item constrain the parameter space of the LED and NSI models.
\end{itemize}

In what follows, we will first recall the main features of  the NSI and LED models. Then, after a brief description of the
statistical technique employed in analyzing the experimental data, we will discuss how the presence of new physics modifies
the correlation among the standard oscillation parameters. Finally, we will give the constraints on the model parameters 
obtained from the joint analysis of the Daya Bay and T2K data.

\section{Basic formalism}
\label{sec:form}

\subsection{The NSI model}

\label{Sec:NSI}

The NSI approach describes a large class of new physics 
models where the neutrino interactions with ordinary matter are parametrized at low energy in terms of effective flavour-dependent
couplings $\varepsilon_{\alpha \beta}$, of ${\mathcal O}(M_{EW} / M_{NP})$, where $M_{EW}$ is the electroweak scale 
and $M_{NP}$ is the scale where new physics sets in. The presence of these new couplings can affect the neutrino
production and detection \cite{Wolf78,Grossman:1995wx}. Many authors have studied the impact of NSI 
on conventional neutrino beams~\cite{Ota:2002na,Kitazawa:2006iq,Blennow:2007pu,Kopp:2007ne,Blennow:2008ym,Kopp:2010qt}, 
and on active or completed reactor experiments~\cite{Ohlsson:2008gx,Leitner:2011aa,Khan:2013hva},
finding that the sensitivity reach of such experiments on NSI parameters can be of the order of ${\mathcal O}(10^{-2})$.

In what follows, we consider the analytic treatment of 
the NSI as described in \cite{Kopp:2007ne}. 
The starting point is to assume that
the neutrino states at the source (s) and at the detector (d) are a superposition of 
the orthonormal flavor eigenstates $|\nu_e \rangle$,  
$|\nu_\mu \rangle$ and $|\nu_\tau \rangle$
\cite{Ohlsson:2013nna,Ohlsson:2012kf,Meloni:2009cg}:
\begin{eqnarray}\label{eq:s}
|\nu^{s}_\alpha \rangle & = &  |\nu_\alpha \rangle +
\sum_{\beta=e,\mu,\tau} \varepsilon^{ s}_{\alpha\beta}
|\nu_\beta\rangle   = \big[ (1 + \varepsilon^{ s}) |\nu \rangle \big]_{\alpha} \,, \\
\langle \nu^{d}_\beta| & = &\langle
 \nu_\beta | + \sum_{\alpha=e,\mu,\tau}
\varepsilon^{ d}_{\alpha \beta} \langle  \nu_\alpha  | =  \big[ \langle \nu
|  (1 + \varepsilon^{ d}) \big]_{\beta}
\,. \label{eq:d}
\end{eqnarray}
The oscillation probability can be obtained by squaring the amplitude $\bra{\nu^d_\beta} e^{-i H L} \ket{\nu^s_\alpha}$:
\begin{align}
  P_{\nu^s_\alpha \rightarrow \nu^d_\beta}
    &= |\bra{\nu^d_\beta} e^{-i H L} \ket{\nu^s_\alpha}|^2 \nonumber\\
    &= \big| (1 + \eps^d)_{\gamma\beta} \, \big( e^{-i H L} \big)_{\gamma\delta}
             (1 + \eps^s)_{\alpha\delta} \big|^2 .         
               \label{eq:P-ansatz}
\end{align}
Since the parameters $\eps_{e\alpha}^s$ and  $\eps_{\alpha e}^{d}$ receive contributions from 
the same higher dimensional operators \cite{Kopp:2007ne}, 
one can constrain them by the relation:
\be
\eps_{e\alpha}^s = \eps_{\alpha e}^{d*} 
\equiv  \eps_{e\alpha} e^{\text{i}  \phi_{e\alpha}}\;,
\label{eq:asseps}
\ee
$\eps_{e\alpha}$ and $\phi_{e\alpha}$ being the modulus and 
 the argument of $\eps_{e\alpha}^s$.
For $\eps_{\alpha\beta}$ 
there exist model independent bounds derived in \cite{enrique}, which at 90\% C.L. read:
\bea
\label{limitienr}
& \varepsilon_{ee} < 0.041, \quad \varepsilon_{e\mu} < 0.025, 
\quad \varepsilon_{e\tau} < 0.041 \;, \nonumber \\
& |\varepsilon_{\mu e}^{s,d}| < 0.026, \quad |\varepsilon_{\mu \mu}^{s,d}| < 0.078, 
\quad |\varepsilon_{\mu \tau}^{s,d}| < 0.013 \;, 
\eea
%
whereas for the related CP violating phases $\phi_{\alpha \beta}$  
no constraints have been obtained so far. 
These bounds can be improved by future reactor neutrino experiments \cite{Ohlsson:2013nna} and, in particular,  
at neutrino factories \cite{Coloma:2011rq}, where
the non-diagonal couplings are expected to be constrained at the level of ${\mathcal O}(10^{-3})$.
\\
The Hamiltonian of the system is given by:
\begin{align}
  H_{\alpha\beta} &= \frac{1}{2E_{\nu}} \left[
           U_{\alpha j} \begin{pmatrix}
                          0 &      & \\
                            & \sdm & \\
                            &      & \ldm
                        \end{pmatrix}_{jk}
          (U^\dag)_{k \beta} 
       \right],
  \label{eq:H-ansatz}
\end{align}
where $U$ is the Pontecorvo, Maki, Nakagawa and Sakata (PMNS) matrix \cite{01-BPont67}, 
for which we assume the standard parameterization \cite{Agashe:2014kda}.
For the analysis of the Daya Bay data, $P(\overline \nu_e \rightarrow \overline \nu_e)$ can be obtained from
Eq.~(\ref{eq:P-ansatz}) and Eq.~(\ref{eq:H-ansatz}), expanding for small $\eps$
and neglecting terms of order $\mathcal{O}(\sdm L /(4 E_{\nu}))$ and $\mathcal{O}(\eps^2)$:
\be
\begin{split}
P(\overline \nu_e \rightarrow \overline \nu_e) & = 1 
 - \sin^2 2\theta_{13} \sin^2  \left[ \frac{\ldm \, L}{4 E_{\nu}} \right ] 
 + 4 \eps_{ee} \cos \phi_{ee}  \left[1 - \sin^2 \left( \frac{\ldm \, L}{4 E_{\nu}} \right ) \sin^2 2 \theta_{13} \right] \\
& -4 \eps_{e\mu}  \sin 2\theta_{13} \sin \theta_{23} \cos 2 \theta_{13}   \cos(\delta - \phi_{e\mu}) \sin^2  \left[ \dfrac{\Delta m^2_{31} \, L}{4 E_{\nu}} \right ] \\
&  -4 \eps_{e\tau} \sin 2\theta_{13} \cos \theta_{23}  \cos 2 \theta_{13}  \cos(\delta - \phi_{e\tau} ) \sin^2 \left[ \dfrac{\Delta m^2_{31} \, L}{4 E_{\nu}} \right ]  \\
&   + \mathcal{O}(\eps^2) + \mathcal{O}(\sdm L /(4 E_\nu))\;.\\
\end{split}
\label{Eq:PNSIlimit}
\ee
On the other hand, $P(\nu_{\mu} \rightarrow \nu_e)$ and
$P(\nu_{\mu} \rightarrow \nu_{\mu})$ (relevant for the analysis of the T2K data), cannot be evaluated neglecting
terms of $\mathcal{O}(\sdm L /(4 E_{\nu}))$, otherwise the correct
dependence on the standard CP phase $\delta$ would be lost.
Thus, $P(\nu_{\mu} \rightarrow \nu_e)$ can be written as:
\be
\label{pmue}
\begin{split}
P(\nu_{\mu} \rightarrow \nu_e) & \simeq \sin^2\theta_{23} \sin^22\theta_{13} \sin^2 \left(\frac{\Delta m^2_{31}\, L}{4E_\nu}\right) \\
& - 2 \sin 2 \theta_{12} \sin 2 \theta_{23} \sin \left(\frac{\Delta m^2_{21}\, L}{4E_\nu}\right)
\sin \theta_{13} \cos^2 \theta_{13}  \sin^2 \left(\frac{\Delta m^2_{31}\,L}{4E_\nu}\right) \sin \delta + P_0 + P_1 \;,\\
\end{split}
\ee
where $P_0$ and $P_1$ are the zero and the first order contributions of the NSI expanded for small $\sdm L /(4 E_{\nu})$, respectively.
Using the constraints in Eq.(\ref{eq:asseps}) and defining $\eps^{s,d}_{\alpha\beta} = |\eps^{s,d}_{\alpha\beta}|
\exp(i \phi^{s,d}_{\alpha\beta})$, one finds:
\be
\begin{split}
P_0 = &-4 |\eps^s_{\mu e}| \sin \theta_{13} \sin \theta_{23} \cos (\delta + \phi^s_{\mu e} )  \sin^2 \left( \dfrac{\Delta m^2_{31} \, L}{4 E_{\nu}} \right ) \\
& -4 |\eps^s_{\mu e}| \sin \theta_{13} \sin \theta_{23} \sin (\delta + \phi^s_{\mu e} )  \sin \left( \dfrac{\Delta m^2_{31} \, L}{4 E_{\nu}} \right )   \cos \left( \dfrac{\Delta m^2_{31} \, L}{4 E_{\nu}} \right ) \\
& -4 \eps_{e \mu} \sin \theta_{13} \sin \theta_{23}  \cos (\delta - \phi_{e \mu} ) \cos 2 \theta_{23} \sin^2 \left( \dfrac{\Delta m^2_{31} \, L}{4 E_{\nu}} \right )  \\
& -4 \eps_{e \mu} \sin \theta_{13}  \sin \theta_{23}  \sin (\delta - \phi_{e \mu} ) \sin \left( \dfrac{\Delta m^2_{31} \, L}{4 E_{\nu}} \right )   \cos \left( \dfrac{\Delta m^2_{31} \, L}{4 E_{\nu}} \right )  \\
& + 8 \eps_{e \tau} \sin \theta_{13} \sin^2 \theta_{23} \cos \theta_{23} \cos (\delta - \phi_{e \tau} ) \sin^2 \left( \dfrac{\Delta m^2_{31} \, L}{4 E_{\nu}} \right ) + \mathcal{O}(\eps^d_{ee} \sin^2 \theta_{13})  \\
& + \mathcal{O}( \eps^s_{\mu \tau}  \sin^2 \theta_{13}) + \mathcal{O}(\eps^s_{\mu \mu} \sin^2 \theta_{13}) + \mathcal{O}(\eps^s_{\mu e} \sin^3 \theta_{13}) + \mathcal{O}(\eps_{e \mu (e \tau )} \sin^3 \theta_{13}) + \mathcal{O}(\eps^2) \;,\\
\label{Eq:PNSIlimitP0}
\end{split}
\ee
and
\be
\begin{split}
P_1 = & -|\eps^s_{\mu e}| \sin 2 \theta_{12} \cos \theta_{23} \sin \phi^s_{\mu e} \frac{\sdm L}{2 E_{\nu}}  \\
& + 2 \eps_{e \mu} \sin 2 \theta_{12} \sin^2 \theta_{23} \cos\theta_{23}  \cos \phi_{e \mu} \frac{\sdm L}{4 E_{\nu}} \sin \left( \dfrac{\Delta m^2_{31} \, L}{2 E_{\nu}} \right ) \\
& +  \eps_{e \mu} \sin 2 \theta_{12} \cos \theta_{23} \sin \phi_{e \mu} \frac{\sdm L}{2 E_{\nu}} \left[1 - 2 \sin^2 \theta_{23} \sin^2 \left( \dfrac{\Delta m^2_{31} \, L}{4 E_{\nu}} \right ) \right]\\
& + 2 \eps_{e \tau} \sin 2 \theta_{12} \sin \theta_{23} \cos^2 \theta_{23} \cos \phi_{e \tau}   \frac{\sdm L}{4 E_{\nu}} \sin\left( \dfrac{\Delta m^2_{31} \, L}{2 E_{\nu}} \right )\\
& -  2 \eps_{e \tau} \sin 2 \theta_{12} \sin \theta_{23} \cos^2 \theta_{23} \sin \phi_{e \tau} \frac{\sdm L}{2 E_{\nu}} \sin^2\left( \dfrac{\Delta m^2_{31} \, L}{4 E_{\nu}} \right)  \\
& + \mathcal{O}\left(\eps \sin \theta_{13} \frac{\sdm L}{4 E_{\nu}} \right)  + \mathcal{O}(\eps^2) \,.\\
\label{eq:NSIP1}
\end{split}
\ee
\\
Note that Eqs.~(\ref{Eq:PNSIlimit})-(\ref{eq:NSIP1}) coincide with those derived in \cite{Girardi:2014kca} for $\delta = 0$.
Within the same approximations $P(\nu_{\mu} \rightarrow \nu_\mu)$ reads:
\\
\be
\begin{split}
P(\nu_{\mu} \rightarrow \nu_\mu) & = P^{SM}(\nu_{\mu} \rightarrow \nu_\mu) +  2 |\eps^s_{\mu \mu}| \cos \phi^s_{\mu \mu} + 2 |\eps^d_{\mu \mu}| \cos \phi^d_{\mu \mu} \\
& - \left[ 2 |\eps^s_{\mu \mu}| \cos \phi^s_{\mu \mu} + 2 |\eps^d_{\mu \mu}| \cos \phi^d_{\mu \mu} \right ]  \sin^2 2 \theta_{23} \sin^2\left[ \dfrac{\Delta m^2_{31} \, L}{4 E_{\nu}} \right ] \\
& - 2 \left( |\eps^s_{\mu \tau}| \cos \phi^s_{ \mu \tau} +  |\eps^d_{\tau \mu}| \cos \phi^d_{\tau \mu} \right) \cos 2 \theta_{23} \sin 2 \theta_{23} \sin^2\left[ \dfrac{\Delta m^2_{31} \, L}{4 E_{\nu}} \right ] \\
& + \left( |\eps^s_{\mu \tau}| \sin \phi^s_{\mu \tau} +  |\eps^d_{\tau \mu}| \sin \phi^d_{\tau \mu} \right) \sin 2 \theta_{23}  \sin \left[ \dfrac{\Delta m^2_{31} \, L}{4 E_{\nu}} \right ] + \mathcal{O}\left( \frac{\sdm}{\ldm} \right) + \mathcal{O}(\sin \theta_{13} \eps) + \mathcal{O}(\eps^2) \,,
\label{eq:Pmumu}
\end{split}
\ee
where the approximate formula for $P^{SM}(\nu_{\mu} \rightarrow \nu_\mu)$ can be found in \cite{Donini:2005db}:
\bea
P^{SM}(\nu_\mu \to \nu_\mu) & = & 1-  \left [ \sin^2 2 \theta_{23} -\sin^2 \theta_{23} \sin^2 2 \theta_{13} \cos
    2\theta_{23} \right ]\, \sin^2\left({\frac{\Delta m^2_{23}\, L}{4 E_\nu}}\right) \cr
& - & \left({\frac{\Delta m^2_{21}\, L}{4E_\nu}}\right) [\sin^2 \theta_{12} \sin^2 2 \theta_{23} + \tilde{J} 
\sin^2 \theta_{23} \cos \delta] \, \sin{\left(\frac{\Delta m^2_{23}\, L}{2E_\nu}\right)} \cr
& - & \left({\frac{\Delta m^2_{21}\, L}{4E_\nu}}\right)^2 \left[\cos^4 \theta_{23} \sin^2 2\theta_{12}+
\sin^2 \theta_{12} \sin^2 2\theta_{23} \cos{\left(\frac{\Delta m^2_{23}\, L}{4E_\nu}\right)}\right] ,
\label{eq:probdismu}
\eea
with $\tilde{J}=\cos \theta_{13} \sin 2\theta_{12}\sin 2\theta_{13}\sin 2\theta_{23}$.
We use Eq.~(\ref{eq:Pmumu}) to compute the theoretical predictions for the number of events at both near and
far detectors in the disappearance channel at the T2K experiment.

\begin{figure*}[t!]
\subfigure{%
\hspace{-0.1cm}
 \includegraphics[height=7cm]{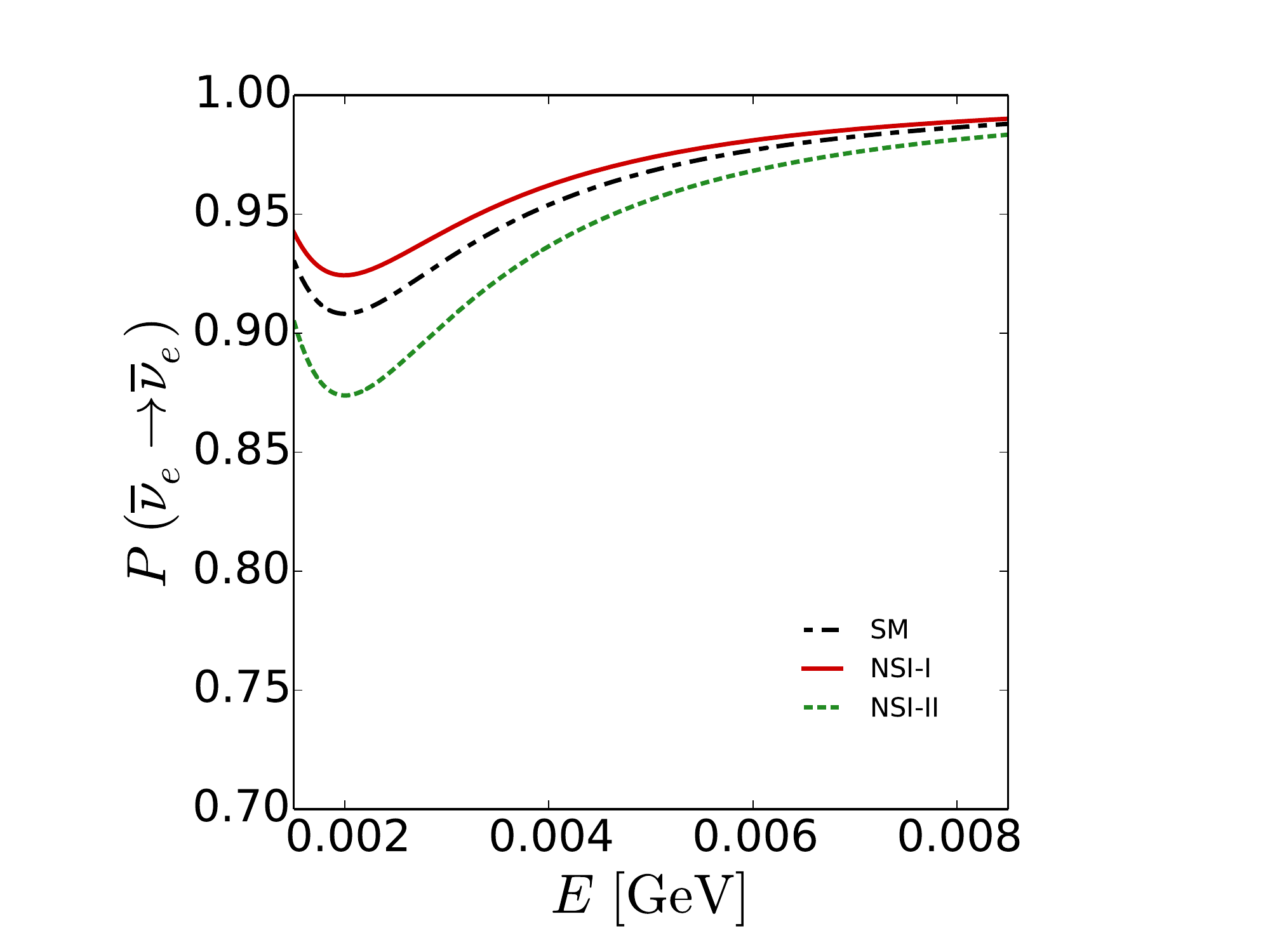}}%
\subfigure{%
\hspace{-2.8cm}
   \includegraphics[height=7cm]{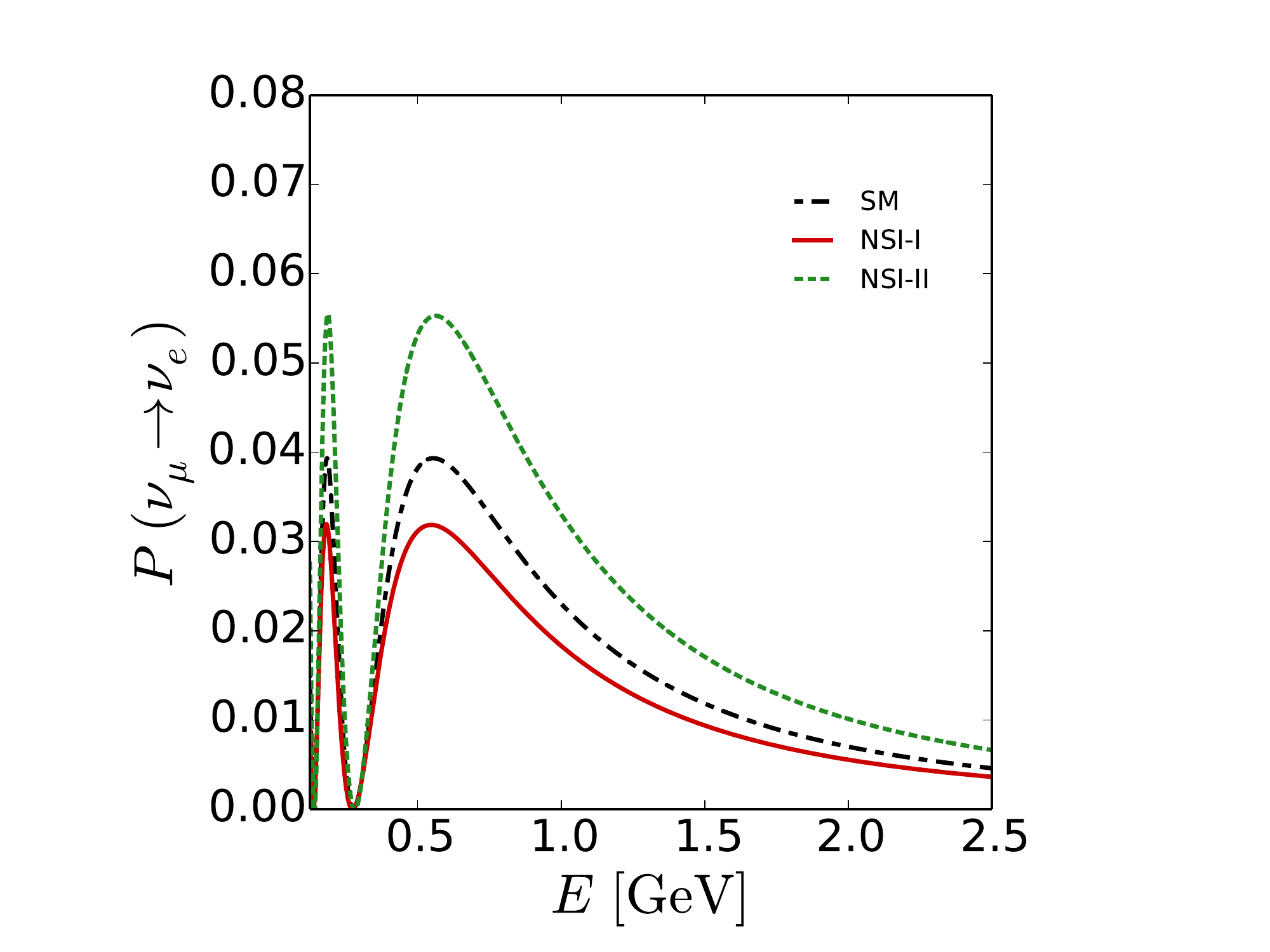}}
    \caption{ \it $P(\overline \nu_e \rightarrow \overline \nu_e)$ and 
    $P(\nu_{\mu} \rightarrow \nu_e)$ oscillation probabilities
    as a function of the neutrino energy $E$. In both panels the standard oscillation parameters have
    been fixed as follows:
    $\sin^2 2 \theta_{13} = 0.09$, $\sin^2 \theta_{12} = 0.308$,
    $\sin^2 \theta_{23} = 0.437$, $\Delta m^2_{21} = 7.54 \times 10^{-5} \, {\rm eV^2}$,
    $\Delta m^2_{31} = 2.5  \times  10^{-3} \, {\rm eV^2}$ and $\delta = 0$.
    For the case  NSI-I, we fixed $\varepsilon^s_{\mu e} = \varepsilon_{e \mu} = \varepsilon_{e \tau} = 0.01$,
    $\phi_{e \mu} = \phi_{e \tau} = \pi$, $\phi^s_{\mu e} = 0$, whereas for the  NSI-II case
    we fixed $\varepsilon_{e\tau} = 0.04$, $\phi_{e \tau} = 0$,  all the other NSI
    parameters being equal to zero. In the left (right) panel
    $L = 1$ km ($L = 275$ km). The dot-dashed, solid and dashed lines correspond
    to SM, NSI-I and NSI-II cases, respectively.
    }
\label{fig:NSIProb}
\end{figure*}

In Fig.~\ref{fig:NSIProb} we show the $P(\overline \nu_e \rightarrow \overline \nu_e)$ and 
    $P(\nu_{\mu} \rightarrow \nu_e)$ oscillation probabilities, with the standard oscillation parameters
    fixed as follows: $\sin^2 2 \theta_{13} = 0.09$, $\sin^2 \theta_{12} = 0.308$,
    $\sin^2 \theta_{23} = 0.437$, $\Delta m^2_{21} = 7.54 \times 10^{-5} \, {\rm eV^2}$,
    $\Delta m^2_{31}= 2.5 \times  10^{-3} \, {\rm eV^2}$ and $\delta = 0$
    for normal ordering (NO) neutrino mass spectrum. The effect of NSI is shown for a particular choice
    of the parameters, namely, $\varepsilon^s_{\mu e} = \varepsilon_{e \mu} = \varepsilon_{e \tau} = 0.01$,
    $\phi_{e \mu} = \phi_{e \tau} = \pi$, $\phi^s_{\mu e} = 0$ for NSI-I (solid lines) and
    $\varepsilon_{e\tau} = 0.04$, $\phi_{e \tau} = 0$ for NSI-II (dashed lines).

\subsection{The LED model}
In LED models, sterile neutrinos can propagate, as well as gravity, in a larger than three dimensional space
whereas the SM left-handed neutrinos are confined to a 4-D space-time brane \cite{Barbieri,Mohapatra,Davoudiasl:2002fq}.
This framework has been introduced to explain the weakness of gravity, since it propagates in the higher dimensional space, 
and, at the same time, to solve the hierarchy problem \cite{ADD}.
In order to avoid the strong constraints on scenarios with one extra dimension only \cite{Agashe:2014kda}, 
we assume to work in a $d + 4$ dimensional space in which one of the $d$ extra dimensions is much larger than the others, 
so we can use an effective five-dimensional formalism to compute the neutrino oscillation probabilities. 
The effects of, for instance, a sixth dimension much smaller than the fifth one can be safely neglected in our study:
in fact, the mass separation among the KK states is proportional to $1/R$, being $R$ the largest radius of the
compactified extra dimensions, so, in the case of $R_{d=2} \ll R_{d=1}$ 
the "new" KK excitations would be at a much higher scale and then 
energetically less accessible by the oscillation phenomenon.
The strongest bound on the radius $R$ of the largest extra dimension at $95\%$ C.L. \cite{Adelberger:2009zz} is:
\bea
\label{Rlimits}
R < 44\ \mu\mathrm{m},
\eea
reached in experiments based on the torsion pendulum instrument testing deviations 
from the Newtonian theory of gravity.

\begin{figure*}[t!]
\subfigure{%
\hspace{-0.1cm}
 \includegraphics[height=7cm]{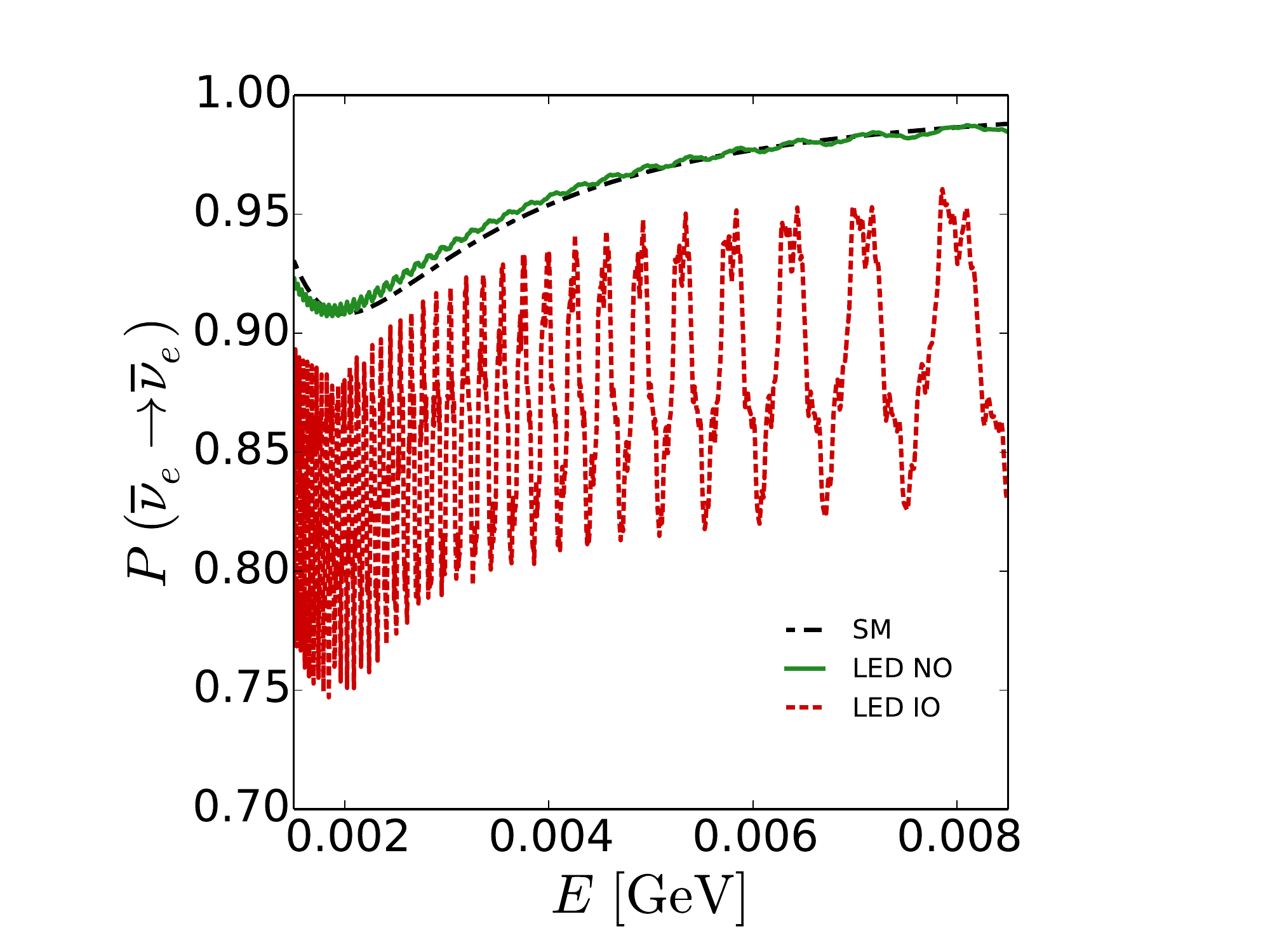}}%
\subfigure{%
\hspace{-2.8cm}
   \includegraphics[height=7cm]{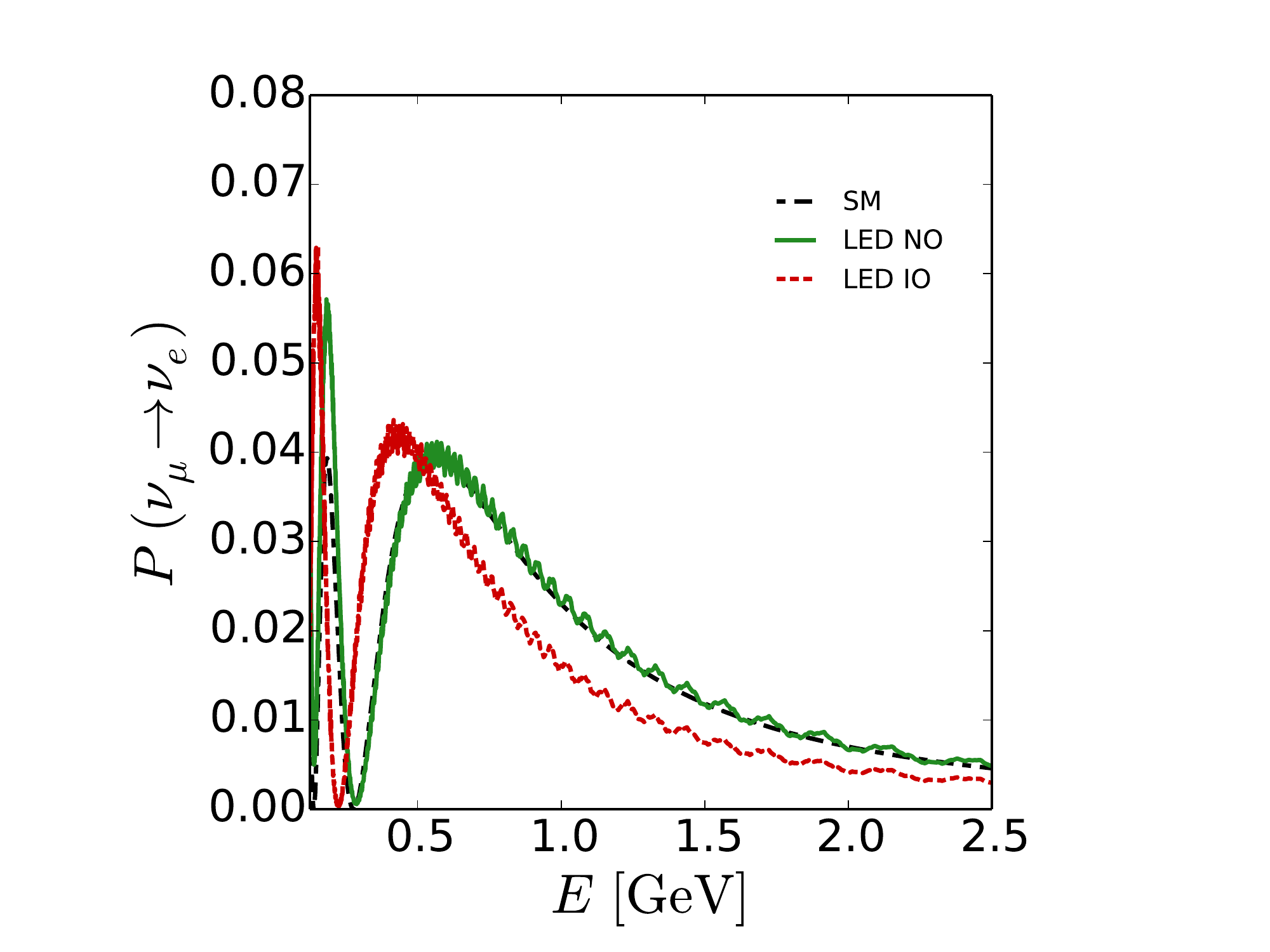}}
    \caption{\it Same as in Fig.~\ref{fig:NSIProb} but for the LED scenario, Eq.~(\ref{eq:ProbAmpNSI}), 
    and fixing $R = 0.5 \, \mu m$, $m_0 = 0$ and $\delta = 0$.
    The sum in Eq.~(\ref{eq:ProbAmpNSI}) has been numerically performed 
    over the first five Kaluza-Klein modes.
    The dot-dashed, solid, and dashed lines correspond to
    the SM, LED NO and LED IO cases, respectively. For NO the values 
    of the mixing parameters are the same as those used in 
    Fig.~\ref{fig:NSIProb}, whereas for the IO we fixed 
    $\sin^2 \theta_{23} = 0.455$ and 
    $|\Delta m^2_{31}|= 2.3 \times  10^{-3} \, {\rm eV^2}$.}
\label{fig:LEDProb}
\end{figure*}

In this paper we are interested in scenarios in which 3 bulk sterile neutrinos
give a Dirac mass term for the 3 active ones. This model is often indicated
as the $(3,3)$ LED model. 
The action of 5-dimensional massless bulk neutrinos $\Psi_i(x_{\mu},y)$, 
interacting with the standard left-handed neutrinos $\nu^{\alpha}_L$, is as follows:
\be
S = \displaystyle \int d^4x \, dy \, i \, \ov \Psi^{\alpha} \, \Gamma_A \partial^A \Psi^{\alpha} + 
\int d^4x \left( i \, \ov \nu^{\alpha}_L \gamma_{\mu} \partial^{\mu} \nu_L^{\alpha} + 
\lambda_{\alpha \beta} \, H \, \ov \nu^{\alpha}_L \, \psi^{\beta}_R (x_{\mu},0) + \rm h.c. \right) \;,
\ee
where $\Gamma^A$ are the Dirac matrices in five dimensions, $\lambda_{\alpha \beta}$ the Yukawa couplings and $H$ the Higgs doublet. 
After the EWSB the neutrino mass matrix can be extracted from the following Lagrangian \cite{Davoudiasl:2002fq}:
\be
\hskip -0.5cm
\mathcal{L_{\rm{eff}}} = \displaystyle
\sum_{\alpha,\beta}m_{\alpha\beta}^{D}\left[\overline{\nu}_{
    R}^{\alpha \left(0\right)}\,\nu_{L}^{\beta}+\sqrt{2}\,
  \sum_{n=1}^{\infty}\overline{\nu}_{
    R}^{\alpha \left(n\right)}\,\nu_{
    L}^{\beta}\right]  
 +  \sum_{\alpha}\sum_{n=1}^{\infty}\displaystyle
\frac{n}{R}\, \overline{\nu}_{R}^{\alpha \left(n\right)} \,
\nu_{L}^{\alpha \left(n\right)} + \rm h. c. \;,
\label{Eq:Lnumass}
\ee
where $\alpha$, $\beta = e,\;\mu,\;\tau$, $m^D_{\alpha \beta}$ 
is a Dirac mass matrix and $\overline{\nu}_{R}^{\alpha \left(0\right)}$, 
$ \overline{\nu}_{R}^{\alpha \left(n\right)}$ and
$ {\nu}_{L}^{\alpha \left(n\right)}$ are linear combinations of the bulk 
fermions. The diagonalization of the mass matrix allows to find
the neutrino mass eigenstates and then to compute the oscillation probability in vacuum \cite{Davoudiasl:2002fq}:
\be
\begin{split}
& P(\nu_{\alpha} \rightarrow \nu_{\beta}) = \left | \mathcal A_{\alpha \beta} (L) \right|^2 \;, \\
& \mathcal A_{\alpha \beta} (L) = \sum_{i = 1}^3 \sum_{n = 0}^{\infty}U^{\alpha i} U^{* \beta i} \left[ U_i^{0n} \right]^2 \exp \left( i \frac{\lambda_i^{(n)2} L}{2 E_{\nu} R^2}\right)\;,\\
\end{split}
\label{eq:ProbAmpNSI}
\ee
where $U^{\alpha i}$ is the matrix element of the PMNS matrix $U$ and $U_i^{0n}$ the matrix element that connects the zero mode, 
i.e. the usual standard neutrinos, with the $n$-th KK mode \footnote{In Appendix A we give the expressions of the unitary 
transformations $U_i^{0n}$ in terms of the radius $R$ and 
the neutrino masses.}. The value of $U_i^{0n}$ can be computed by solving the eigenvalues 
problem for the mass matrix defined in Eq.~(\ref{Eq:Lnumass}). The dimensionless parameters $\lambda_j^{(n)}$
are defined as $\lambda_j^{(n)} \equiv m_j^{(n)} R$ and  can be calculated in a perturbative scheme, as briefly 
reported in Appendix A.
In the case of reactor experiments, the above-mentioned procedure allows to calculate the LED contribution
to the amplitude $\mathcal A_{ee}^{(LED)}$.
Introducing the expansion parameter $\xi_i \equiv \sqrt{2} m_i R$, this reads \cite{Machado:2011jt}:
\begin{align}
\mathcal A_{ee}^{(LED)} \sim & \, \xi_1^2 \, \left|U_{e1}\right|^2 + \xi_2^2 \,\left|U_{e2}\right|^2+ \xi_3^2 \,\left|U_{e3}\right|^2  \, \nonumber \\
\sim & \, \xi_1^2 \cos^2 \theta_{12} \cos^2 \theta_{13} + \xi_2^2 \cos^2 \theta_{13} \sin^2 \theta_{12} + \xi_3^2 \sin^2 \theta_{13}\,.
\label{eq:approxAee}
\end{align}
In the normal ordering (NO) case ($m_3 > m_2 > m_1=m_0$), $\mathcal A_{ee}^{(LED)}$ is dominated by the last term and thus suppressed by the small reactor angle.
For the inverted ordering (IO) case ($m_2> m_1 > m_3=m_0$)  the first two terms dominate the amplitude and no suppressing factor is at work.
We then expect the IO scenario to give better constraints on $R$ and $m_0$ than the NO case.

The situation is quite different for the $\nu_{\mu} \rightarrow \nu_{\mu}$
 and $\nu_{\mu} \rightarrow \nu_e$ probabilities. Indeed for the disappearance channel
we have:
\begin{align}
\mathcal A_{\mu\mu}^{(LED)} \sim & \, \xi_1^2 \, \left| U_{\mu 1} \right|^2 + \xi_2^2 \, \left| U_{\mu 2} \right|^2 + \xi_3^2 \, \left| U_{\mu 3} \right|^2 \nonumber \\
\sim & \, \xi_1^2 \cos^2\theta_{23} \sin^2 \theta_{12} + \xi_2^2 \cos^2 \theta_{12} \cos^2 \theta_{23} + \xi_3^2 \cos^2 \theta_{13} \sin^2 \theta_{23} +  \label{Amumu} \\
 & \, 2 (\xi_1^2 - \xi_2^2) \cos \theta_{12} \cos \theta_{23} \sin \theta_{12} \sin \theta_{13} \sin \theta_{23} \cos \delta + \mathcal{O}(\sin^2 \theta_{13}) \,,\nonumber
\end{align}
and, due to the absence of the $\sin \theta_{13}$ suppression in the $\xi^2_3$ term, we do not expect significant difference in sensitivity between NO and IO. 
This channel is also expected to give better constraints than the 
$\nu_{\mu} \rightarrow \nu_e$ appearance one. Indeed in the latter case the amplitude reads:
\begin{align}
\mathcal A_{\mu e}^{(LED)}\sim & \, \xi_1^2 \, U_{e 1} \, U_{\mu 1}^* +\xi_2^2 \, U_{e 2} \, U_{\mu 2}^* +\xi_3^2 \, U_{e 3} \, U_{\mu 3}^* \nonumber \\
\sim & \, (\xi_2^2 -\xi_1^2) \cos \theta_{12} \sin \theta_{12} \cos \theta_{13} \cos \theta_{23} + \xi_3^2 \sin \theta_{13} \cos \theta_{13} \sin \theta_{23} e^{-i \delta} + 
\, \label{Amue} \\
& \, -\sin \theta_{13} \sin \theta_{23} \cos \theta_{13} e^{-i \delta} (\xi_1^2 \cos^2 \theta_{12} + \xi_2^2 \sin^2 \theta_{12}) \nonumber \,,
\end{align}
and every term is suppressed by either 
$\sdm$ or $\sin \theta_{13}$.

In Fig.~\ref{fig:LEDProb} we show the $P(\overline \nu_e \rightarrow \overline \nu_e)$ (for $L = 1$ km) and 
    $P(\nu_{\mu} \rightarrow \nu_e)$ (for $L = 275$ km) oscillation probabilities fixing $R = 0.5 \, \mu m$, $m_0 = 0$ and $\delta = 0$.
The mixing parameters have been fixed to the same values used in Fig.~\ref{fig:NSIProb} for NO,
     whereas for the IO we fixed 
    $\sin^2 \theta_{23} = 0.455$ and 
    $|\Delta m^2_{31}|= 2.3 \times  10^{-3} \, {\rm eV^2}$.

\section{Neutrino facilities and details of the statistical analysis}
The Daya Bay experimental setup that we take into account consists of six reactors~\cite{An:2013uza}, emitting
antineutrinos $\bar \nu_e$ whose spectra have been recently
estimated in Refs.~\cite{Mueller:2011nm,Huber:2011wv}.
The total flux of arriving $\bar \nu_e$ at the six antineutrino detectors has been estimated
using the convenient parametrization discussed in Ref.~\cite{Mueller:2011nm} and taking into account
all the distances between the detectors and the reactors (summarised in Tab.~2 of Ref.~\cite{An:2013uza}).
For this analysis we use the data set accumulated during 217 days 
extracted from Fig.~2 of Ref.~\cite{An:2013zwz}, with a 1.5 MeV threshold in the positron energy. 
The antineutrino energy $E$ is reconstructed
by the prompt energy deposited by the positron $E_{ \rm prompt}$ using
the approximated relation \cite{An:2013uza}
$E \simeq E_{\rm prompt} + 0.8 \; {\rm MeV}$.
The energy resolution
function is a Gaussian function, parametrized according to:
\begin{eqnarray}
\label{gaussf}
\sigma(E) [\rm MeV]=
\begin{cases}
\gamma \sqrt{E/\rm MeV - 0.8} \, , \; \mbox{for } E > 1.8 \; \rm MeV \, ,\\
\gamma \, , \; \mbox{for } E \leq 1.8 \; \rm MeV\,,\
\end{cases}
\end{eqnarray}
with $\gamma = 0.08$ MeV \cite{An:2013zwz}. 
The antineutrino cross section for the inverse beta decay (IBD) process has been taken
from \cite{Vogel:1999zy}. 
The statistical analysis of the data has been performed using a modified version of the GLoBES software
\cite{GLOB2} with the $\chi^2$ function defined as follows \cite{An:2013uza}:
\begin{eqnarray}
 \label{eqn:chispec}
&& \chi^2_{DB}(\theta,\Delta m^2, \vec S,\alpha_r,\varepsilon,\varepsilon_d,\eta_d) = \nonumber 
 \sum_{d=1}^{6}\sum_{i=1}^{26}
 \frac{\left[M_i^d -T_i^d \cdot \left(1 + \varepsilon + 
 \sum_r\omega_r^d\alpha_r + \varepsilon_d  \right) 
  + \eta_d \right]^2}
 {M_i^d + B_i^d } \nonumber \\
&& + \frac{\varepsilon^2}{\sigma^2_{\varepsilon}} + \sum_r\frac{\alpha_r^2}{\sigma_r^2}
 + \sum_{d=1}^{6}\left[
\frac{\varepsilon^2_d}{\sigma^2_d}
 + \frac{\eta_d^2}{\sigma_{B_d}^2}
 \right]   + \mbox{Priors} \,,
\end{eqnarray}
where $\vec S$ is a vector containing the new physics parameters,
$M^d_i$ are the measured IBD events of the d-th detector ADs
in the i-th bin, $B^d_i$ the corresponding background and $T^d_i = T_i(\theta,\Delta m^2, \vec S)$ are the
theoretical predictions for the rates (the $\sum_i$ is over the bins in prompt reconstructed energy).
The parameter $\omega_r^d$ is the fraction
of IBD contribution of the r-th reactor to the d-th detector AD, determined by
the approximate relation $\omega_r^d \sim L_{rd}^{-2} / (\sum_{r = 1}^6 1/L_{rd}^2 )$,
where $L_{rd}$ is the distance between the d-th detector and the r-th reactor. 
The parameter $\sigma_{\varepsilon}$ is the reactor flux uncertainty ($\sigma_{\varepsilon} \simeq 3\%$), $\sigma_d$
is the uncorrelated detection uncertainty ($\sigma_d = 0.2$\%) and $\sigma_{B_d}$ is the background uncertainty of the d-th detector 
obtained using the information given in \cite{An:2013zwz}:
$\sigma_{B_1} = \sigma_{B_2} = 8.21$, $\sigma_{B_3} = 5.95$, $\sigma_{B_4} = \sigma_{B_5} = \sigma_{B_6} = 1.15$. Finally,
$\sigma_r = 0.8$\% are the uncorrelated reactor uncertainties. 
The corresponding pull parameters are $\varepsilon,\varepsilon_d,\eta_d$ and $\alpha_r$.
With this choice of nuisance parameters we are able to reproduce the 1$\sigma$, 2$\sigma$ and 3$\sigma$ confidence level 
results presented in Fig.~3 of Ref. \cite{An:2013zwz} with high accuracy. 
The differences are at the level of few percent (see Tab. I and Tab. II of Ref. \cite{Girardi:2014gna}).

The T2K experiment \cite{Abe:2013hdq} consists of two separate detectors, 
both of which are 2.5 degrees off axis
of the neutrino beam. The far detector is located at 
$L_F = 295$ km from the source, the ND280 near detector 
is $L_N = 280$ meters from the target.
%
In our analysis we used
the public data in \cite{Abe:2013hdq,Abe:2014ugx}, which
reported 28 events in the appearance channel
and 120 events in the disappearance one (constrained with the 17369 CC$\pi$0 events at the ND280 near detector). 
The neutrino flux has been estimated from \cite{Abe:2013jth}.
We fixed the fiducial mass of the near and the far detector as 
$FM_{\rm ND280} = 1529$ Kg and $FM_{\rm SK} = 22.5$ Kton \cite{Meloni:2012fq}, respectively; bin
to bin normalization coefficients have been introduced in order to reproduce the 
T2K best fit events \cite{Abe:2013hdq}.
For the energy resolution function we adopt a Gaussian function as in Eq.~(\ref{gaussf}), 
with $\gamma =0.085$ GeV (see, e.g., \cite{Huber:2002mx}).
 
The $\chi^2_{T2K}$ is defined as:
\be
\begin{split}  \label{eqn:chispec2}
\chi^2_{T2K} (\theta,\Delta m^2, \vec S,\rho_d,\Omega_d,\alpha_d,\alpha_N) & = 
\sum_{d=1}^{2}\sum_{i=1}^{n_{bins}^d}
 2 \left[M_i^d  -T_i^d \cdot \left(1 +
 \rho_d + \Omega_d  \right) 
  + M_i^d \log \frac{M_i^d}{T_i^d \cdot \left(1 +
 \rho_d + \Omega_d  \right)} \right]
  \\
  & + \sum_{i=1}^{n_{bins}^N}
 2 \left[M_i^N  -T_i^N \cdot \left(1 +
 \rho_1 + \rho_2 + \Omega_N  \right) 
  + M_i^N \log \frac{M_i^N}{T_i^N \cdot \left(1 +
 \rho_1 + \rho_2 + \Omega_N  \right)} \right]
  \\
 & 
 + \sum_{d=1}^{2} \left( \frac{\rho_d^2}{\sigma^2_{\rho_d}} + 
\frac{\Omega_d^2}{\sigma_{\Omega_d}^2} \right) + \frac{\Omega_N^2}{\sigma_{\Omega_N}^2} + {\rm Priors}
 \,. \\
\end{split}
\ee
In the previous formula, $\vec S$ is a vector containing the new physics 
parameters, $M_i^d$ are the measured events,
including the backgrounds, 
of the d-th channel of the far detector
in the i-th bin, $T_i^d = T_i^d(\theta,\Delta m^2,\vec S,\alpha_d)$
are the theoretical predictions for the rates, $\theta$ and $\Delta m^2$ are respectively 
the mixing angles and the squared mass differences contained in the oscillation
probability, $n_{bins}^d$ is the number of bins for the d-th channel of the far detector
(the $\sum_i$ is over the bins in prompt reconstructed energy).
With obvious notation, $M_i^N$ and $T_i^N = T_i^N(\theta,\Delta m^2,\vec S,\alpha_N)$ are the measured and theoretical 
event rates at the near
 detector, respectively.
The parameter $\sigma_{\rho_d}$ contains the systematic uncertainties in the d-th channel: 
$(\sigma_{\rho_1}, \sigma_{\rho_2}) = (8.8 \% , 8.1 \%)$
which are extracted from Table II of \cite{Abe:2013hdq} and Table I of \cite{Abe:2014ugx}; $\sigma_{\Omega_d}$ are
the fiducial mass uncertainties for the d-th detector ($\sigma_{\Omega_d}$ and $\sigma_{\Omega_N}$ have been estimated 
of the order of $1 \%$ for the far
and the near detectors similarly to \cite{Huber:2003pm}), $\alpha_d$ and  $\alpha_N$ are free parameters which represent 
the energy scale for predicted signal events with uncertainty $\sigma_{\alpha_d}$ and $\sigma_{\alpha_N}$, ($\sigma_{\alpha_d}, \sigma_{\alpha_N} = 1 \%$ \cite{Coloma:2012ji}).
The corresponding pull parameters are ($\rho_d,\Omega_d,\Omega_N, \alpha_d,\alpha_N$). The experimental event rates at the near detector have
been estimated rescaling the non-oscillated event rates at the far detector (extracted from \cite{talk:T2K}) using the 
scale factor $L^2_F / L^2_N \times  FM_{\rm ND280} / FM_{\rm SK}$.

Priors in Eqs.~(\ref{eqn:chispec}) and (\ref{eqn:chispec2}) are described at the
beginning of the following section.
The whole Daya Bay and T2K data sample is analyzed using: 
\bea
\label{chitot}
\chi^2_{tot} = \chi^2_{DB} + \chi^2_{T2K}\,. 
\eea

\section{Numerical results}
In the following plots, unless explicitly stated, all the not shown parameters have been marginalized over. 
In particular, the SM quantities  $\theta_{13}$, $\theta_{23}$, $\delta$ and $\ldm$ are unconstrained, 
since they have to be reconstructed from the 
data themselves, whereas 
for the solar angle and the solar mass difference
we used external best fit points and 1$\sigma$ errors from \cite{Capozzi:2013csa} 
and define the gaussian priors as follows: 
$\sin^2 \theta_{12} = 0.308 \pm 20\%$ and $\sdm = (7.54 \pm 5 \%) \times 10^{-5} \, {\rm eV^2}$.
In the following sections we first discuss the impact
of NSI and LED parameters in the determination
of $\sin^2 \theta_{13}$, $\sin^2 \theta_{23}$, $\delta$ and $\Delta m^2_{31}$,
using the current upper limits, Eqs.~(\ref{limitienr}) and (\ref{Rlimits}), and 
the additional restriction to be in the perturbative regime, e.g., 
$\xi_i  < 0.2$.
Then, we derive the constraints on these parameters which arise
from the Daya and the T2K experiments assuming the LED and
NSI parameters as free parameters, and therefore we do not
impose any constraints on them.

\subsection{Standard Model}
\begin{figure*}[h]
\subfigure{%
\hspace{-1.4cm}
 \includegraphics[height=6cm]{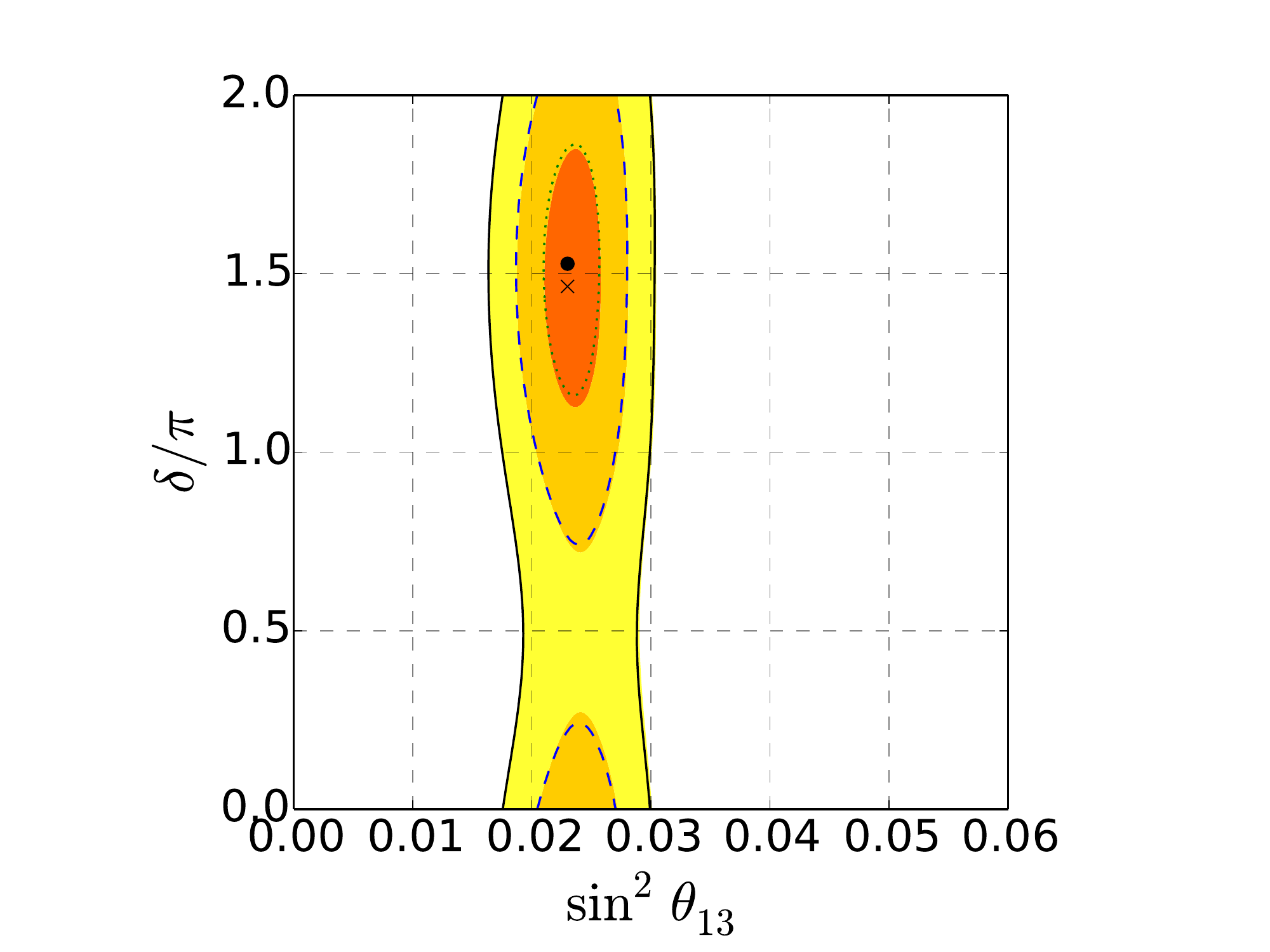}}%
\subfigure{%
\hspace{-2.6cm}
   \includegraphics[height=6cm]{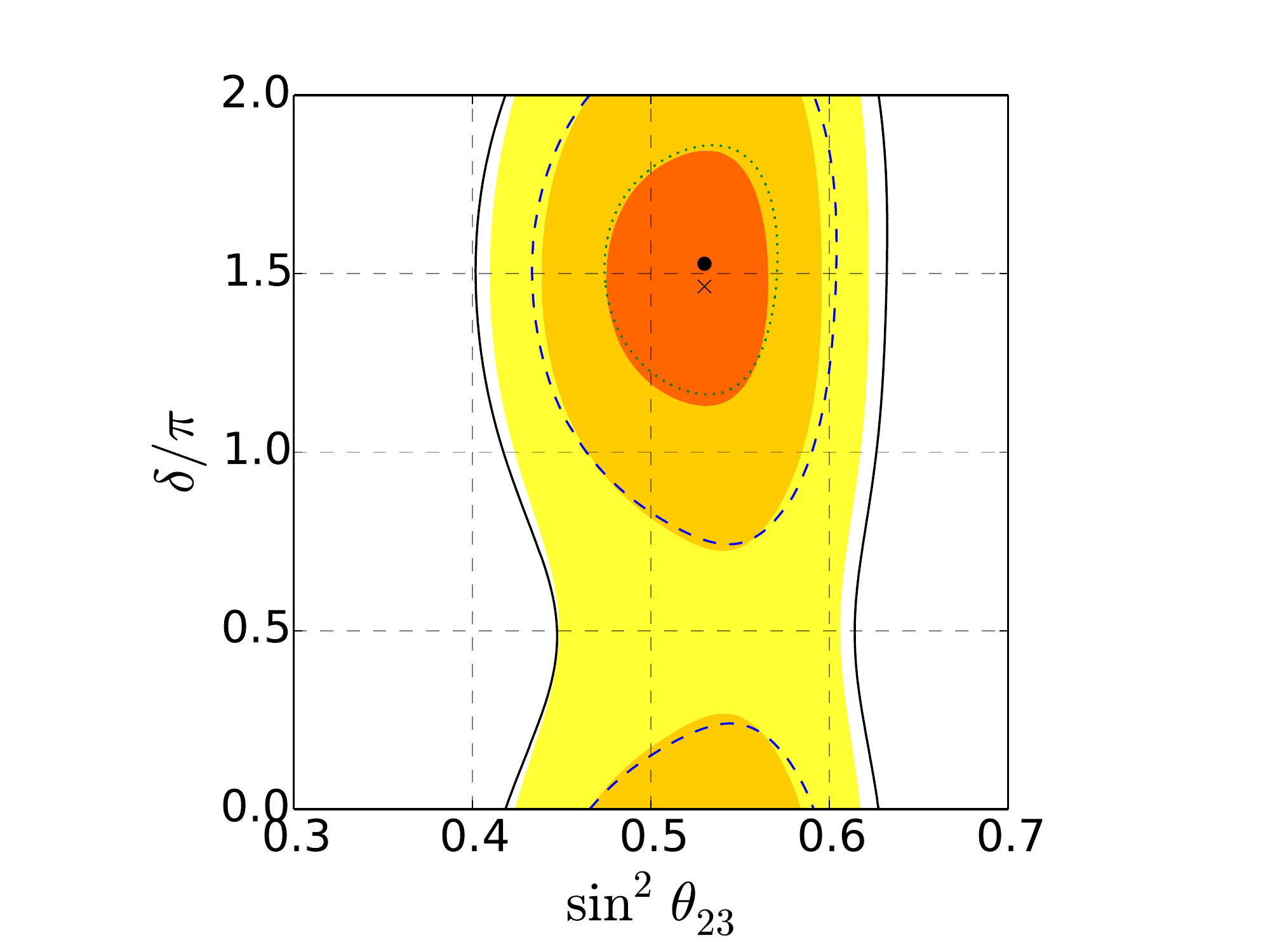}}%
   \subfigure{%
\hspace{-2.4cm}
     \includegraphics[height=6cm]{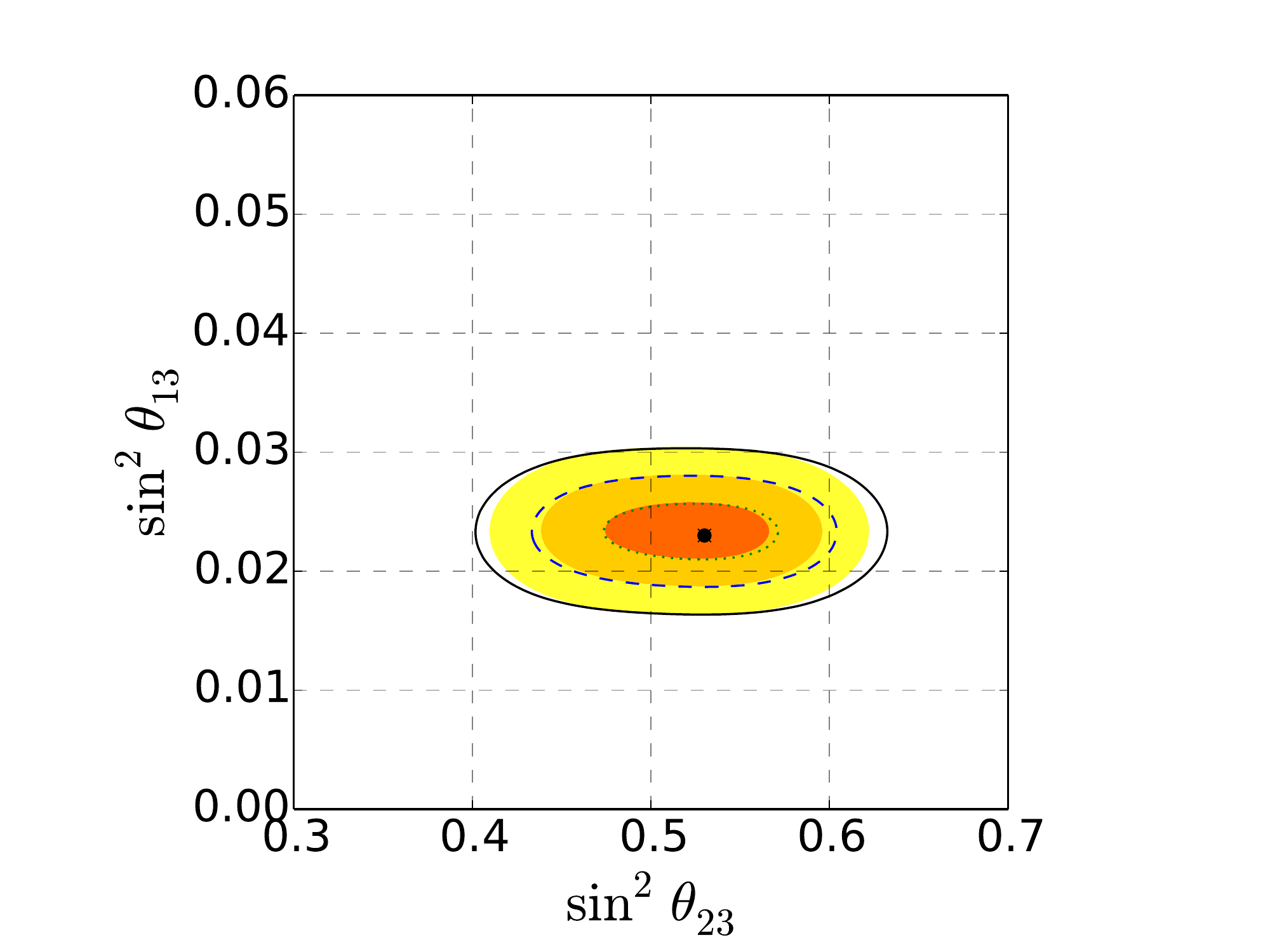}}%
     \vspace{-0.4cm}
     \subfigure{%
  \includegraphics[height=6cm]{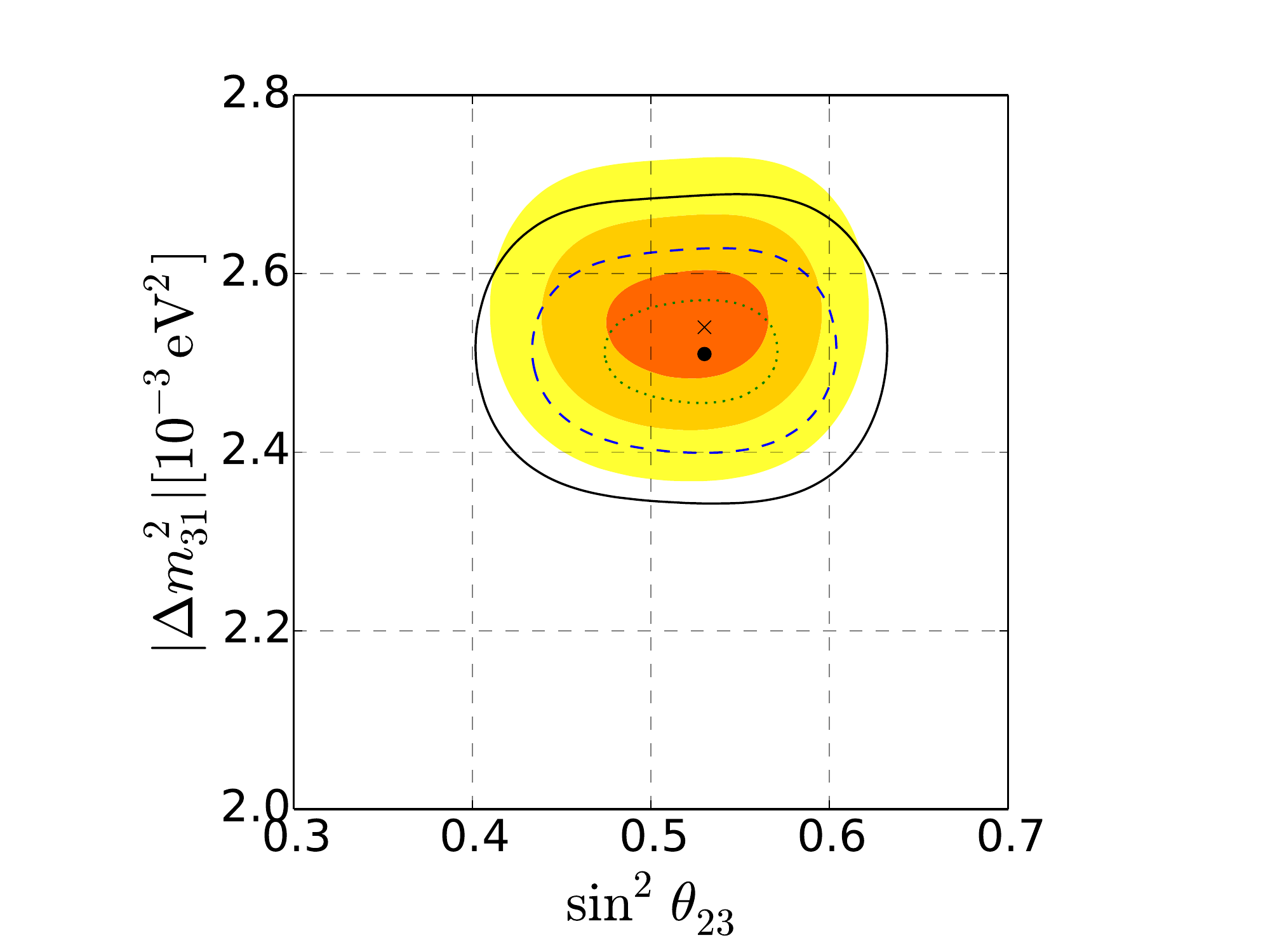}}%
\subfigure{%
\hspace{-2.2cm}
\includegraphics[height=6cm]{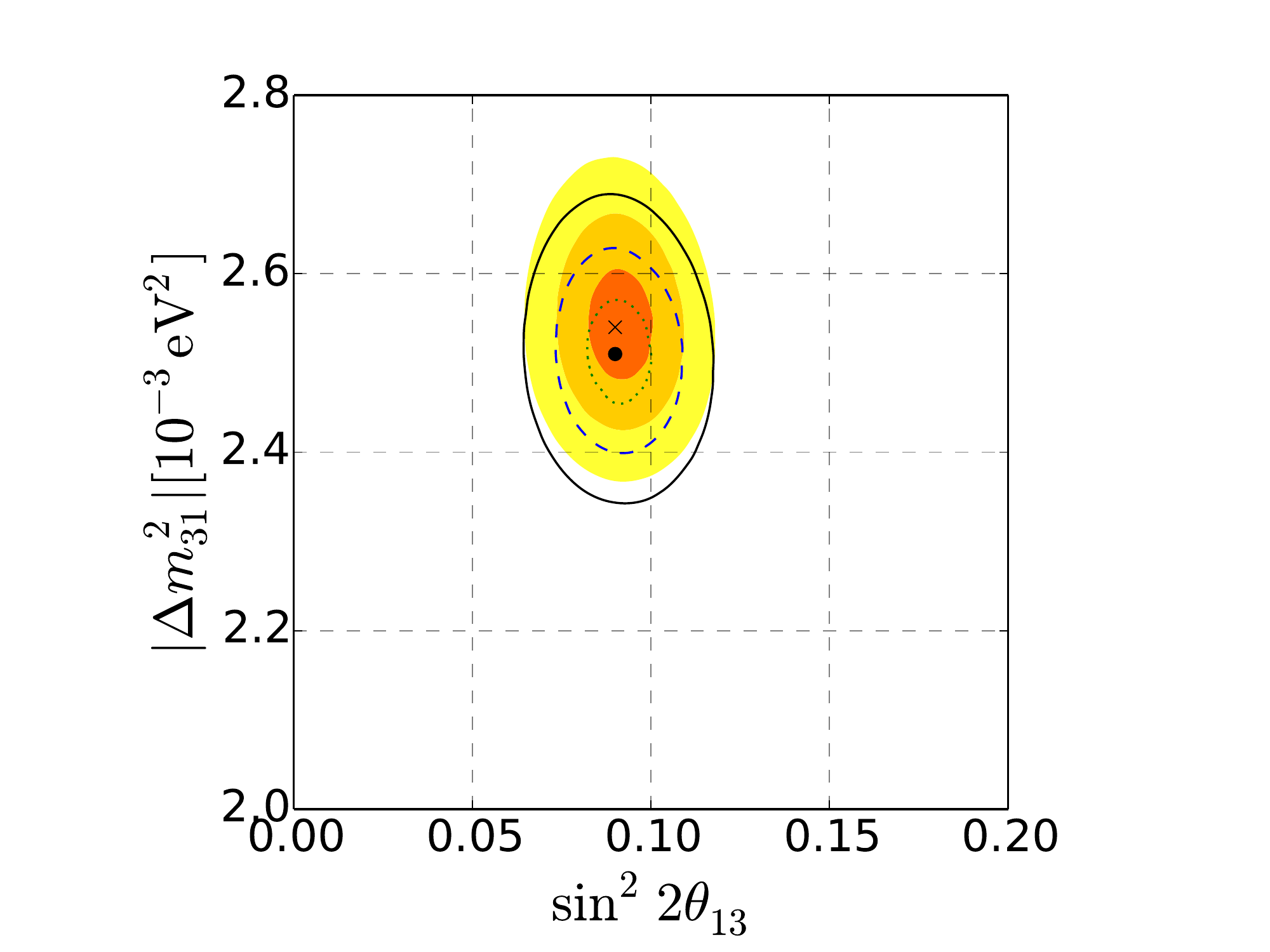}}%
    \caption{ \it 1$\sigma$, 2$\sigma$, 3$\sigma$ confidence
level regions for 1 dof of the
combined fit to the Daya Bay \cite{An:2013zwz} 
and T2K \cite{Abe:2013hdq,Abe:2014ugx} data
using the SM probabilities. 
The case of normal ordering of the neutrino mass
spectrum is represented with dotted, dashed and solid lines respectively, whereas the inverted spectrum
in red (dark-gray), orange (gray) and yellow (light-gray).
}
\label{fig:SM1}
\end{figure*}

We first consider the fit to the data in the standard three-neutrino framework, with the intent to 
make easier the comparison of the standard results with the ones obtained with the contribution of new physics.
In Fig.~\ref{fig:SM1} we show our results in the 
$[\sin^2 \theta_{13},\delta]$, $[\sin^2 \theta_{23},\delta]$,
$[\sin^2 \theta_{23} , \sin^2 \theta_{13}]$, $[\sin^2 \theta_{23},\ldm]$ and
$[\sin^2 2 \theta_{13},\ldm]$ planes.
The curves represent 
the 1$\sigma$, 2$\sigma$, 3$\sigma$ confidence
level regions for 1 degree of freedom (dof).
The case of normal ordering of the neutrino mass
spectrum is represented with dotted, dashed and solid lines  whereas the inverted spectrum
in red (dark-gray), orange (gray) and yellow (light-gray).
The obtained best fit points are indicated with a circle
for NO and with a cross for IO. The figures have been obtained using
the standard oscillation probabilities relevant for the
$\overline \nu_e \rightarrow \overline \nu_e$,
$\nu_{\mu} \rightarrow \nu_e$ and
$\nu_{\mu} \rightarrow \nu_{\mu}$ transitions.
In Tab.~\ref{tabNufitSM} we summarize
the best fit points
and the $1\sigma$, $2\sigma$ and $3\sigma$ confidence level regions.
Our results are in agreement with Ref.~\cite{Capozzi:2013csa} 
as can be observed from Fig.~\ref{fig:SM1} and in Ref.~\cite{Girardi:2014gna}.
\begin{table}[h]
\centering
\renewcommand{\arraystretch}{1.2}
\begin{tabular}{|clccc|}
\hline
 \bf Parameter    &  \bf Best-fit ($\pm 1\sigma$) \phantom{123}  & \phantom{123}  \bf 1$\sigma$ range \phantom{123} & \phantom{123} \bf 2$\sigma$ range \phantom{123} & \phantom{123} \bf 3$\sigma$ range \phantom{123}  \\ \hline
$ \Delta m^{2}_{31}~\text{ (NO)}  \; [10^{-3}\text{ eV}^2]$ \phantom{123} & $2.51^{+0.06}_{-0.06}$  & $2.45-2.57$ & $2.40-2.63$ & $2.34-2.69$  \\
$ |\Delta m^{2}_{31}|~\text{ (IO)}  \; [10^{-3}\text{ eV}^2]$ \phantom{123} & $2.54^{+0.06}_{-0.06}$ & $2.48-2.60$ & $2.42-2.67$ & $2.37-2.73$ \\
$\sin^2\theta_{23}/10^{-1}$   & $5.3^{+0.4}_{-0.6}$ & $4.7-5.7$ & $4.3-6.0$ & $4.0-6.3$  \\
\phantom{$\sin^2\theta_{23}/10^{-1}$}    &  $5.3^{+0.4}_{-0.5}$  & $4.8-5.7$ & $4.4-5.8$ & $4.1-6.2$  \\
$\sin^2\theta_{13}/10^{-2}$   &  $2.3^{+0.3}_{-0.2}$  & $2.1-2.6$ & $1.9-2.8$ & $1.6-3.0$ \\
\phantom{$\sin^2\theta_{13}/10^{-1}$}  & $2.4^{+0.2}_{-0.3}$ & $2.1-2.6$ & $1.9-2.8$ & $1.6-3.0$   \\
$\delta/\pi$   &  $1.53^{+0.33}_{-0.37}$  & $1.16-1.86$ & $0.00-0.24$ $\oplus$ $0.75-2.00$& | \\
\phantom{$\delta/\pi$}    & $1.48^{+0.36}_{-0.35}$ & $1.13-1.84$ & $0.00-0.26$ $\oplus$ $0.73-2.00$ & | \\
\hline
\end{tabular}
\caption{\it Best fit and 1$\sigma$, 2$\sigma$ and 3$\sigma$ errors of the
standard oscillation parameters obtained in the combined fit of the 
Daya Bay and T2K data using the standard oscillation probabilities.
If two values are given, the upper one corresponds to NO
and the lower one  to IO.
}
 \label{tabNufitSM}
\end{table}

\subsection{Effects of including NSI and LED}
\begin{figure*}[h!]
\subfigure{%
\hspace{-1.4cm}
   \includegraphics[height=6cm]{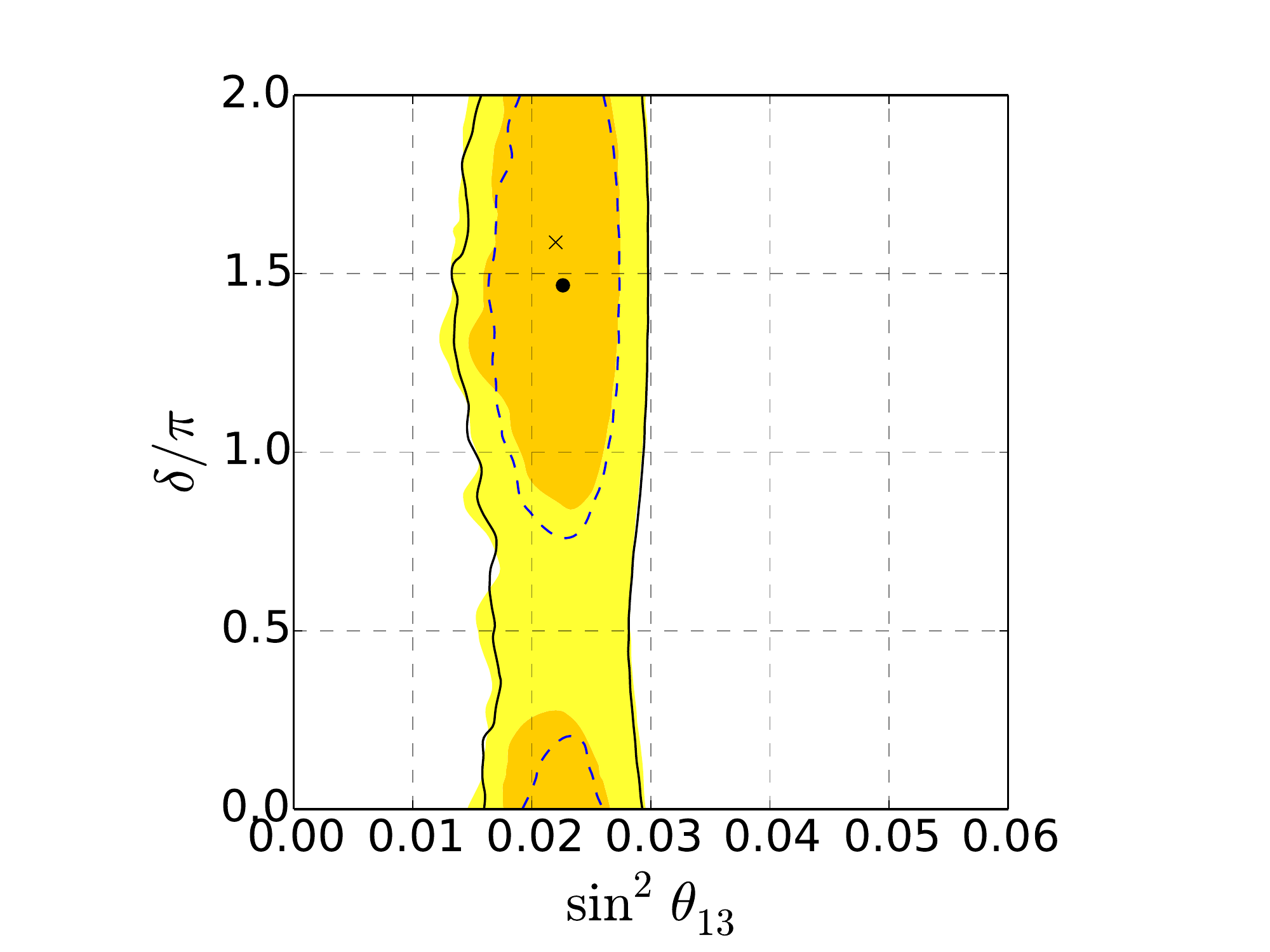}}%
\subfigure{%
\hspace{-2.4cm}
 \includegraphics[height=6cm]{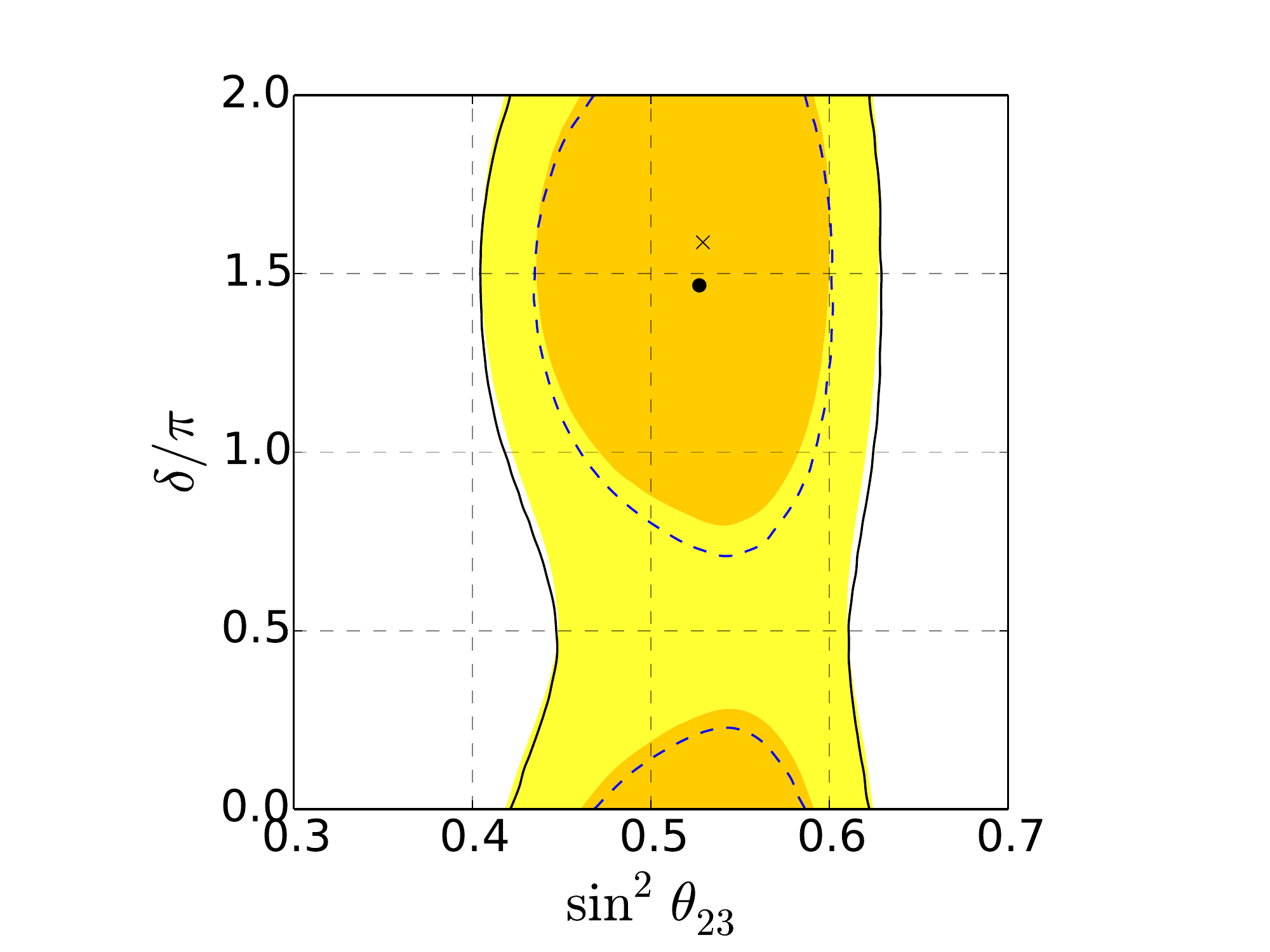}}%
   \subfigure{%
\hspace{-2.4cm}
     \includegraphics[height=6cm]{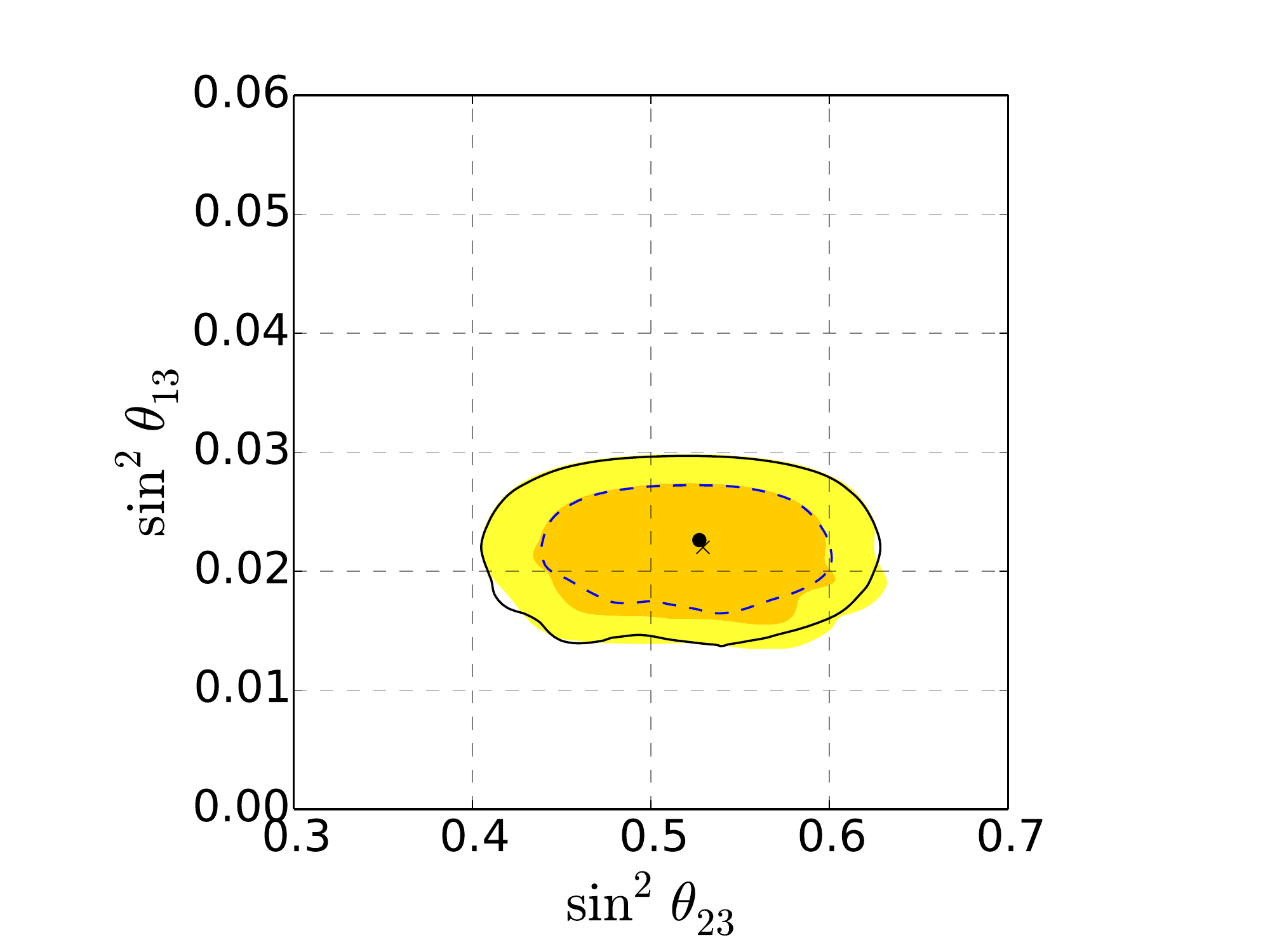}}%
          \vspace{-0.4cm}
        \subfigure{%
\includegraphics[height=6cm]{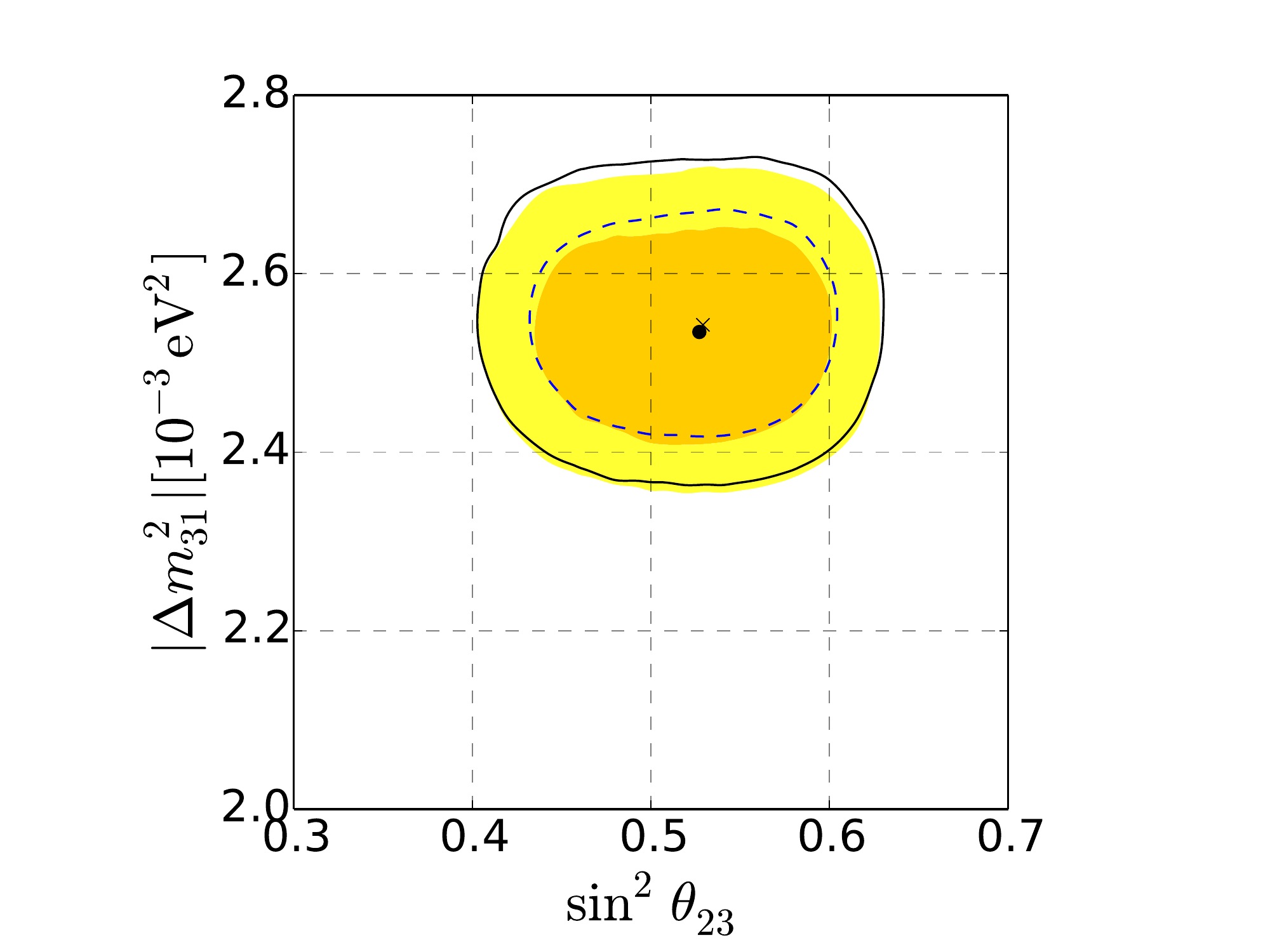}}%
        \subfigure{%
\hspace{-2.4cm}
     \includegraphics[height=6cm]{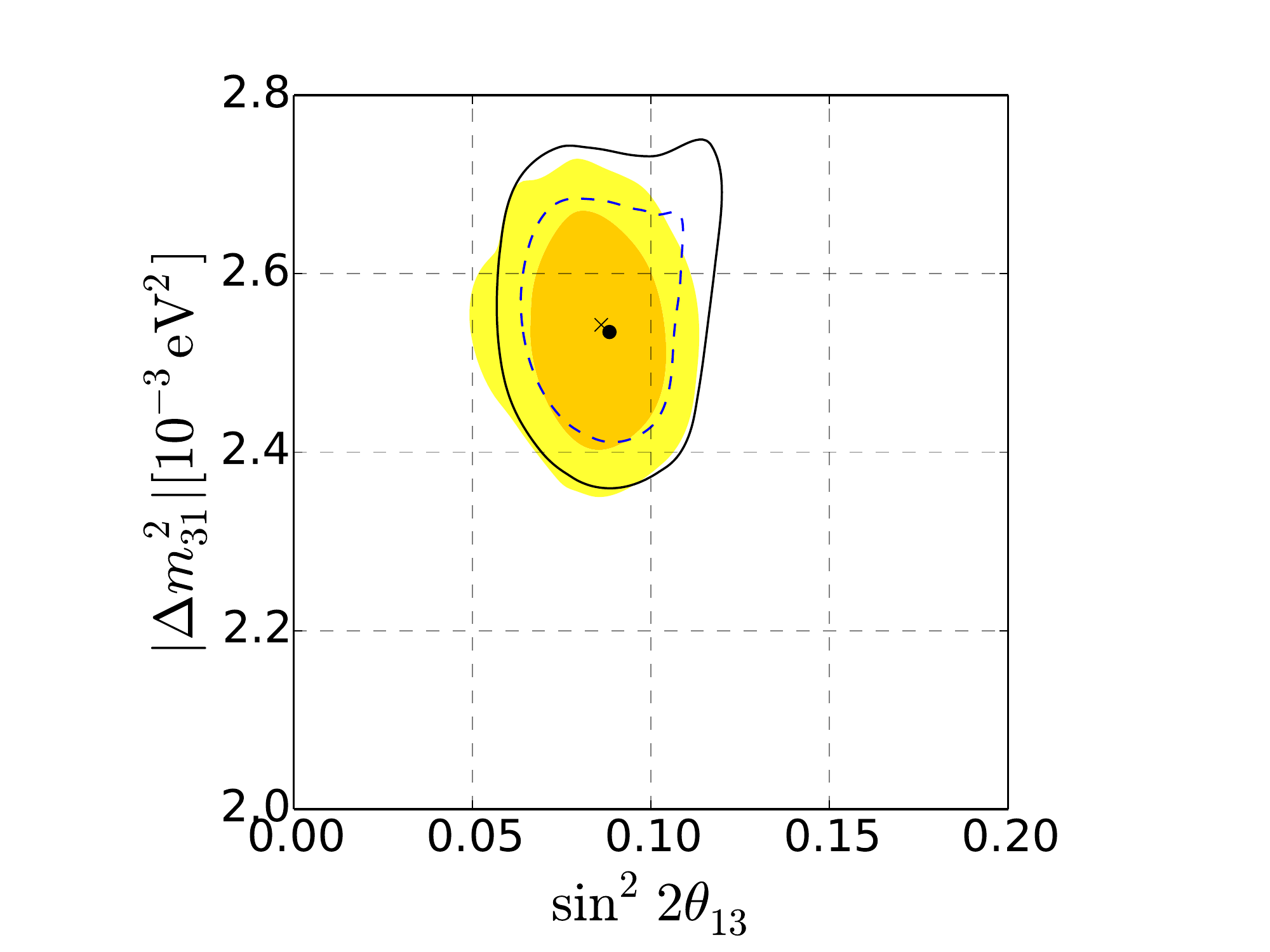}}%
    \caption{ \it 2$\sigma$ and 3$\sigma$ confidence
level regions on the standard neutrino oscillation parameters for 1 dof from 
the combined fit to the Daya Bay and T2K data, obtained using the LED oscillation probabilities. 
Same conventions as in Fig.~\ref{fig:SM1}. 
}
\label{fig:LED1}
\end{figure*}

\begin{figure*}[h]
\subfigure{%
\hspace{-1.4cm}
 \includegraphics[height=6cm]{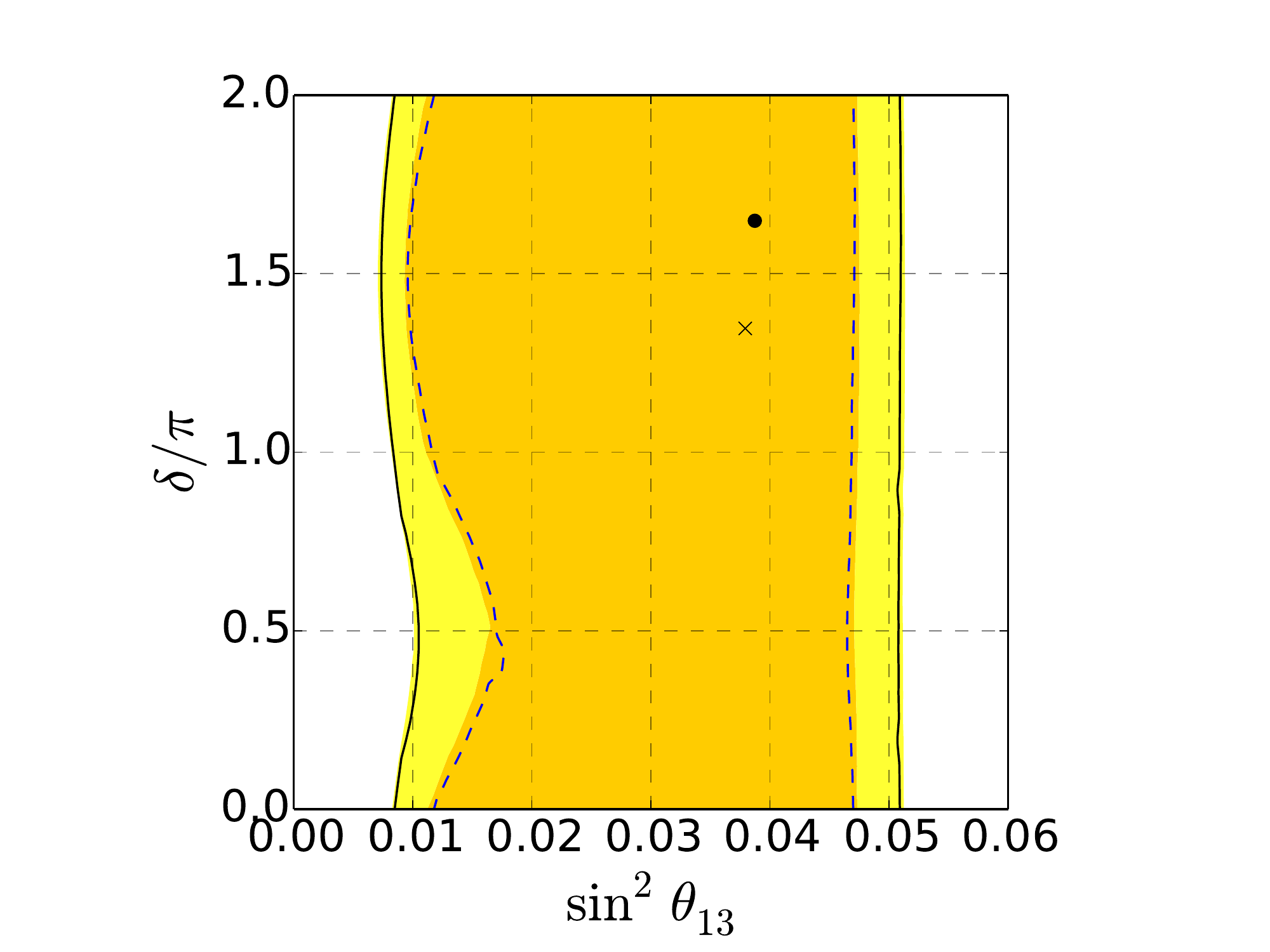}}%
\subfigure{%
\hspace{-2.4cm}
   \includegraphics[height=6cm]{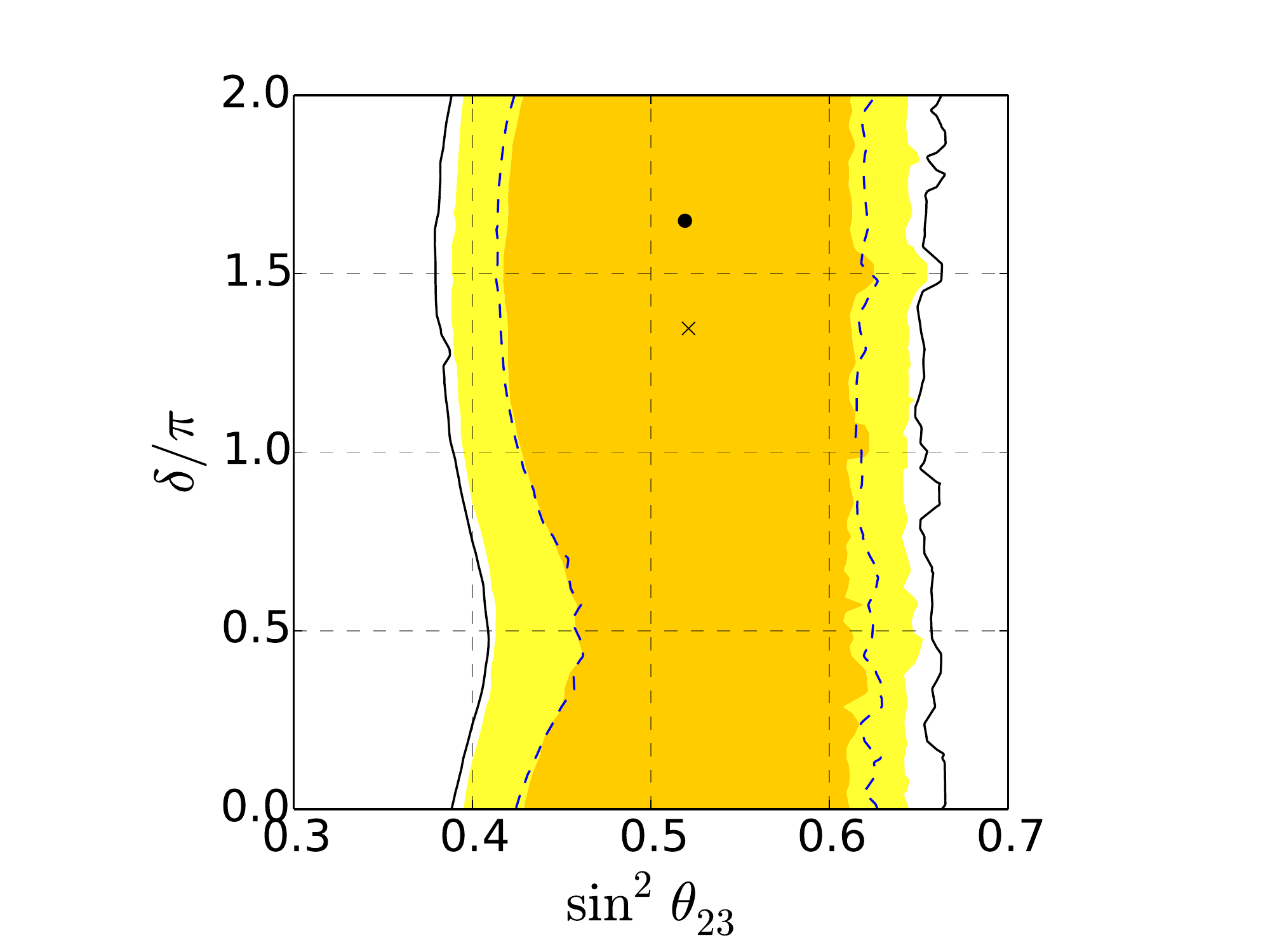}}%
   \subfigure{%
\hspace{-2.4cm}
     \includegraphics[height=6cm]{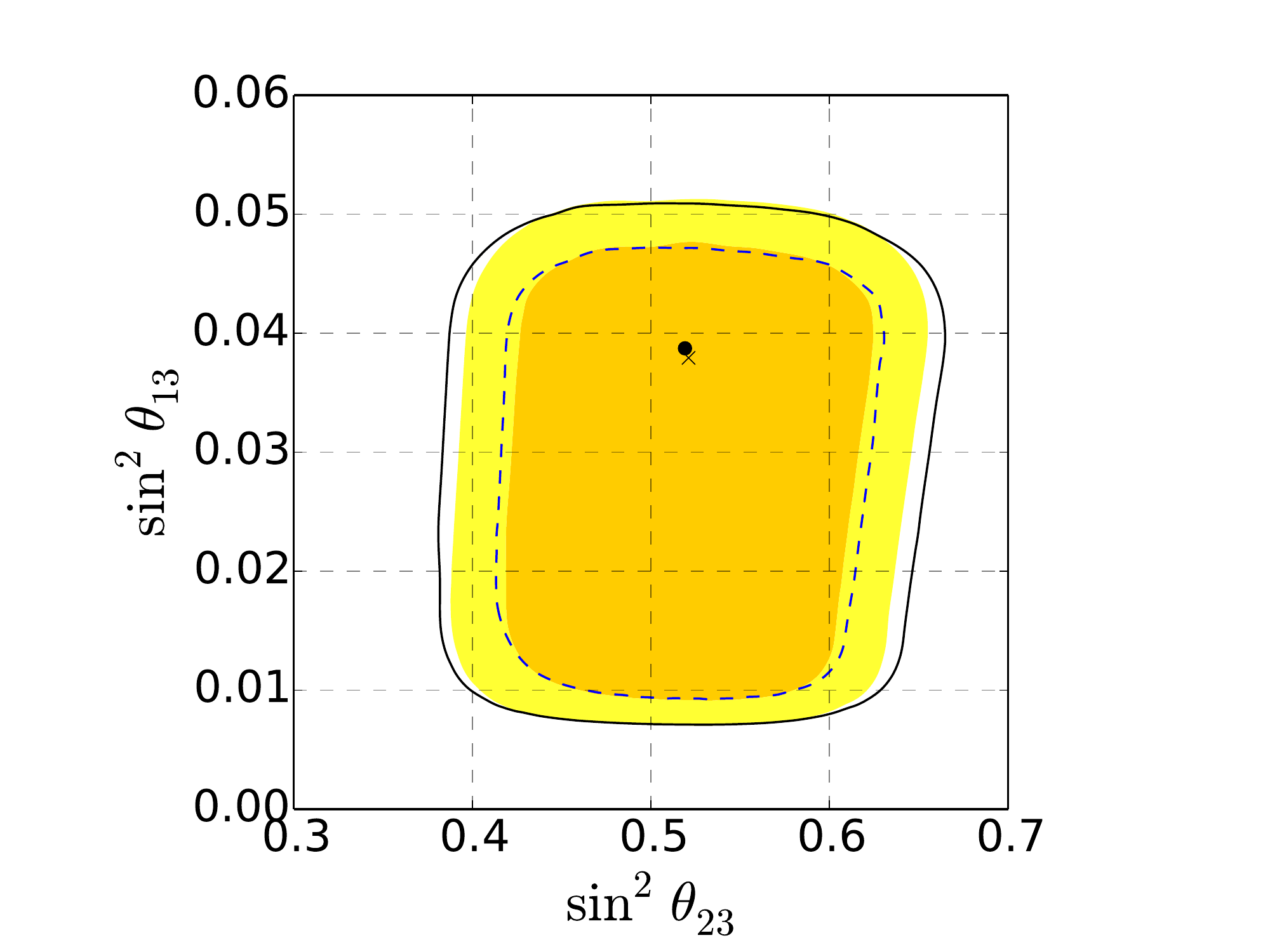}}%
     \vspace{-0.4cm}
        \subfigure{%
\includegraphics[height=6cm]{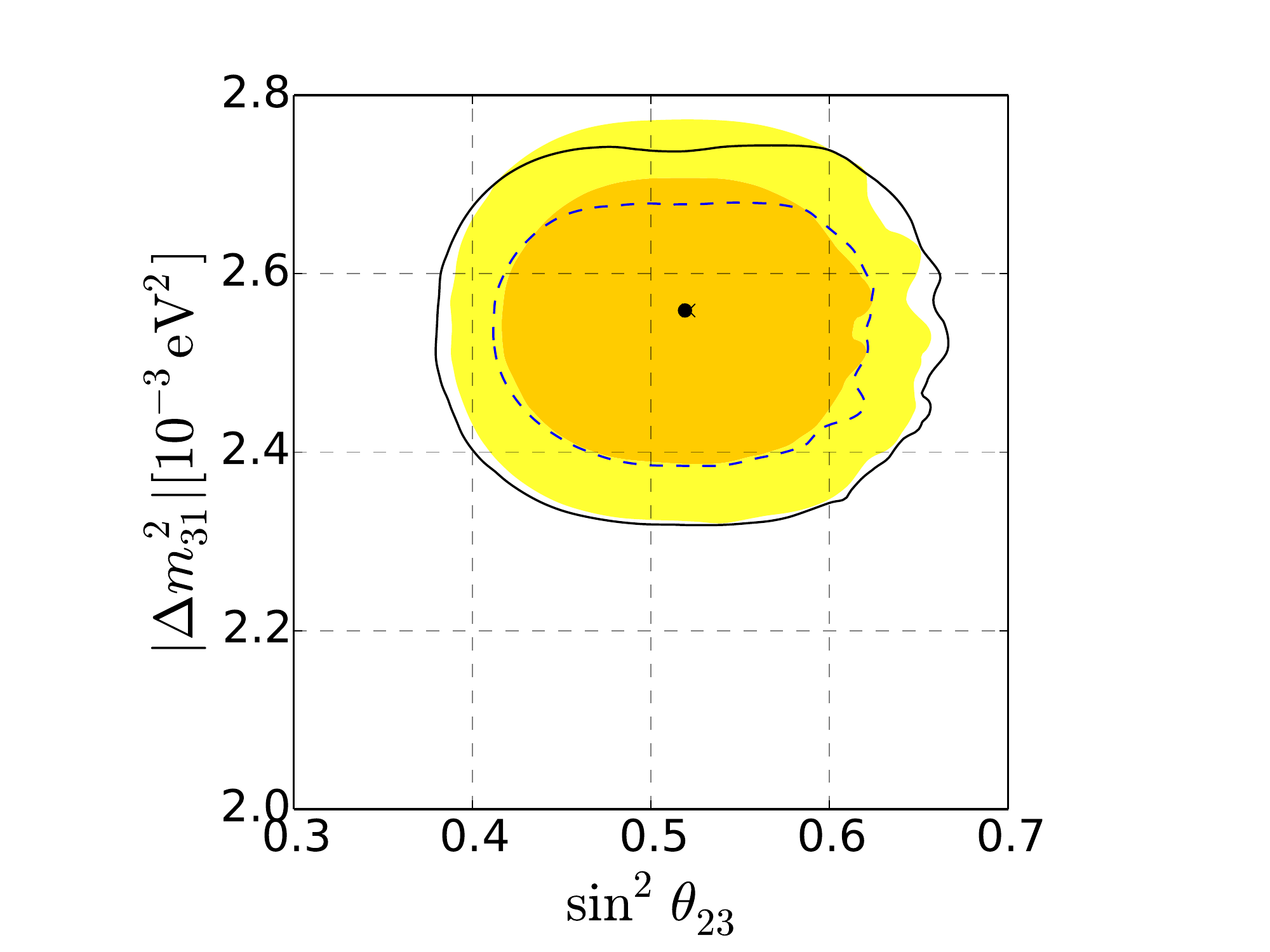}}%
        \subfigure{%
\hspace{-2.4cm}
     \includegraphics[height=6cm]{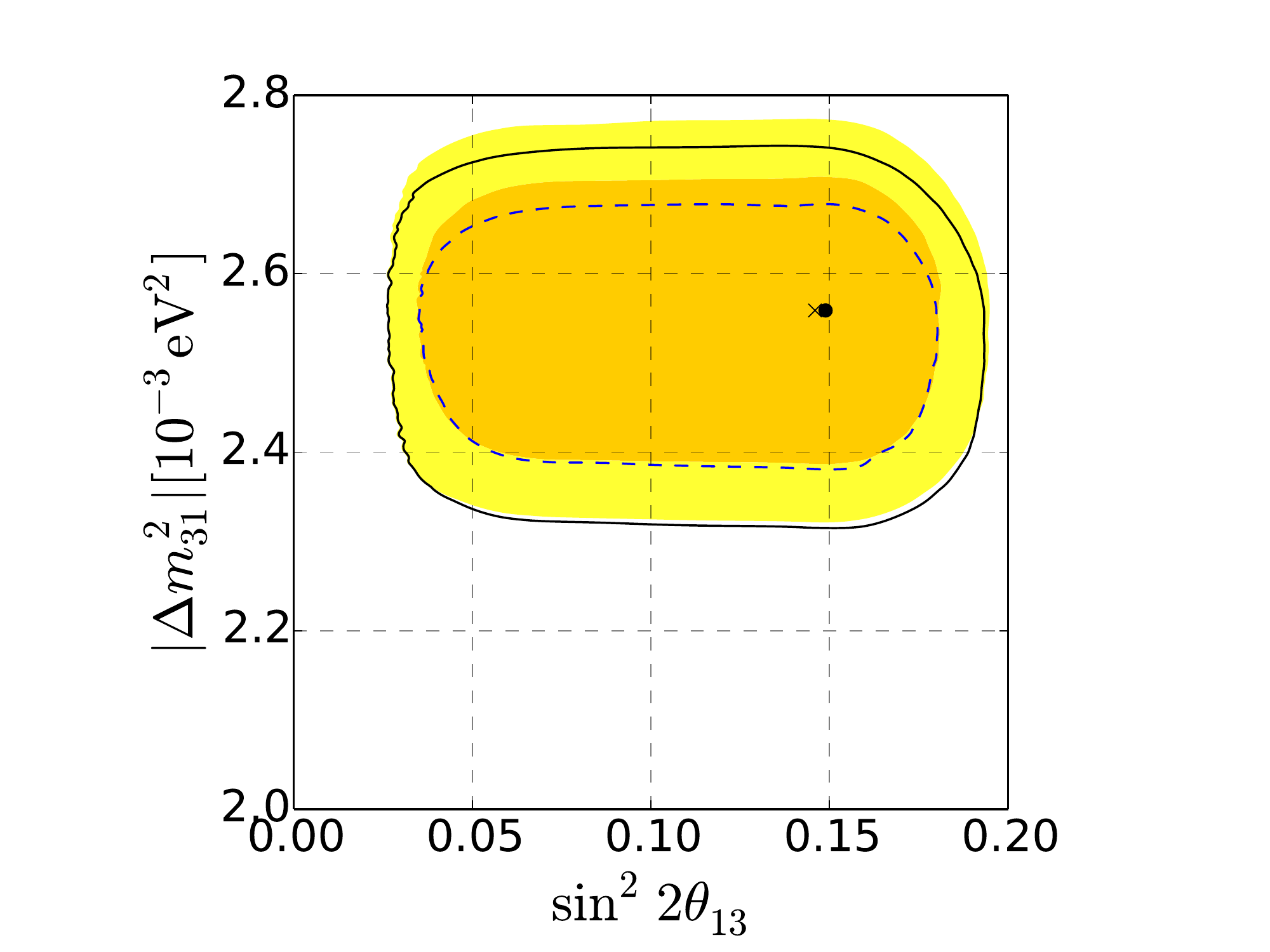}}%
     \caption{ \it Same as in Fig.~\ref{fig:LED1}  but
     using the NSI oscillation probabilities.
}
\label{fig:NSI1}
\end{figure*}

The modification of the relevant transition probabilities due to the presence of NSI and LED parameters  
can result in a distortion of the allowed regions of the standard neutrino mixing parameters.
In order to quantify such effects, we repeat the previous fit on the T2K and Daya Bay data, using the modified 
expressions of the transition probabilities in the
$\overline \nu_e \rightarrow \overline \nu_e$,
$\nu_{\mu} \rightarrow \nu_e$ and
$\nu_{\mu} \rightarrow \nu_{\mu}$ channels, illustrated in Eqs.~(\ref{Eq:PNSIlimit})-(\ref{eq:probdismu}) and Eq.~(\ref{eq:ProbAmpNSI}).
For the sake of a more clear presentation, we limit
ourselves to $2\sigma$ and $3\sigma$ confidence level.
Our results are presented in Fig.~\ref{fig:LED1} for LED and in Fig.~\ref{fig:NSI1} for NSI models.
We use
the same conventions as in Fig.~\ref{fig:SM1} and give
the obtained best fit points and confidence level regions in Tab.~\ref{tabNufitNSI}
for NSI and in Tab.~\ref{tabNufitLED} for LED. 
For completeness, we also show in Appendix B
the one dimensional projections of $\Delta \chi^2 = \chi^2 - \chi^2_{min}$
as a function of the standard oscillation parameters $\sin^2 \theta_{13}$,
$\sin^2 \theta_{23}$, $\ldm$ and $\delta$.

%
%
%
%
The presence of LED parameters in the oscillation formulae does not affect too much
the shape of the contours (Fig.~\ref{fig:LED1}); in this respect,  
the importance of including the T2K data in our analysis is mainly visible in the determination of 
$\ldm$: in fact, in the analysis of the Daya Bay data only performed in \cite{Girardi:2014gna},
the 3$\sigma$ confidence region for $\ldm$ was roughly $5 \%$ larger with respect to the SM determination, whereas in the 
present  analysis this difference is reduced to roughly $1 \%$. 
In the NSI scenario the presence of the new
couplings $\eps_{\alpha \beta}$ enlarges the confidence
regions of the standard oscillation parameters and, in particular, reduces the hints 
for a maximal CP violation since the whole $[0,2\pi]$ range for $\delta$ is allowed at 2$\sigma$
confidence level. This effect is caused by the new sources of the CP violation, encoded in the 
unconstrained phases $\phi_{\alpha\beta}$ of Eq.~(\ref{Eq:PNSIlimit}) and Eqs.~(\ref{Eq:PNSIlimitP0})-(\ref{eq:NSIP1}).
A large effect is also found in the determination of the reactor angle $\sin^2 \theta_{13}$. Indeed, in the NSI case,
the 3$\sigma$ confidence region of $\sin^2 \theta_{13}$ is roughly twice as large as in the SM case 
(as can be observed from Tables~\ref{tabNufitSM}
and \ref{tabNufitNSI}). 
The main reason for such a behavior is the strong correlation among $\theta_{13}$ and the NSI parameters: 
for large enough $\varepsilon^s_{\mu e}$ and/or $\varepsilon_{e \mu,\tau}$ 
(and an appropriate choice of the related CP phases), huge cancellations can occur with the standard part of the 
probability, thus causing an increase of the allowed $\theta_{13}$; the opposite can also happen: positive interferences 
can decrease the expected value of the reactor angle \cite{Girardi:2014gna}.

\begin{table}[h!]
\centering
\renewcommand{\arraystretch}{1.2}
\begin{tabular}{|clccc|}
\hline
 \bf Parameter  & \bf Best-fit ($\pm 1\sigma$) & \phantom{123}  \bf 1$\sigma$ range \phantom{123} & \phantom{123} \bf 2$\sigma$ range \phantom{123} & \phantom{123} \bf 3$\sigma$ range \phantom{123}  \\ \hline
$ \Delta m^{2}_{31}~\text{ (NO)}  \; [10^{-3}\text{ eV}^2]$ \phantom{123} &$2.53^{+0.07}_{-0.05}$ & $2.48-2.60$ & $2.42-2.66$ & $2.37-2.73$ \\
$ |\Delta m^{2}_{31}|~\text{ (IO)}  \; [10^{-3}\text{ eV}^2]$ \phantom{123} &$2.54^{+0.07}_{-0.07}$ & $2.47-2.61$ & $2.41-2.65$ & $2.35-2.71$ \\
$\sin^2\theta_{23}/10^{-1}$ &$5.3^{+0.4}_{-0.5}$ & $4.8-5.7$ & $4.3-6.0$ & $4.1-6.3$\\
\phantom{$\sin^2\theta_{23}/10^{-1}$} & $5.3^{+0.4}_{-0.5}$ & $4.8-5.7$ & $4.4-6.0$ & $4.1-6.3$ \\
$\sin^2\theta_{13}/10^{-2}$ &$2.3^{+0.2}_{-0.4}$ & $1.9-2.5$ & $1.7-2.7$ & $1.4-3.0$ \\
\phantom{$\sin^2\theta_{13}/10^{-1}$} &$2.2^{+0.1}_{-0.1}$ & $2.1-2.3$ & $1.7-2.7$ & $1.3-2.9$ \\
$\delta/\pi$ &$1.47^{+0.35}_{-0.31}$ & $1.16-1.82$ & $0.00-0.21$ $\oplus$ $0.72-2.00$ & | \\
\phantom{$\delta/\pi$} &$1.59^{+0.30}_{-0.36}$ & $1.23-1.89$ & $0.00-0.28$ $\oplus$ $0.78-2.00$& | \\
\hline
\end{tabular}
\caption{\it Best fit and 1$\sigma$, 2$\sigma$ and 3$\sigma$ errors of the
standard oscillation parameters obtained in the combined fit of the 
Daya Bay and T2K data using the LED probabilities.
If two values are given, the upper one corresponds to neutrino
mass spectrum with normal ordering (NO)
and the lower one  to spectrum with inverted ordering (IO)
(see text for further details).
}
 \label{tabNufitLED}
\end{table}
\begin{table}[h!]
\centering
\renewcommand{\arraystretch}{1.2}
\begin{tabular}{|clccc|}
\hline
 \bf Parameter  &  \bf Best-fit ($\pm 1\sigma$)    & \phantom{123}  \bf 1$\sigma$ range \phantom{123} & \phantom{123} \bf 2$\sigma$ range \phantom{123} & \phantom{123} \bf 3$\sigma$ range \phantom{123}  \\ \hline
$ \Delta m^{2}_{31}~\text{ (NO)}  \; [10^{-3}\text{ eV}^2]$ \phantom{123} &$2.56^{+0.06}_{-0.09}$ & $2.47-2.62$ & $2.38-2.68$ & $2.31-2.74$ \\
$ |\Delta m^{2}_{31}|~\text{ (IO)}  \; [10^{-3}\text{ eV}^2]$ \phantom{123} &$2.56^{+0.09}_{-0.08}$ & $2.47-2.64$ & $2.39-2.71$ & $2.32-2.77$ \\
$\sin^2\theta_{23}/10^{-1}$ &$5.2^{+0.6}_{-0.8}$ & $4.6-5.8$ & $4.1-6.2$ & $3.8-6.5$\\
\phantom{$\sin^2\theta_{23}/10^{-1}$} &$5.2^{+0.6}_{-0.8}$ & $4.6-5.8$ & $4.2-6.1$ & $3.9-6.5$ \\
$\sin^2\theta_{13}/10^{-2}$ &$3.9^{+0.4}_{-2.6}$ & $1.3-4.3$ & $0.9-4.7$ & $0.7-5.1$ \\
\phantom{$\sin^2\theta_{13}/10^{-1}$} &$3.8^{+0.5}_{-2.6}$ & $1.2-4.3$ & $0.9-4.7$ & $0.7-5.1$ \\
$\delta/\pi$ &$1.65^{+0.55}_{-0.78}$ & $0.00-0.20$ $\oplus$ $0.87-2.00$ & | & | \\
\phantom{$\delta/\pi$} &$1.35^{+0.75}_{-0.49}$ & $0.00-0.10$ $\oplus$ $0.15-0.20$ & | & | \\
\phantom{$\delta/\pi$} &$\ $ &  $0.74-0.81$ $\oplus$ $0.86-2.00$& $\ $ & $\ $ \\
\hline
\end{tabular}
\caption{\it Best fit and 1$\sigma$, 2$\sigma$ and 3$\sigma$ errors of the
standard oscillation parameters obtained in the combined fit of the 
Daya Bay and T2K data using the NSI probabilities.
If two values are given, the upper one corresponds to neutrino
mass spectrum with normal ordering (NO)
and the lower one  to spectrum with inverted ordering (IO)
(see text for further details).
}
 \label{tabNufitNSI}
\end{table}

\subsection{Bounds on LED parameters}
\label{sec:boundsLED}

\begin{figure}[h!]
 \begin{center}
\subfigure{%
\includegraphics[height=7cm]{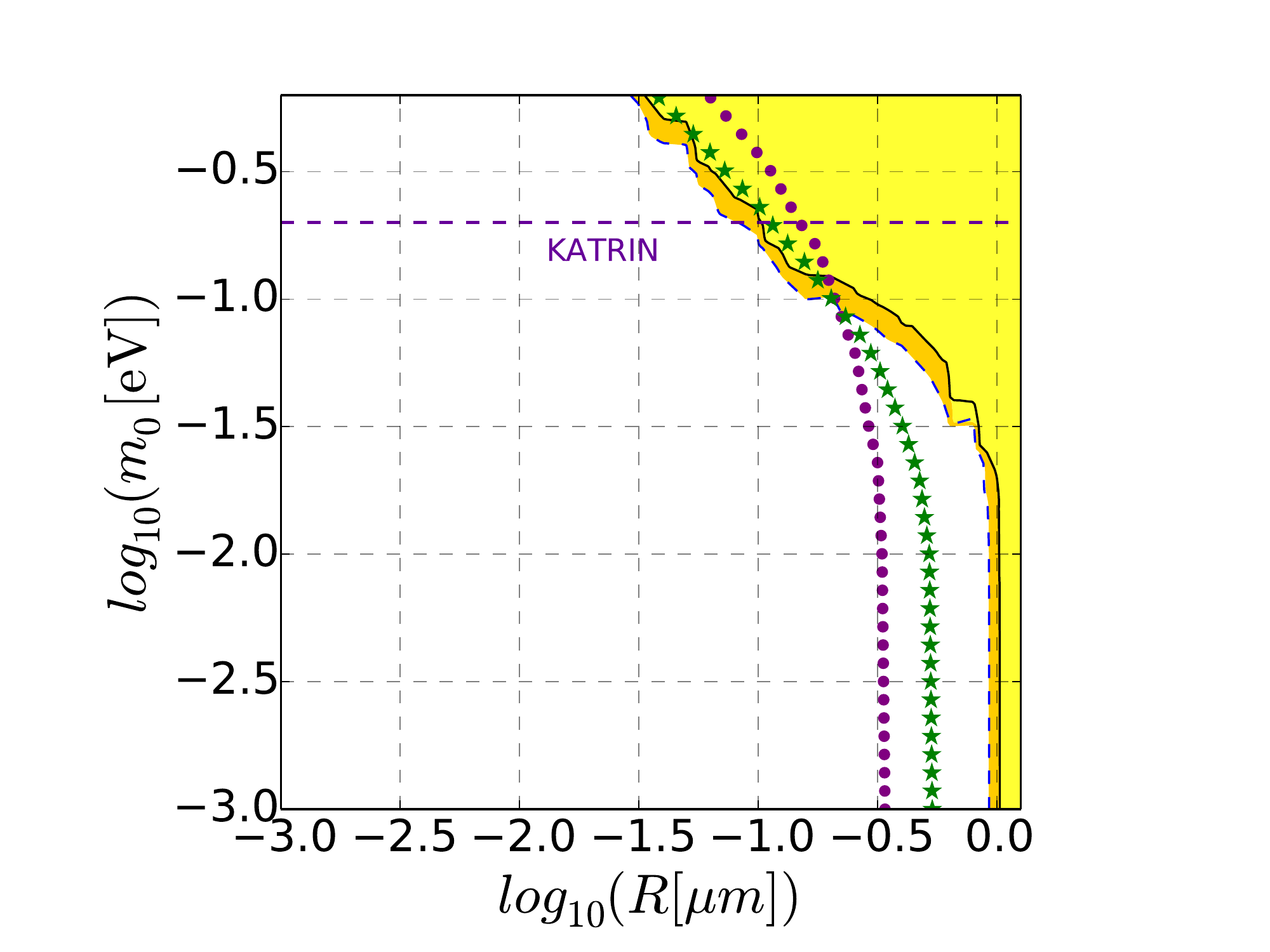}}%
\subfigure{%
\hspace{-2cm}
\includegraphics[height=7cm]{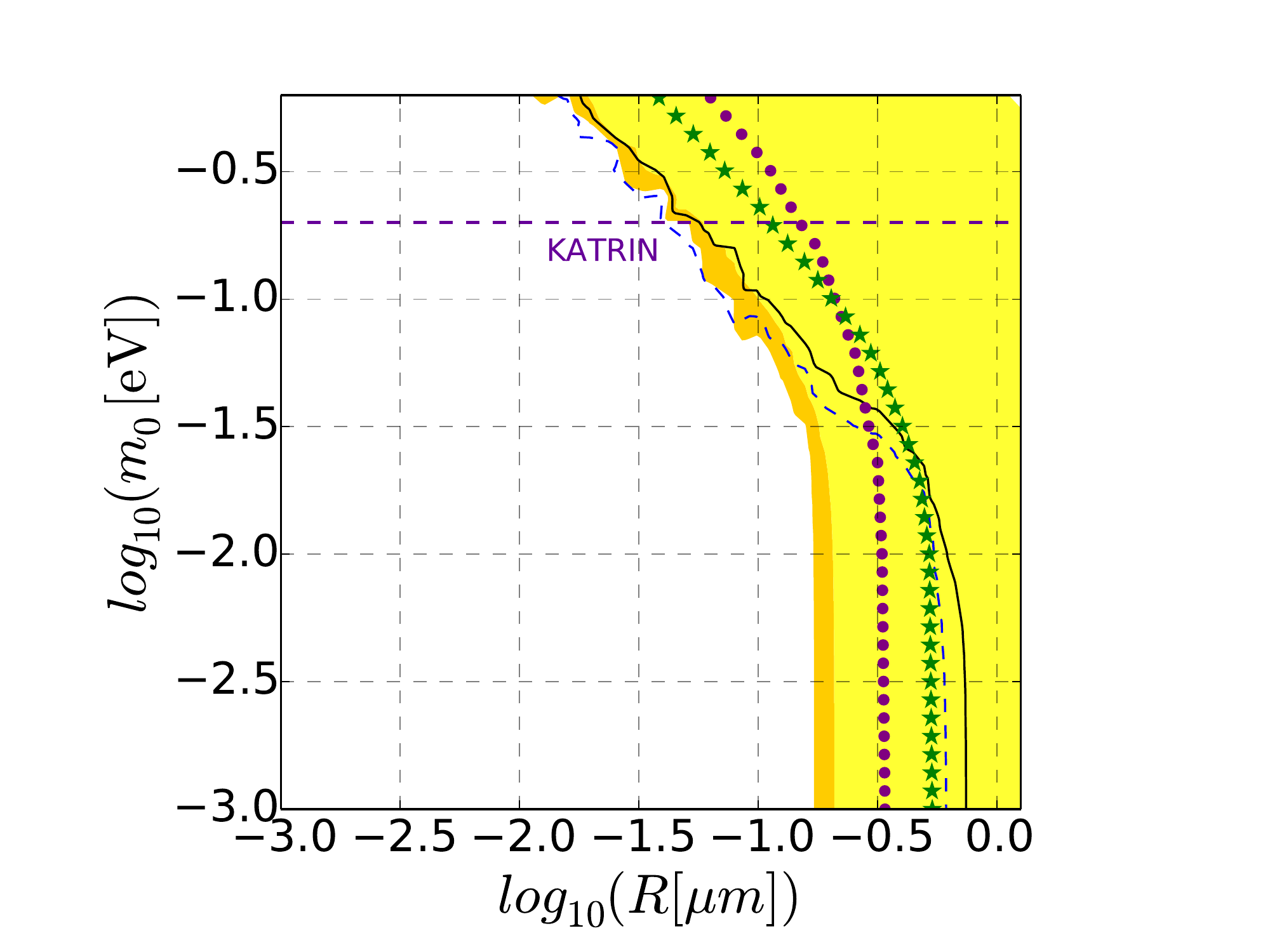}}%
    \end{center}
 \caption{\label{fig:m0R} \it Exclusion regions at 2$\sigma$ and 3$\sigma$ confidence levels for 1 dof 
in the $[\log_{10}( R ),\log_{10}( m_0 )]$-plane (LED model).  
Dashed and solid lines refer to NO, whereas the orange (gray) and yellow (light gray) regions are for IO
neutrino mass spectrum. Left panel: results obtained from the fit to the T2K data only.
Right panel: results obtained from the fit to the T2K $\oplus$ Data Bay data.
The circles and the stars represent the 2$\sigma$ bounds
obtained using the IceCube IC-40 and IC-79 data set \cite{Esmaili:2014esa}, respectively.}
\end{figure}

We now consider the bounds on $m_0$ and on the size of the largest extra dimension $R$. We 
perform a fit on the T2K data only (left panel of Fig.~\ref{fig:m0R}) and also show the results for a combined analysis
of the T2K and the Daya Bay data (right panel of Fig.~\ref{fig:m0R}).
In both the figures, the horizontal dashed line represents the expected sensitivity on the lightest neutrino mass from
KATRIN \cite{Eitel:2005hg}, whereas the
2$\sigma$ and 3$\sigma$ exclusion limits are represented with the dashed and solid lines
for NO and with the orange (gray) and yellow (light gray) regions for IO
neutrino mass spectrum. 
In addition, the circles and the stars indicate the 2$\sigma$ bounds (for 1 dof)
obtained using the IceCube IC-40 and IC-79 data set \cite{Esmaili:2014esa}, respectively,
from which we have the following constraints:
$R<0.54\, (0.34) \mu m$ using the IC-40 (IC-79) data set.

%
%
%
%

Using the T2K data only, we obtain the same upper bound,
$R \leq 0.93 \, (1.0) \ \mu{\rm m}$, for both normal and inverted orderings at 2$\sigma$(3$\sigma$) C.L., in agreement with our 
expectations, Eq.~(\ref{Amumu}).
However, it is not possible to give better constrains on $m_0$, a part from 
a small region for large enough $R$  where $m_0 \lesssim 0.1$ eV.  
The combined T2K and Daya Bay analysis gives better limits on $R$ only:
$R \leq 0.60\ \mu{\rm m}$ for NO (a bit worst than the IceCube limits) and $R \leq 0.17\ \mu{\rm m}$ for IO 
(roughly a factor of 2 better than the IceCube bounds).
\\
Since the combined analysis is clearly dominated by the Daya Bay experiment (due to the 
higher statistics of the $\bar \nu_e$ disappearance channel), the bounds on $R$ 
are similar to the ones given in \cite{Girardi:2014gna}. 

\subsection{Bounds on NSI parameters}
\label{sec:boundsNSI}
\begin{figure}[h!]
 \begin{center}
\hspace{0.1cm}
\subfigure{%
\includegraphics[height=6cm]{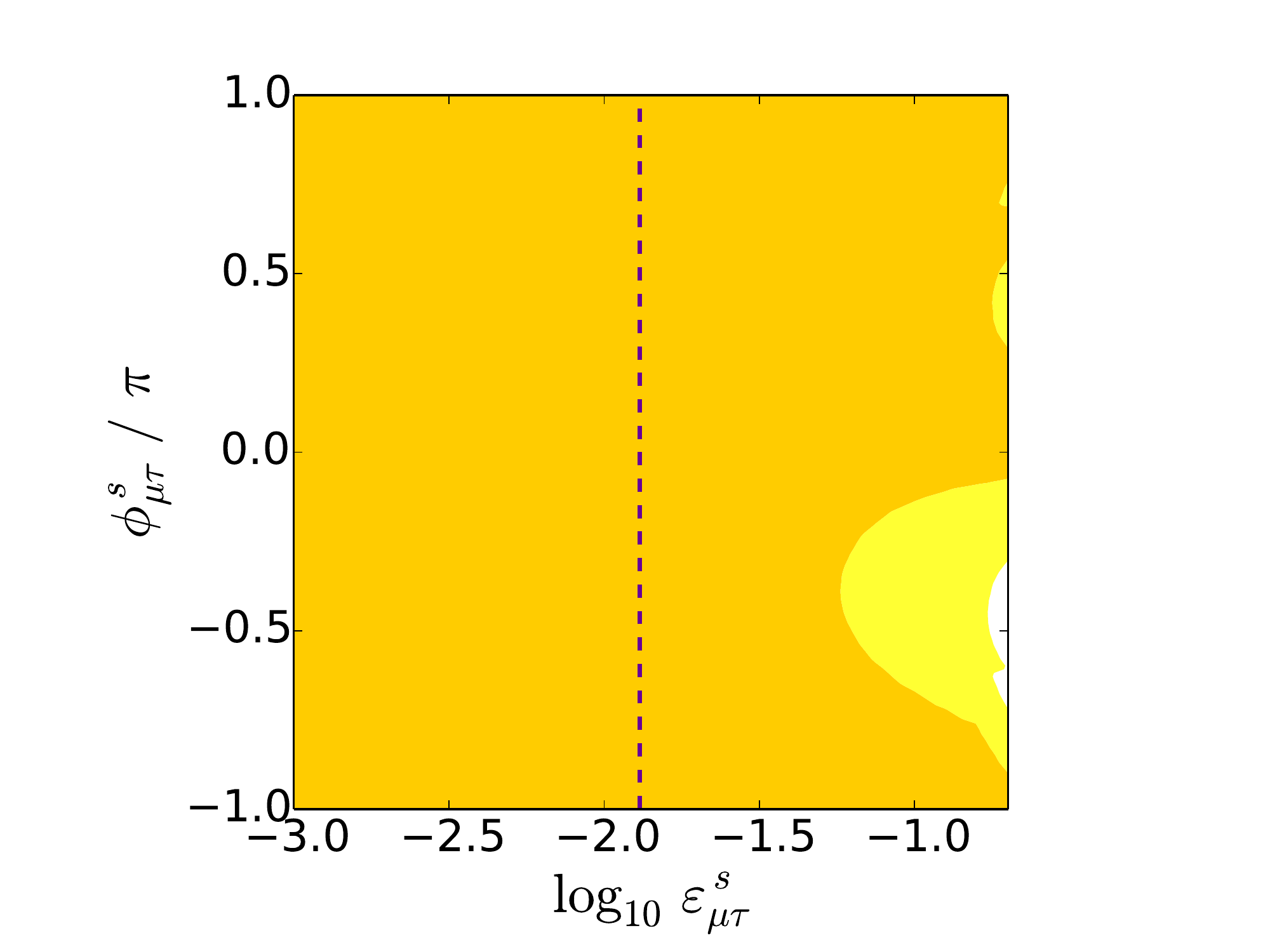}}%
\subfigure{%
\hspace{-2.1cm}
\includegraphics[height=6cm]{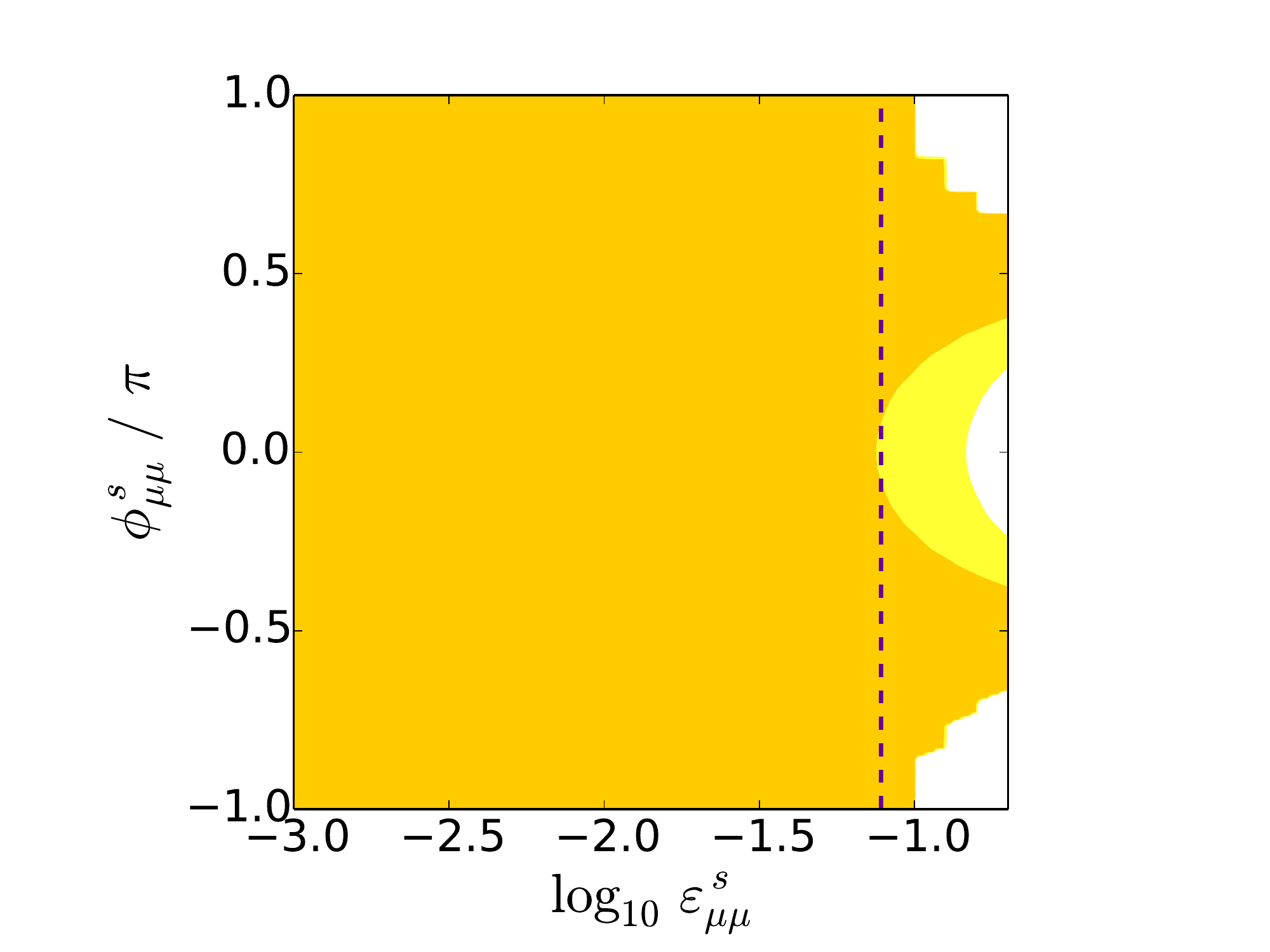}}%
    \end{center}
 \caption{\label{fig:boundsdelta0} \it Left panel: allowed regions in the $[\eps^s_{\mu \tau}, \phi^s_{\mu \tau}]$-plane at
 2$\sigma$ and 3$\sigma$ confidence level for 1 dof (the standard CP violation phase $\delta$ and the not shown NSI parameters are set to zero). 
The dashed vertical line at $\eps^s_{\mu \tau} = 0.013$ corresponds to the current constraint on $\eps^s_{\mu \tau}$
given in Eq.~(\ref{limitienr}).
 Right panel: the same as in the left panel but in the $[\eps^s_{\mu \mu}, \phi^s_{\mu \mu}]$-plane.
The dashed vertical line at $\eps^s_{\mu \mu} = 0.078$ corresponds to the current constraint on $\eps^s_{\mu \mu}$
given in Eq.~(\ref{limitienr}).
}
\end{figure}

Finally, we analyze the bounds on the new couplings $\eps_{\alpha \beta}$ but $\varepsilon_{ee}$, which has been already discussed in 
details in \cite{Girardi:2014gna} and for which the T2K data do not add any additional constraints.
Given the large numbers of new moduli and phases affecting the transition probabilities,
sensitive limits can only be put under the supplementary hypotheses of fixed parameters. We have checked that
no relevant bounds can be obtained on the various $|\eps_{\alpha \beta}|$ if we marginalize the $\chi^2$ function over 
all the other parameters.
In Fig.~\ref{fig:boundsdelta0} we show the 2$\sigma$ and
3$\sigma$ confidence regions for the $\eps^s_{\mu\tau}$ and $\eps^s_{\mu\mu}$ parameters, which 
enter the $\nu_\mu \to \nu_\mu$ probability, 
obtained setting to zero the standard CP phase and all the NSI parameters not shown in the plots 
(results obtained for other fixed values of $\delta$
are similar to the case of $\delta = 0$, due to its $\theta_{13}$-suppressed dependence 
shown in Eq.~(\ref{eq:Pmumu})).
As we can see, the obtained bounds are weaker than those of Eq.~(\ref{limitienr}) and
so not particularly interesting.

\begin{figure}[h!]
 \begin{center}
\subfigure{%
\hspace{-1.4cm}
\includegraphics[height=6cm]{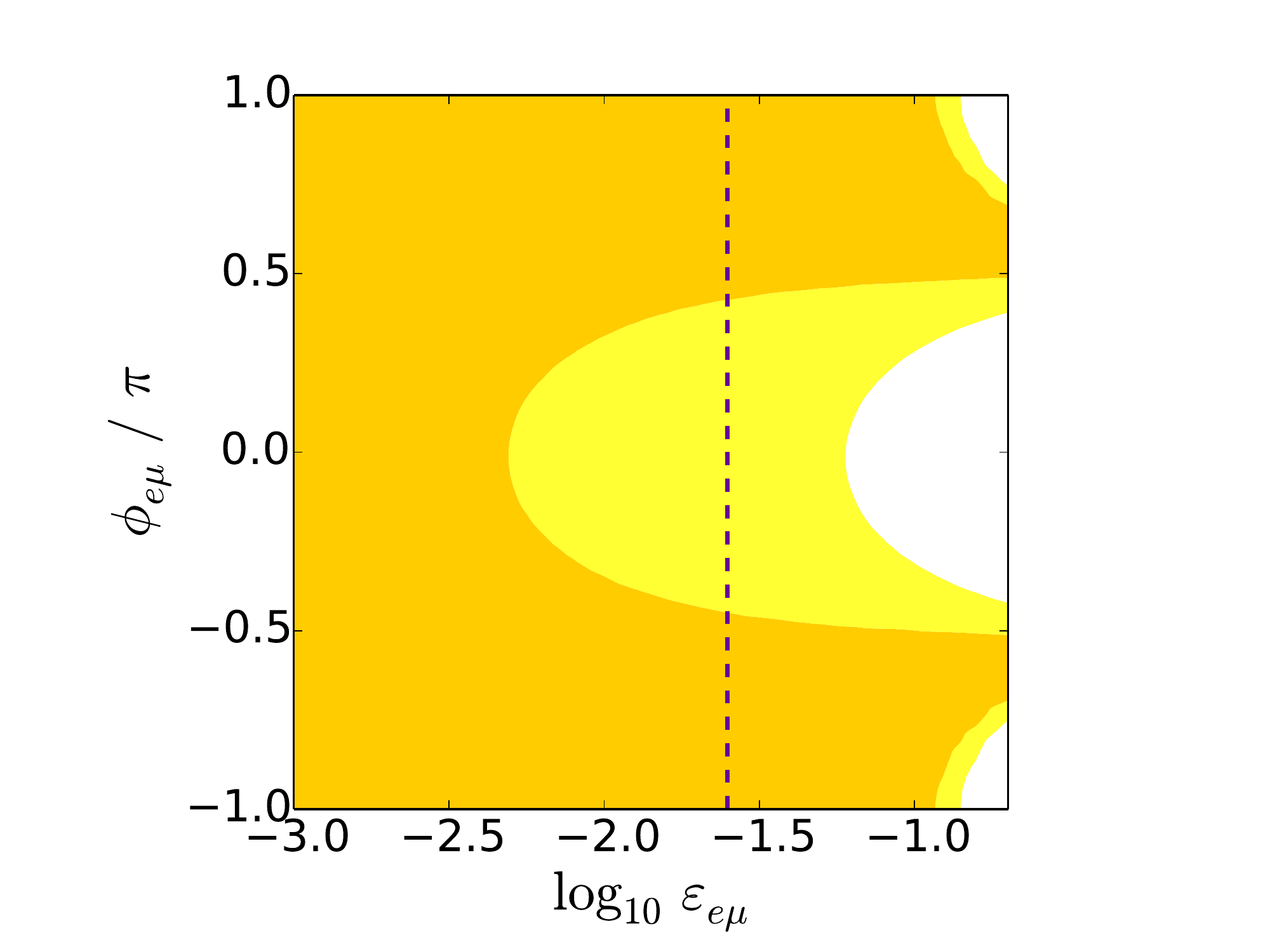}}%
\subfigure{%
\hspace{-2.2cm}
\includegraphics[height=6cm]{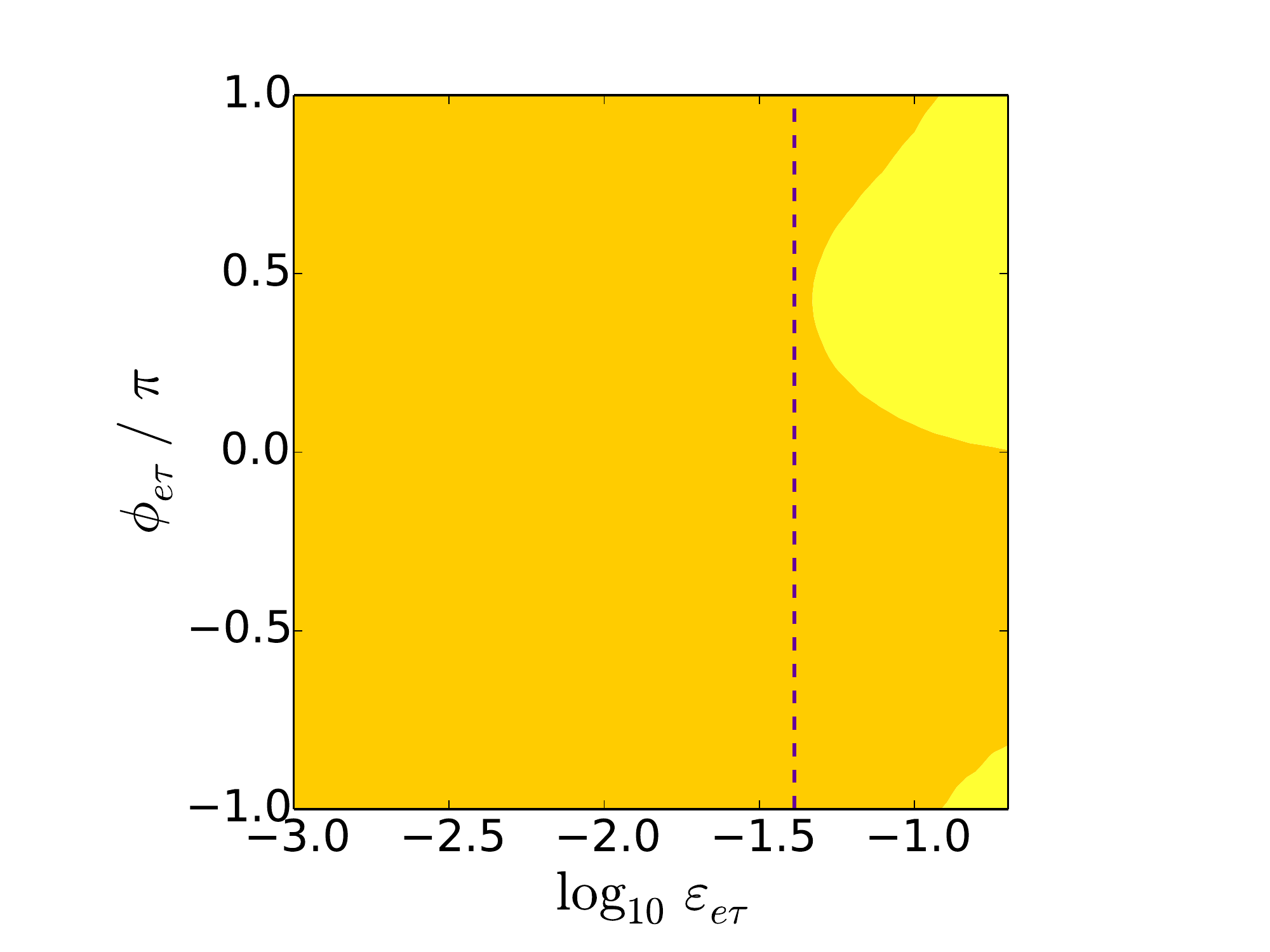}}%
\subfigure{%
\hspace{-2.2cm}
\includegraphics[height=6cm]{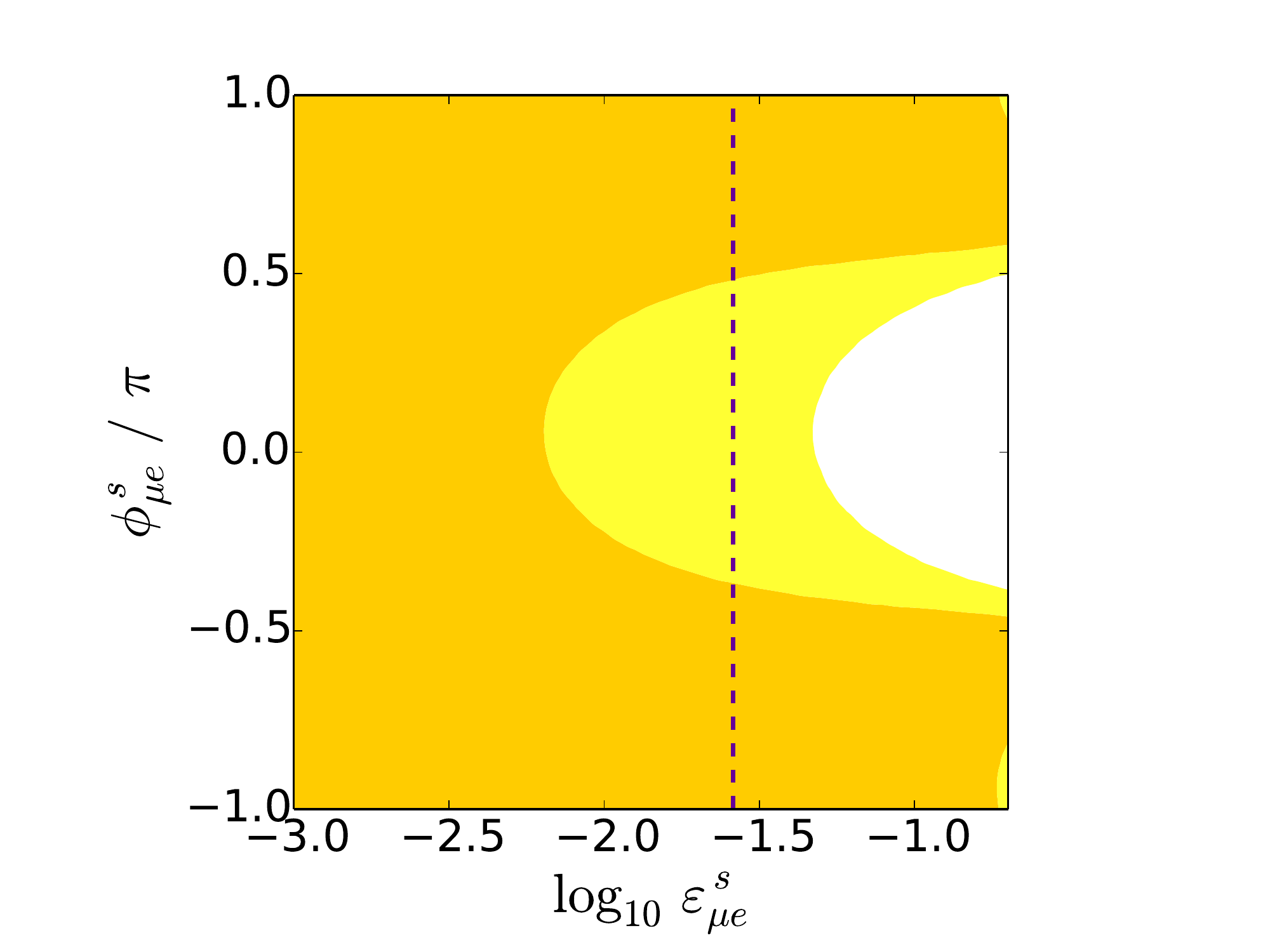}}%
    \vspace{-0.4cm}
    \subfigure{%
\hspace{-1.4cm}
\includegraphics[height=6cm]{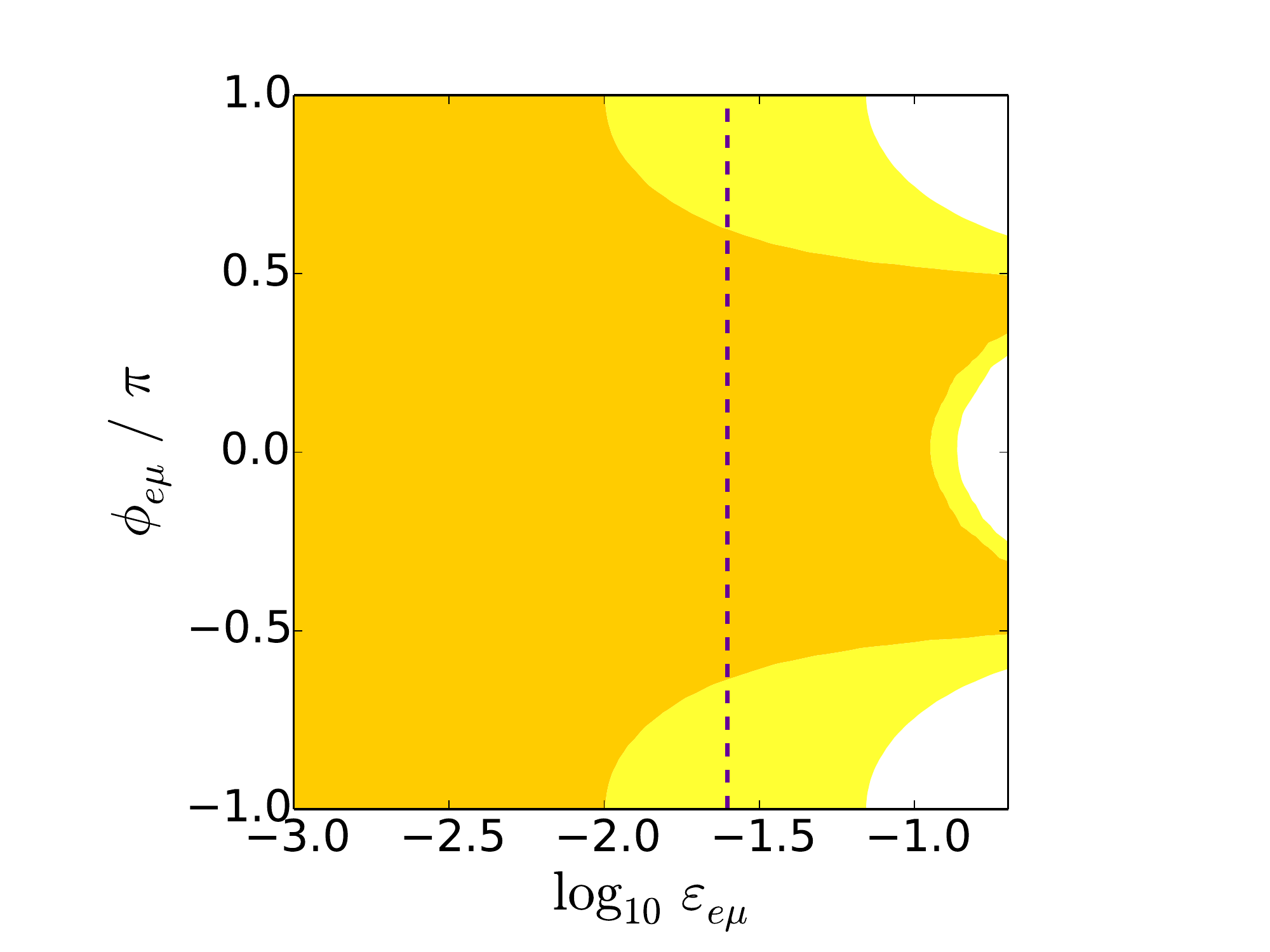}}%
\subfigure{%
\hspace{-2.2cm}
\includegraphics[height=6cm]{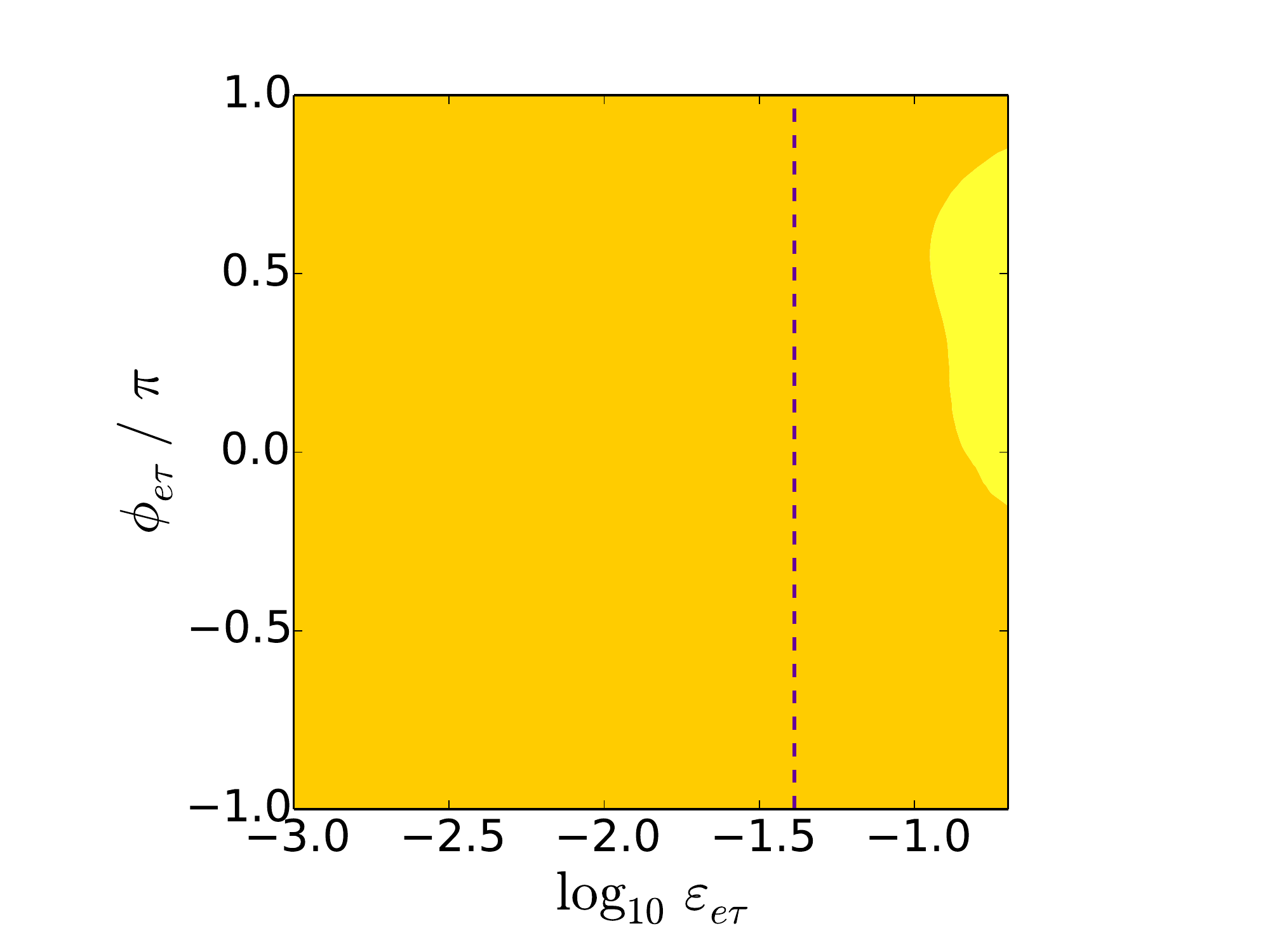}}%
\subfigure{%
\hspace{-2.2cm}
\includegraphics[height=6cm]{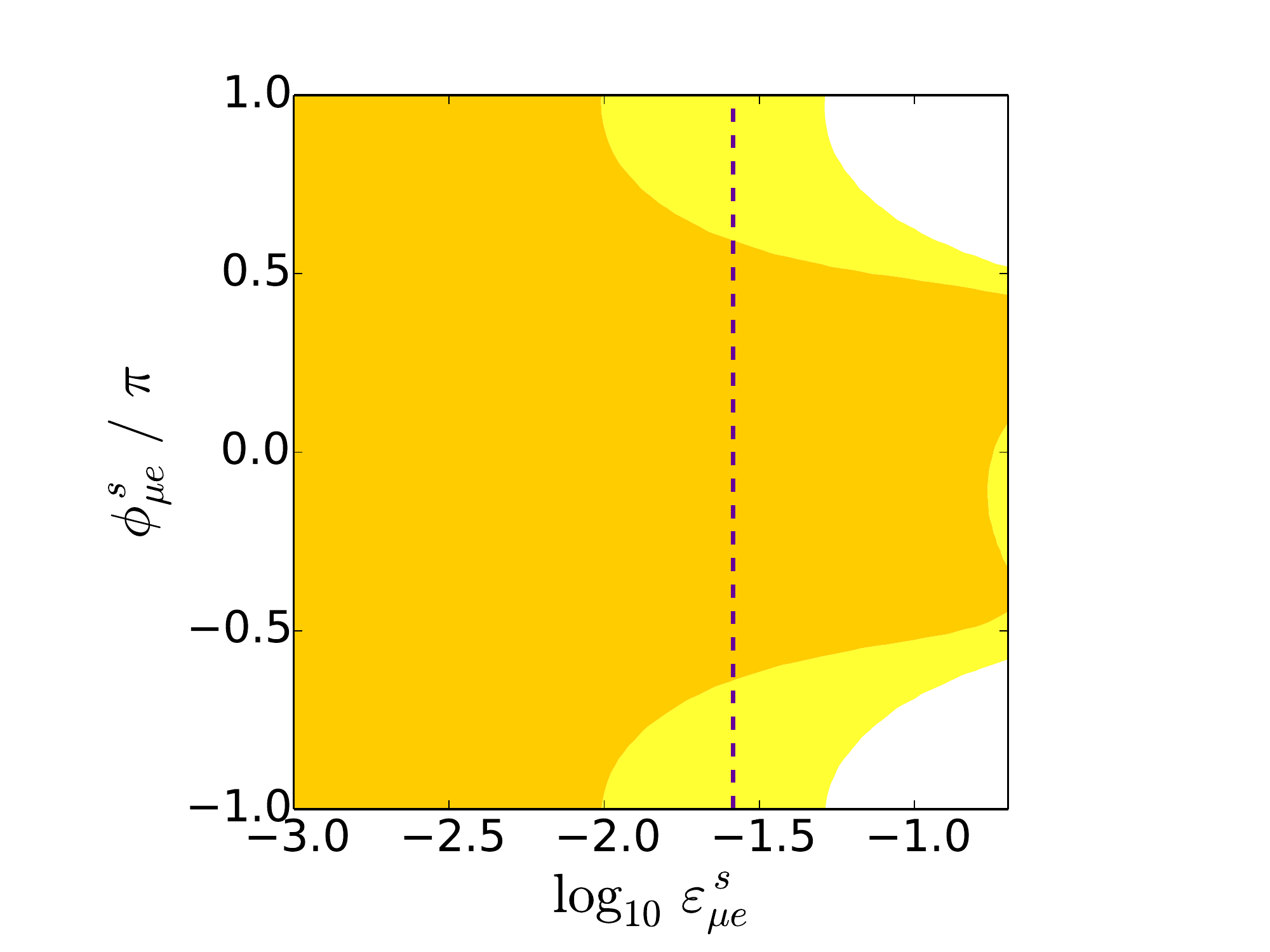}}%
\vspace{-0.4cm}
    \subfigure{%
\hspace{-1.4cm}
\includegraphics[height=6cm]{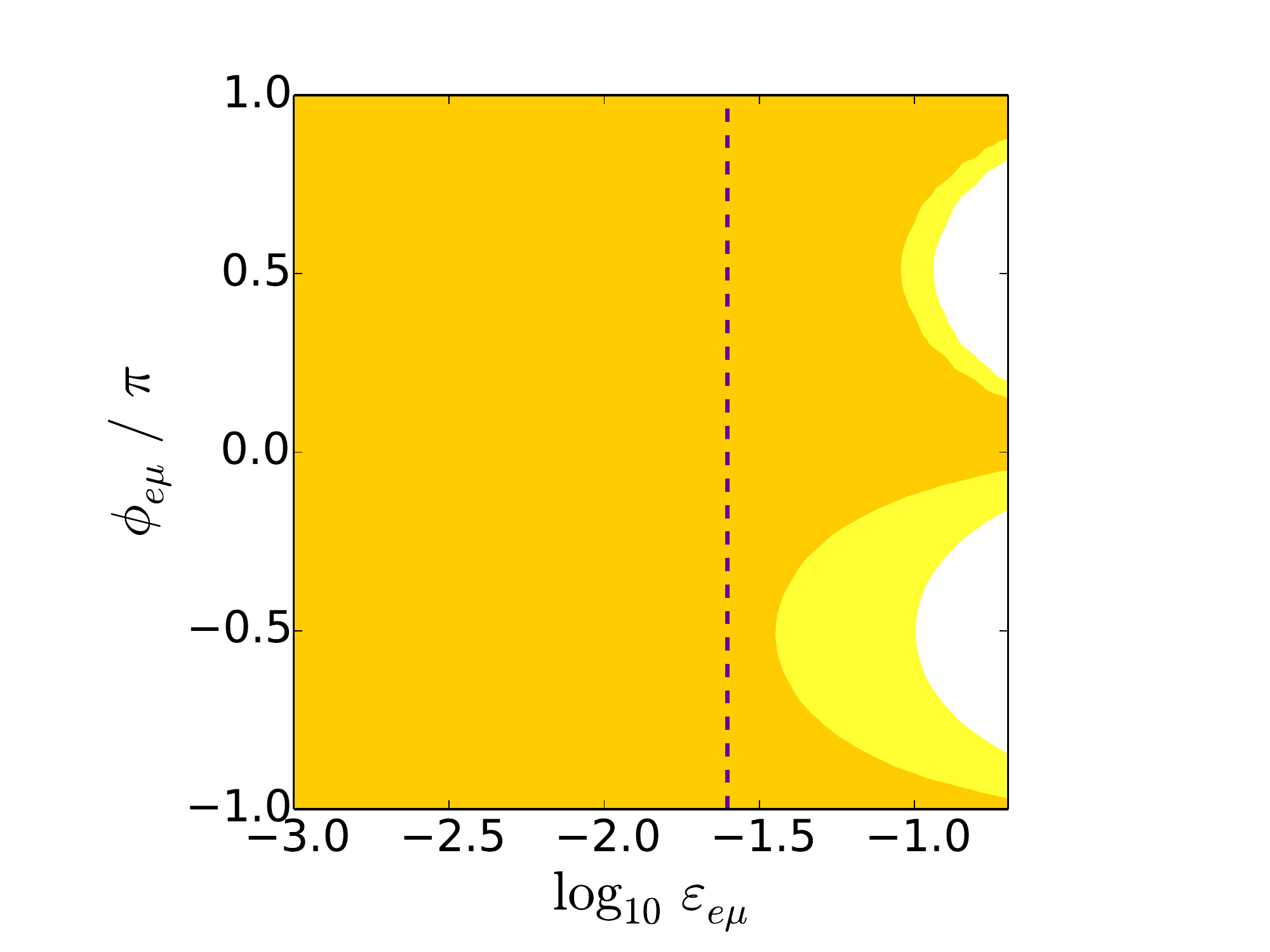}}%
\subfigure{%
\hspace{-2.2cm}
\includegraphics[height=6cm]{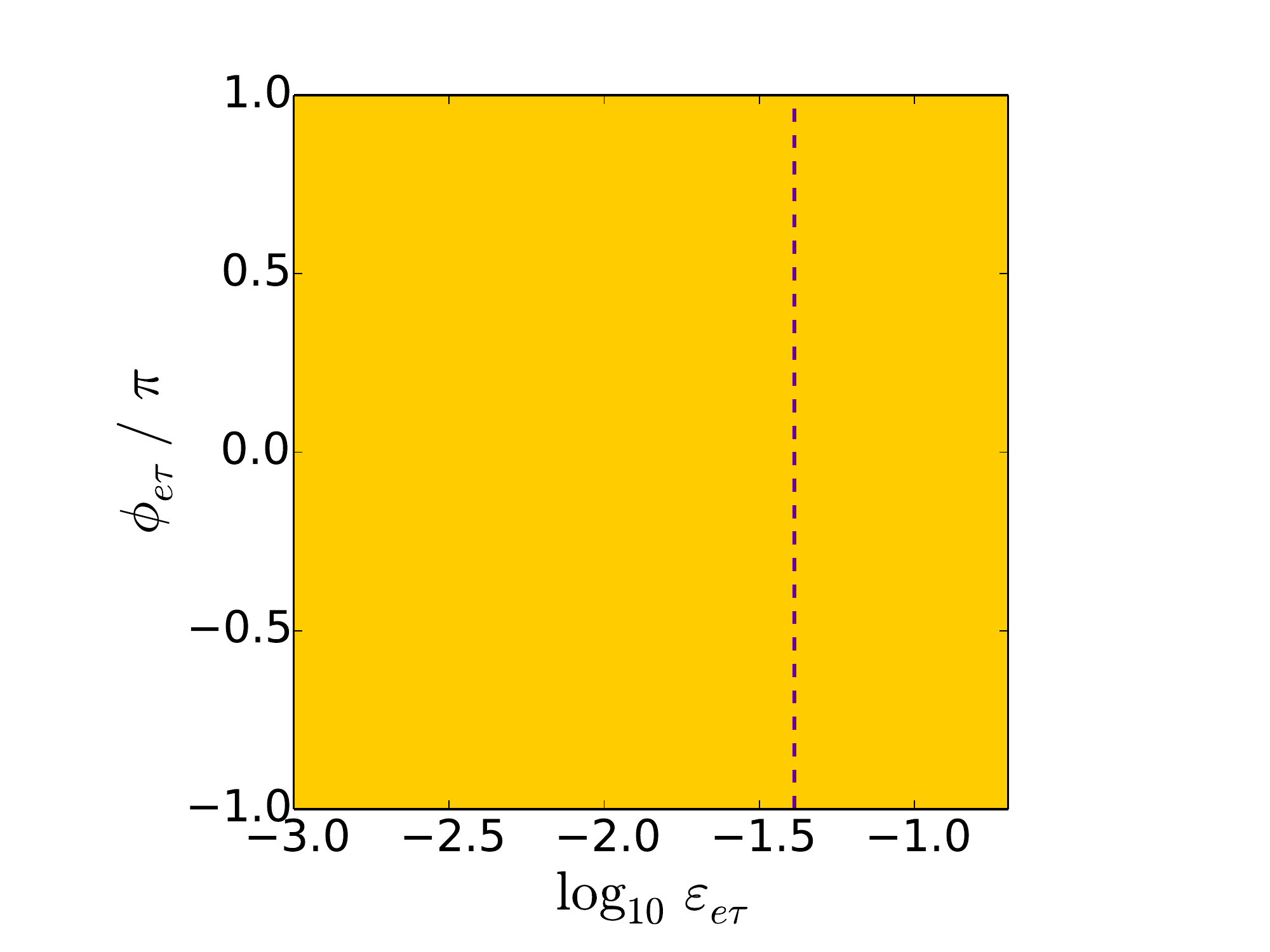}}%
\subfigure{%
\hspace{-2.2cm}
\includegraphics[height=6cm]{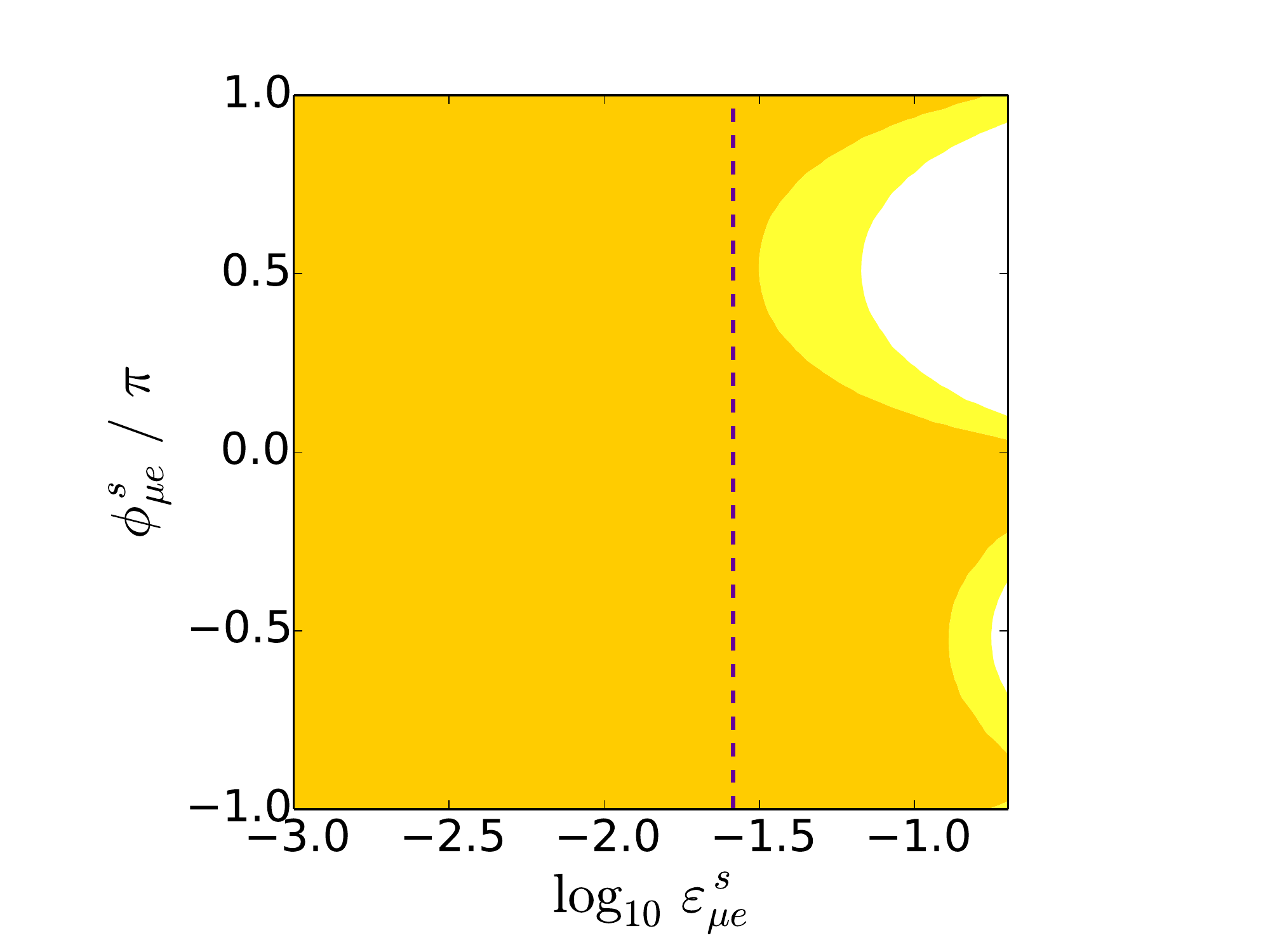}}%
    \end{center}
 \caption{\label{fig:boundsNSI} \it Allowed regions in the $[\eps, \phi]$-plane at
 2$\sigma$ and 3$\sigma$ confidence level for 1 dof and
 $\delta = 0 , \pi , 3 \pi / 2$ in the upper, middle and lower panels, respectively.
 The vertical lines are $\eps_{e\mu} = 0.025$, $\eps_{e\tau} = 0.041$
 and $\eps^s_{\mu e} = 0.026$ (the not shown NSI parameters are set to zero)
 correspond to the current constraints given in Eq.~(\ref{limitienr}).}
\end{figure}

The bounds on $\eps_{e\mu}, \eps_{e\tau}$
 and $\eps^s_{\mu e}$ are  shown in Fig.~\ref{fig:boundsNSI},  for 
$\delta = 0$ (upper panel), $\delta = \pi$ (middle panel) and $\delta =  3\pi/2$ (lower panel).
As it can be seen, the obtained bounds for the absolute values of $\varepsilon_{\alpha \beta}$  
are modulated by the relative phases $\phi_{\alpha \beta}$:
for example, for $\delta = 0$ and $\phi_{e \mu}/\pi \sim 0$
we get $\varepsilon_{e \mu} < \mathcal{O}(10^{-3})$ at $2\sigma$ C.L.,
a bit stronger than the model independent limit derived in \cite{enrique},
whereas for  $\phi_{e \mu}/\pi \sim 1/2$ we have no bound whatsoever.
In Table \ref{tab:boundsNSI} we summarize the bounds
on the NSI parameters obtained for particular choices of the
phases $\phi_{\alpha \beta}$ and $\delta$.
Since the parameter $\eps_{e\tau}$ cannot be better constrained by the current data, we 
do not present the obtained upper bounds in the table. 
\begin{table}[h!]
\centering
\renewcommand{\arraystretch}{1.2}
\begin{tabular}{|lcc|lcc|}
\hline
$\phi_{\alpha \beta} / \pi$ &  \phantom{123} $\delta$ \phantom{123} &  \bf Upper bound ($2\sigma$ C.L.) & $\phi_{\alpha \beta} / \pi$ &  \phantom{123} $\delta$ \phantom{123}  &  \bf Upper bound ($2\sigma$ C.L.)\\
\hline
$\phi_{e\mu}/\pi \sim 0$  & $0$    & $4.85 \times 10^{-3} $ & $\phi^s_{\mu e}/\pi \sim 0$ & $0$ & $6.28 \times 10^{-3} $\\
$\phi_{e\mu}/\pi \sim 1.0$ & $\pi$   & $9.94 \times 10^{-3} $ & $\phi^s_{\mu e}/\pi \sim 1.0$ & $\pi$ & $9.96 \times 10^{-3} $\\
$\phi_{e\mu}/\pi \sim -0.5$ & $3\pi/2$  & $3.50 \times 10^{-2} $ & $\phi^s_{\mu e}/\pi \sim 0.5$ & $3\pi/2$ & $3.12 \times 10^{-2} $\\
\hline
\end{tabular}
\caption{\it Upper bounds on the parameters $\eps_{e\mu}$ and $\eps^s_{\mu e}$ at $2\sigma$ C.L. for
particular choices of the phases $\phi_{e\mu}$, $\phi^s_{\mu e}$ and $\delta$,
obtained from Fig.~\ref{fig:boundsNSI}.
}
 \label{tab:boundsNSI}
\end{table}

\section{Summary and Conclusions}

In this paper we have analyzed the recent appearance \cite{Abe:2013hdq} and disappearance \cite{Abe:2014ugx} data
of T2K experiment and the disappearance data \cite{An:2013zwz} of the Daya Bay reactor experiment to constrain the parameter space of
two models of physics beyond the SM, namely the non standard neutrino interactions 
and large extra dimensions models, and to quantify the impact of this kind of new physics on the
determination of the standard oscillation parameters.
\\
While the impact of LED on the best fit values and 1$\sigma$ errors of the standard oscillation parameters is 
almost negligible (the largest difference 
is found for $\sin^2\theta_{13}$ where the 3$\sigma$ LED confidence region is almost $10 \%$ larger 
than the standard model),  this is not the case for 
the NSI scenario, where  particularly the allowed values of $\theta_{13}$ and $\delta$ are different from the 
standard determination. Indeed the 1$\sigma$ confidence region
for $\theta_{13}$ is roughly six times larger than the standard model analysis.
The situation is similar for the phase $\delta$, where the presence of new phases from the NSI
complex couplings  $\eps_{\alpha \beta}$  reduces the sensitivity with respect to the standard physics. 
In fact, although the best fit is still around the standard solution $\delta \sim 3 \pi / 2$ 
(as found in \cite{Capozzi:2013csa, GonzalezGarcia:2012sz}), the presence of  NSI effects 
makes this value statistically less significant.

As for the  bounds 
on the parameters of the LED and NSI models, we have found the following interesting results:
\begin{itemize}
 \item the strongest 2$\sigma$ C.L. bounds are for $\varepsilon_{e\mu} < 4.85 \times 10^{-3}$ 
(for $\delta \sim 0$) and $\varepsilon_{\mu e}^s < 6.28 \times 10^{-3}$ (for $\delta \sim 0$);
\item for the radius of the largest extra dimension $R$ we obtain $R \leq 0.60\ \mu{\rm m}$ (NO) 
and $R \leq 0.17\ \mu{\rm m}$ (IO) at 2$\sigma$ confidence level, similarly to \cite{Girardi:2014gna}.
\end{itemize}

Following the discussion of the previous section, the current  bounds on the NSI parameters are expected to be improved after a 
better determination of the standard CP phase $\delta$. For the LED parameters, an effort must be done in order to constrain 
the absolute mass $m_0$ and, consequently, the value of $R$.

\section{Acknowledgments}

We thank A. Esmaili, O. L. G. Peres and Z. Tabrizi for providing us the $\chi^2$ function of their LED analysis with
the IceCube data.
We acknowledge MIUR (Italy) for
financial support under the program Futuro in Ricerca 2010 (RBFR10O36O) (D.M. and A.D.I.). 
This work was supported in part by the INFN program on
``Astroparticle Physics'' and  by the European Union FP7-ITN INVISIBLES
(Marie Curie Action PITN-GA-2011-289442-INVISIBLES) (I.G.).

\section{Appendix A: approximate formulae LED}
\label{AppendixA}
For the sake of completeness,  we give here the approximate formulae
of the eigenvalues and the rotation matrices $U_i^{0n}$ for the LED
model.
Following the results of \cite{Davoudiasl:2002fq},
we can write the eigenvalues equation for $\lambda_i = m_i^{(n)} R$ as:
\be
\lambda_i - \frac{\pi}{2} \xi_i^2 \cot(\pi \lambda_i) = 0,
\ee
with $\xi_i \equiv \sqrt{2} m_i^D R$. In the case of small $\xi_i$ ($R^{-1} \gg m_j^D$), we  get:
\be
\lambda_i^{(0)} = \frac{1}{\sqrt{2}} \xi_i - \frac{1}{12 \sqrt{2}} \pi^2 \xi_i^3 + \frac{1}{90 \sqrt{2}} \frac{11}{16} \pi^4 \xi_i^5 + \mathcal{O}(\xi_i^7) \;,
\label{eq:lambda0}
\ee
\be
\lambda_i^{(k)} =k + \frac{1}{2 k} \xi_i^2 - \frac{1}{4 k^3} \xi_i^4 + \mathcal{O}(\xi_i^6) \;.
\label{eq:lambdan}
\ee
Given the rotation matrix $U_i^{0n}$\cite{Davoudiasl:2002fq}: 
\be
\left( U_i^{0n} \right)^2 = \frac{2}{1 + \pi^2 \xi_i^2 / 2 + 2 \lambda_i^{(n)2} / \xi_i^2} \;,
\ee
we can easily derive their expressions in terms of $\xi_i$:
\be
\begin{split}
\left(U_i^{00}\right)^2 & = 1 - \frac{\pi^2}{6} \xi_i^2 + \frac{\pi^4}{60}\xi_i^4 + \mathcal{O}(\xi_i^6)\;, \\
\left( U_i^{0k} \right)^2 & = \frac{\xi_i^2}{k^2} - \frac{3}{2} \frac{\xi_i^4}{k^4} + \frac{(10 - k^2 \pi^2)}{4 k^6} \xi_i^6 + \mathcal{O}(\xi_i^7) \mbox{ for $k>0$}\;.\\
\end{split}
\ee
Notice that the expansion parameter $\xi_i$ depends on the mass ordering and thus affects
both the neutrino eigenvalues and the rotation matrices. 
     
\section{Appendix B: One dimensional projections}
\label{AppendixB}
\begin{figure*}[h!]
\subfigure{%
 \includegraphics[height=6cm]{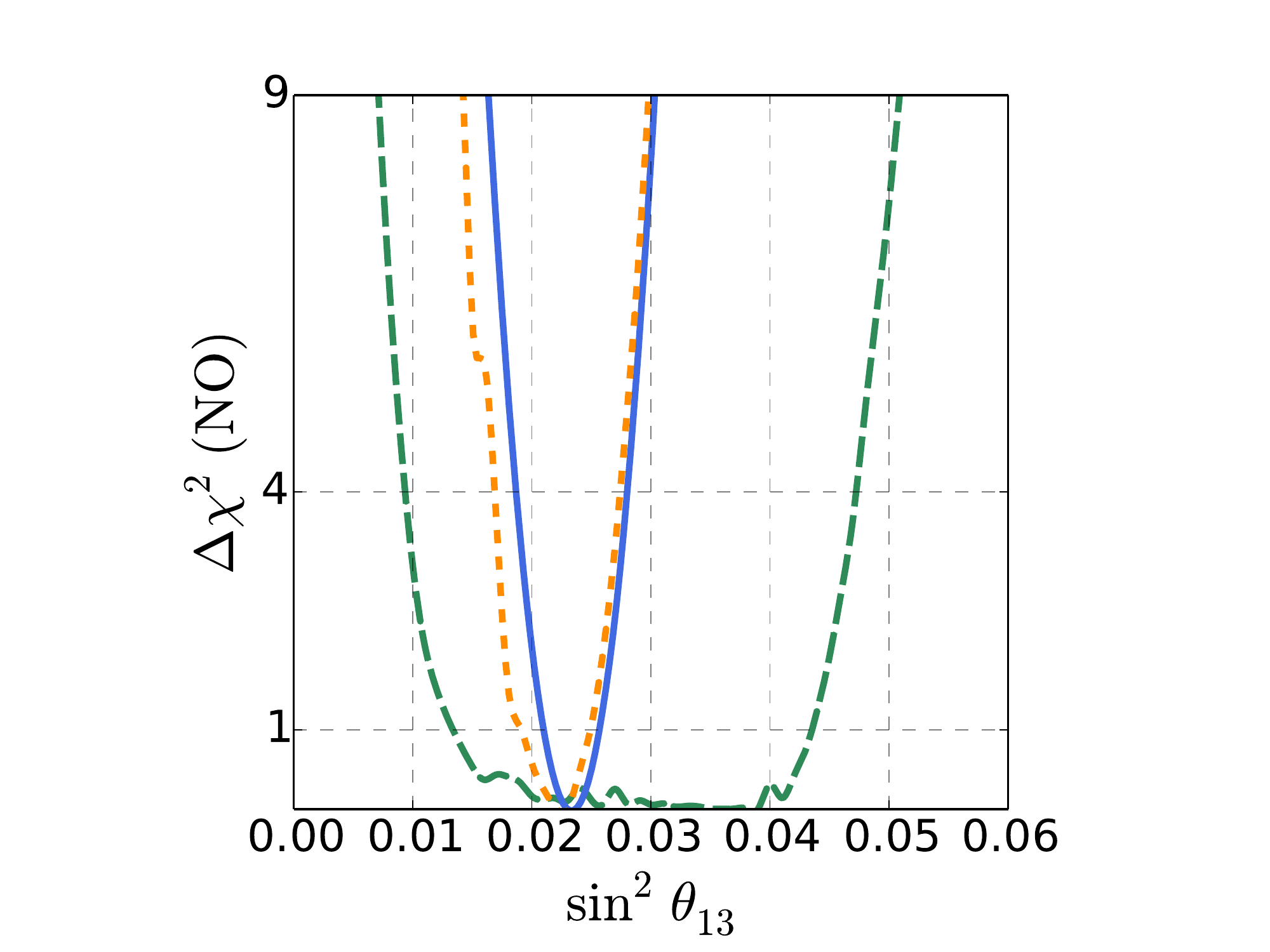}}%
\subfigure{%
\hspace{-2.6cm}
   \includegraphics[height=6cm]{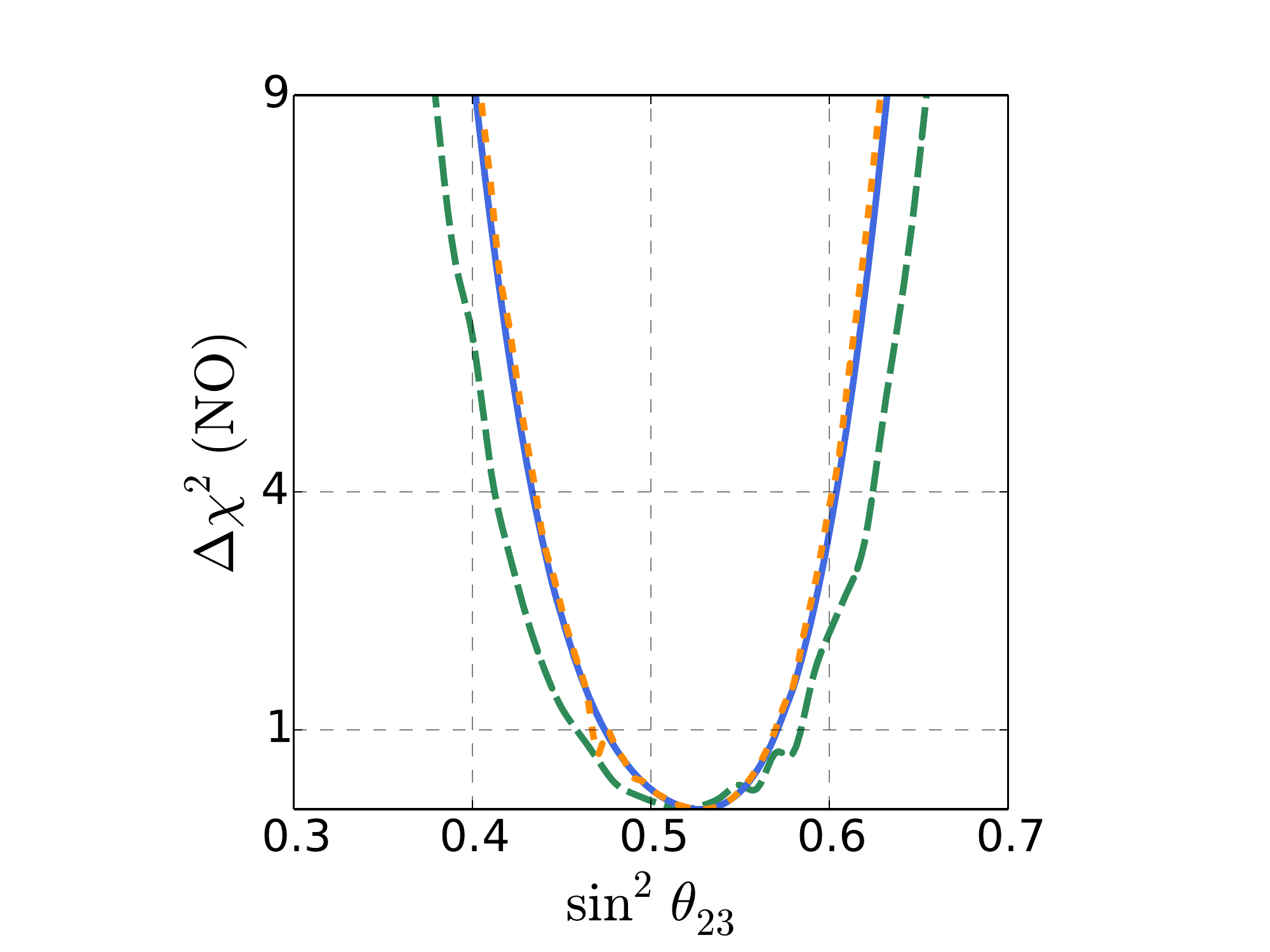}}%
        \vspace{-1cm}
   \subfigure{%
     \includegraphics[height=6cm]{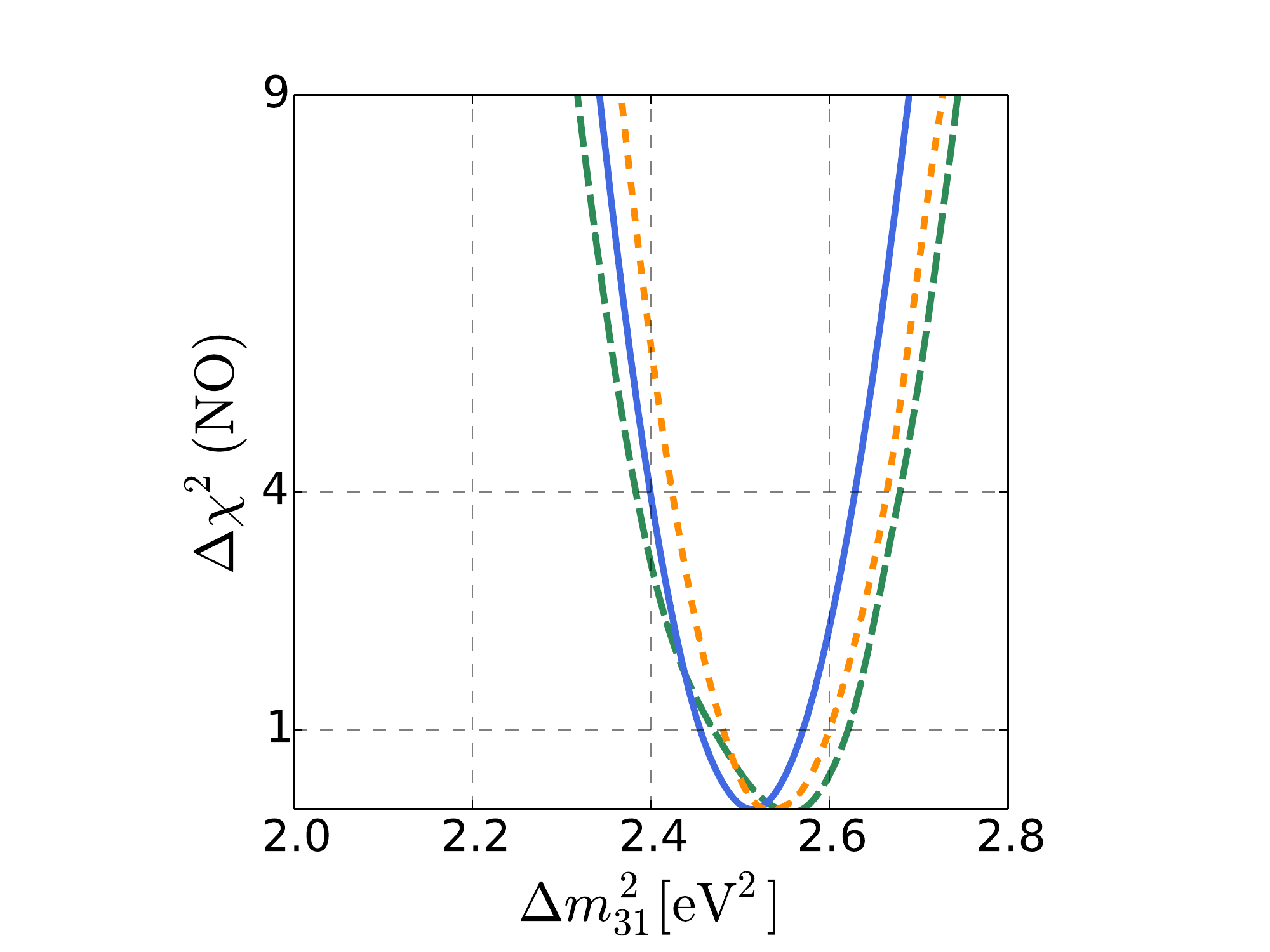}}%
     \subfigure{%
     \hspace{-2.6cm}
     \includegraphics[height=6cm]{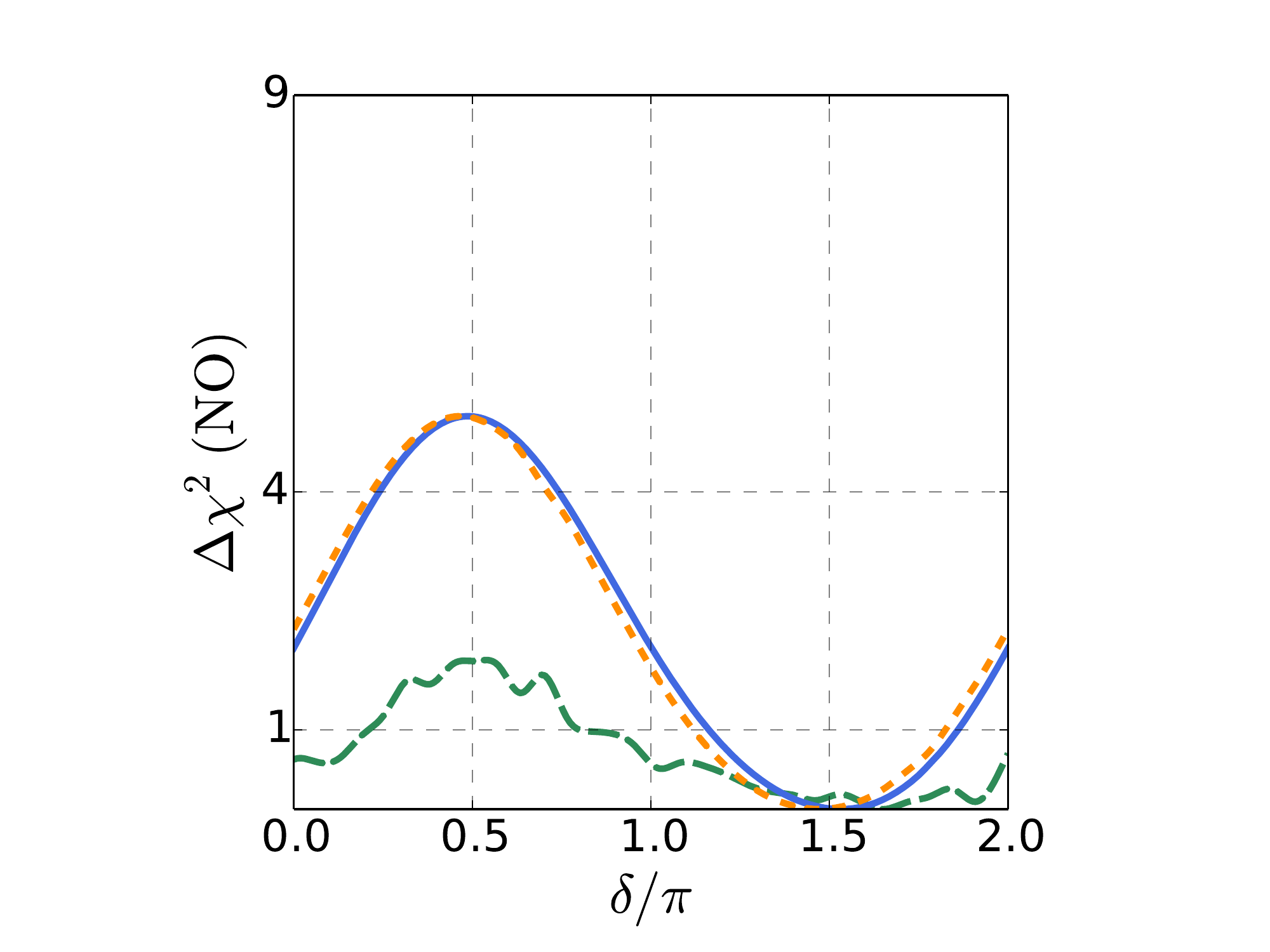}}%
    \caption{ \it Upper panels. $\Delta \chi^2$ as a function of $\sin^2 \theta_{13}$ (left panel)
    and $\sin^2 \theta_{23}$ (right panel) using the SM (solid blue line), LED (small dashed orange line)
    and NSI (large dashed green line) oscillation probabilities for NO neutrino mass spectrum.
    Lower panels. Same as in the upper but for $\Delta \chi^2$ as a function of $\Delta m^2_{31}$ (left panel)
    and $\delta$ (right panel).
}
\label{fig:1dim}
\end{figure*}

\begin{figure*}[h!]
\subfigure{%
 \includegraphics[height=6cm]{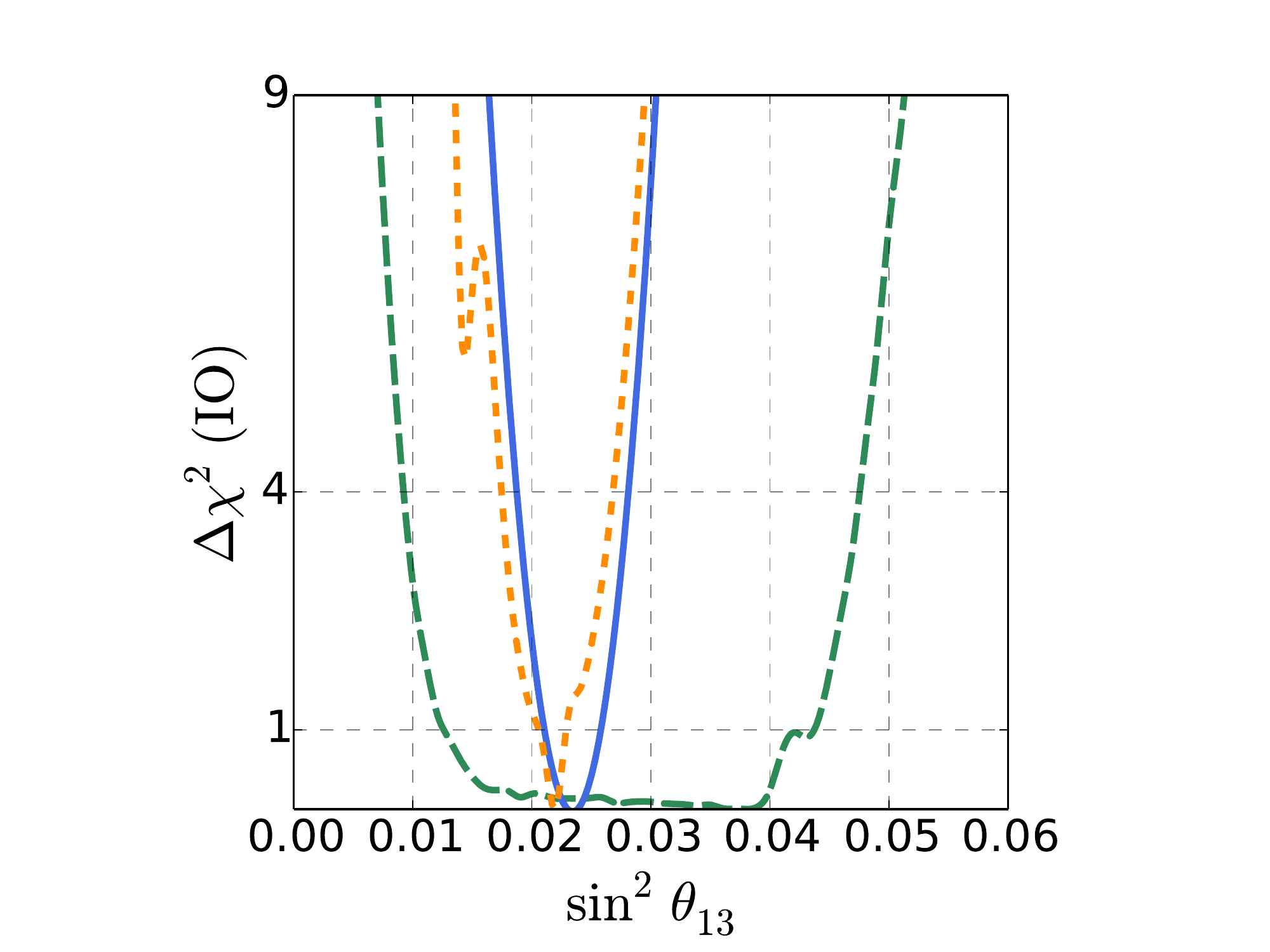}}%
\subfigure{%
\hspace{-2.6cm}
   \includegraphics[height=6cm]{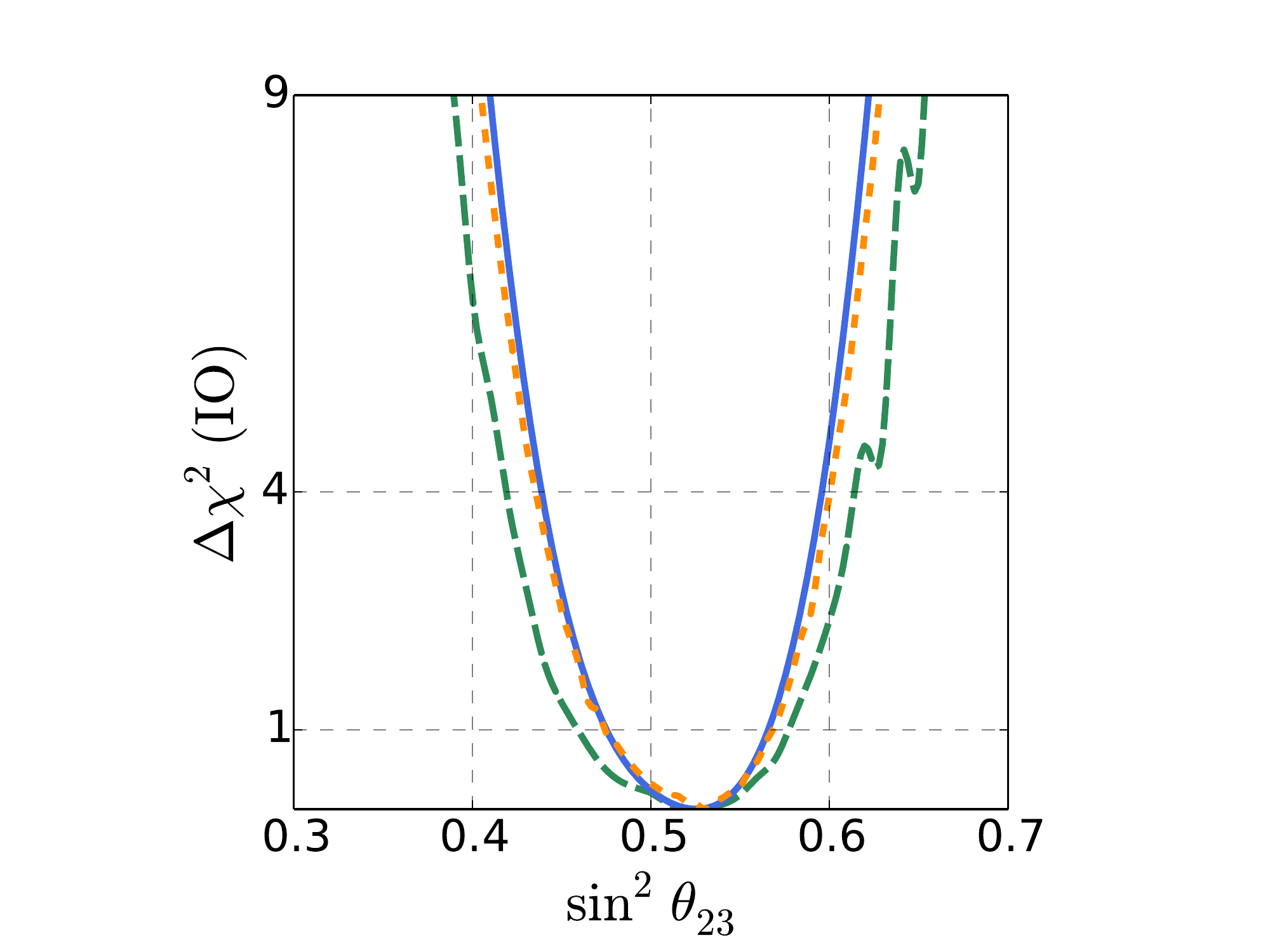}}%
        \vspace{-1cm}
   \subfigure{%
     \includegraphics[height=6cm]{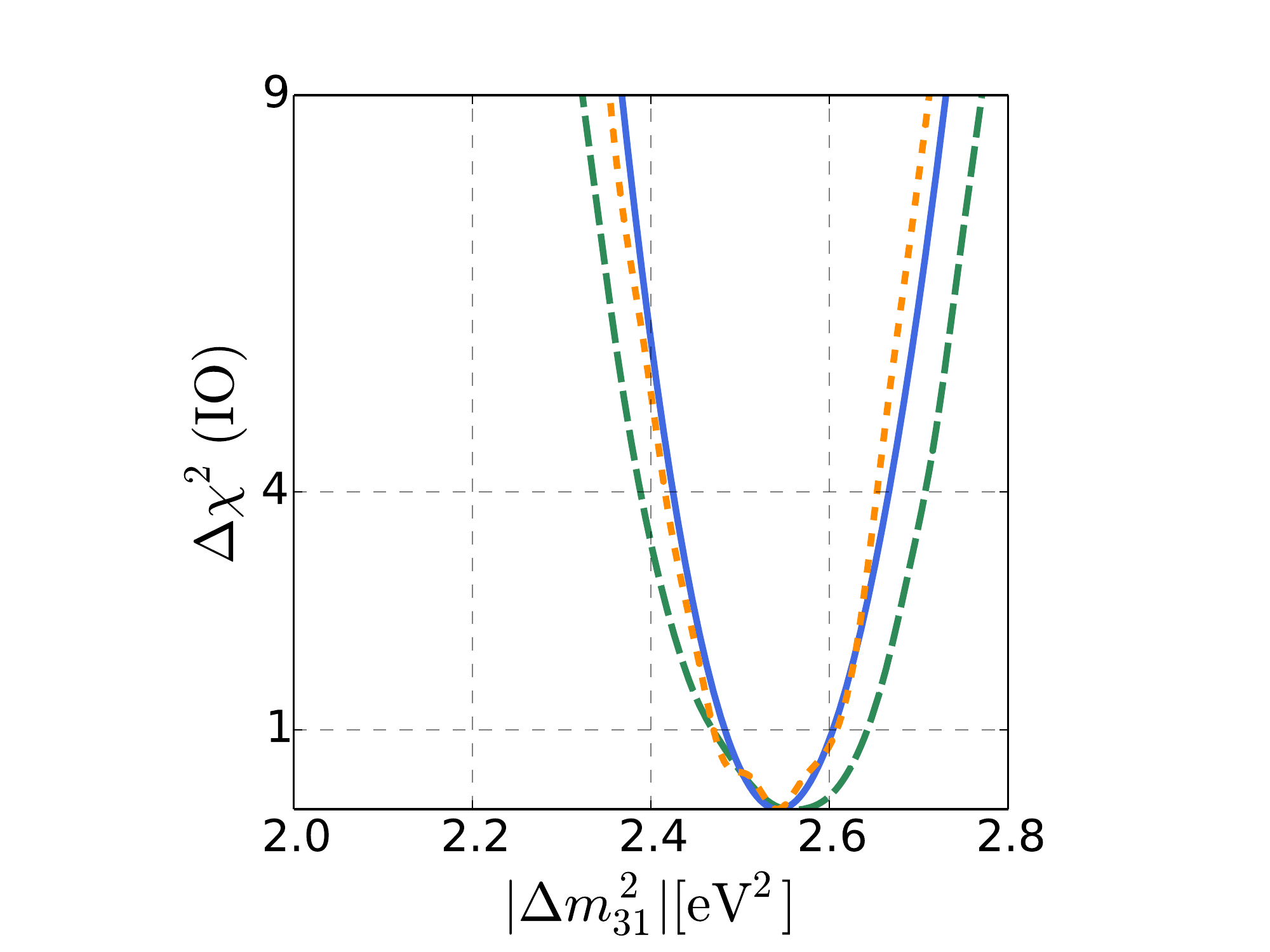}}%
     \subfigure{%
     \hspace{-2.6cm}
     \includegraphics[height=6cm]{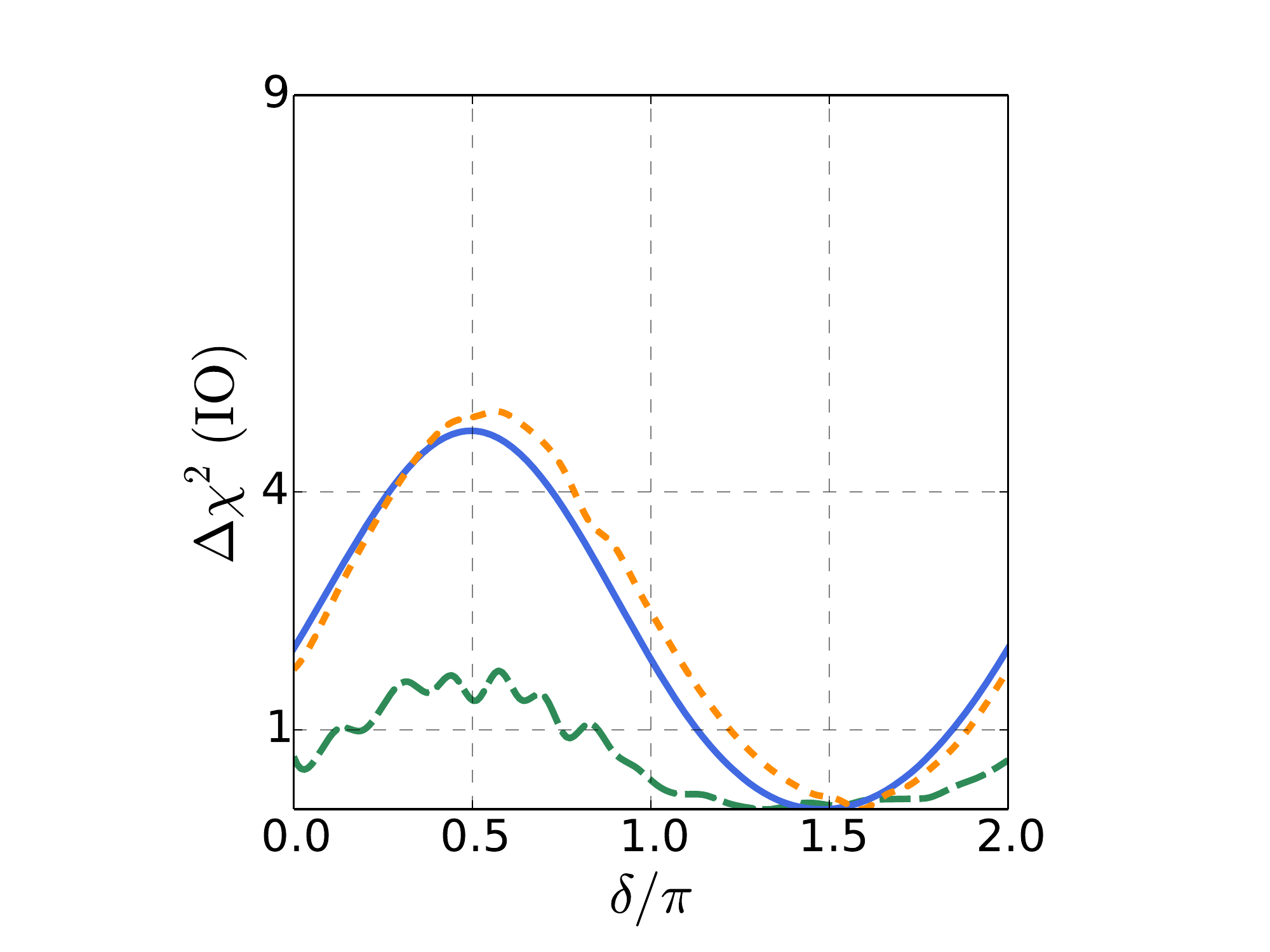}}%
    \caption{ \it Same as Fig. \ref{fig:1dim} but for the IO case.
}
\label{fig:1dimIO}
\end{figure*}

In this section we give the one dimensional projections of $\Delta \chi^2 = \chi^2 - \chi^2_{min}$ as a function of $\sin^2 \theta_{13}$, 
$\sin^2 \theta_{23}$, $\Delta m^2_{31}$ and the CP phase $\delta$. 
The case of NO is shown in Fig.~\ref{fig:1dim}, where the  values are given for the SM (solid blue line), 
the LED case (small dashed orange line) and the NSI case (large dashed green line). 
Notice that, as expected, the $\chi^2$  for $\sin^2 \theta_{13}$ in the NSI case is almost
flat in the 1$\sigma$ allowed region.
The case for the IO is given in Fig.~\ref{fig:1dimIO}.

\expandafter\ifx\csname natexlab\endcsname\relax\def\natexlab#1{#1}\fi
\expandafter\ifx\csname bibnamefont\endcsname\relax
  \def\bibnamefont#1{#1}\fi
\expandafter\ifx\csname bibfnamefont\endcsname\relax
  \def\bibfnamefont#1{#1}\fi
\expandafter\ifx\csname citenamefont\endcsname\relax
  \def\citenamefont#1{#1}\fi
\expandafter\ifx\csname url\endcsname\relax
  \def\url#1{\texttt{#1}}\fi
\expandafter\ifx\csname urlprefix\endcsname\relax\def\urlprefix{URL }\fi
\providecommand{\bibinfo}[2]{#2}
\providecommand{\eprint}[2][]{\url{#2}}


\begin{thebibliography}{27}

 \bibitem[{\citenamefont{F.~P.~An}(2013)}]{An:2013uza}
\bibinfo{author}{\bibfnamefont{F.}~\bibfnamefont{P.}~\bibnamefont{An}} \bibnamefont{et~al.}
  (\bibinfo{collaboration}{Daya Bay Collaboration}), \bibinfo{journal}{Chin.
  Phys. C.} \textbf{\bibinfo{volume}{37}}, \bibinfo{pages}{011001}
  (\bibinfo{year}{2013}). 
  
  
   \bibitem[{\citenamefont{J.~K.~Ahn}(2012)}]{Ahn:2012nd}
\bibinfo{author}{\bibfnamefont{J.}~\bibfnamefont{K.}~\bibnamefont{Ahn}} \bibnamefont{et~al.}
  (\bibinfo{collaboration}{RENO Collaboration}), \bibinfo{journal}{Phys.
  Rev. Lett.} \textbf{\bibinfo{volume}{108}}, \bibinfo{pages}{191802}
  (\bibinfo{year}{2012}). 
  
  
  
\bibitem[{\citenamefont{K.~Abe}(2014)}]{Abe:2013hdq}
\bibinfo{author}{\bibfnamefont{K.}~\bibnamefont{Abe}} \bibnamefont{et~al.}
  (\bibinfo{collaboration}{T2K Collaboration}), \bibinfo{journal}{Phys.
  Rev. Lett.} \textbf{\bibinfo{volume}{112}}, \bibinfo{pages}{061802}
  (\bibinfo{year}{2014}). 
  
  
  \bibitem[{\citenamefont{D.~S.~Ayres}(2004)}]{Ayres:2004js}
\bibinfo{author}{\bibfnamefont{D.}~\bibnamefont{S.}~\bibnamefont{Ayres}} \bibnamefont{et~al.}
  (\bibinfo{collaboration}{NOvA Collaboration}), 
  \eprint{arXiv:0503053}.
  
  
  \bibitem[{\citenamefont{K.~Abe}(2011)}]{Abe:2011ts}
\bibinfo{author}{\bibfnamefont{K.}~\bibnamefont{Abe}} \bibnamefont{et~al.}
(\bibinfo{collaboration}{Hyper-Kamiokande letter of intent}), \eprint{arXiv:1109.3262}.
  
  
  
  \bibitem[{\citenamefont{T.~Akiri}(2011)}]{Akiri:2011dv}
\bibinfo{author}{\bibfnamefont{T.}~\bibnamefont{Akiri}} \bibnamefont{et~al.}
  (\bibinfo{collaboration}{LBNE Collaboration}), \eprint{arXiv:1110.6249}.
  

  
  
       \bibitem[{\citenamefont{C.~Giganti}(2014)}]{Giganti:NOW2014}
\bibinfo{author}{\bibfnamefont{C.}~\bibnamefont{Giganti}, NOW 2014, Neutrino Oscillation Workshop, Conca Specchiulla (Otranto, Lecce, Italy),
September 7-14, 2014, \url{http://www.ba.infn.it/~now/now2014/web-content/index.html}}.


  \bibitem[{\citenamefont{F.~Capozzi}(2013)}]{Capozzi:2013csa}
\bibinfo{author}{\bibfnamefont{F.}~\bibnamefont{Capozzi}, \bibfnamefont{G.}~\bibfnamefont{L.}~\bibnamefont{Fogli}, \bibfnamefont{E.}~\bibnamefont{Lisi},
\bibfnamefont{A.}~\bibnamefont{Marrone}, \bibfnamefont{D.}~\bibnamefont{Montanino} and \bibfnamefont{A.}~\bibnamefont{Palazzo}},
 \bibinfo{journal}{Phys. Rev. D} \textbf{\bibinfo{volume}{89}}, \bibinfo{pages}{093018}
  (\bibinfo{year}{2014}).
  
  
 
 
   \bibitem[{\citenamefont{D.~V.~Forero}(2014)}]{Forero:2014bxa}
\bibinfo{author}{\bibfnamefont{D.}~\bibfnamefont{V.}~\bibnamefont{Forero}, \bibfnamefont{M.}~\bibnamefont{Tortola},
and \bibfnamefont{J.}~\bibfnamefont{W.}~\bibfnamefont{F.}~\bibnamefont{Valle},
 \bibinfo{journal}{Phys. Rev. D} \textbf{\bibinfo{volume}{90}}, \bibinfo{pages}{093006}}
  (\bibinfo{year}{2014}).
 

 


  \bibitem[{\citenamefont{M.~C.~Gonzalez-Garcia}(2012)}]{GonzalezGarcia:2012sz}
\bibinfo{author}{\bibfnamefont{M.}~\bibfnamefont{C.}~\bibnamefont{Gonzalez-Garcia}, \bibfnamefont{M.}~\bibnamefont{Maltoni},
and \bibfnamefont{T.}~\bibnamefont{Schwetz}}, 
 \bibinfo{journal}{JHEP} \textbf{\bibinfo{volume}{1411}}, \bibinfo{pages}{052}
  (\bibinfo{year}{2014}).


  
  
  \bibitem[{\citenamefont{K.~Abe}(2014)}]{Abe:2014tzr}
\bibinfo{author}{\bibfnamefont{K.}~\bibnamefont{Abe}} \bibnamefont{et~al.}
  (\bibinfo{collaboration}{T2K Collaboration}), \eprint{arXiv:1409.7469}.
 
 

 \bibitem[{\citenamefont{Y.~Grossman}(1995)}]{Grossman:1995wx}
\bibinfo{author}{\bibfnamefont{Y.}~\bibnamefont{Grossman}}, 
  \bibinfo{journal}{Phys. Lett. B} \textbf{\bibinfo{volume}{359}}, \bibinfo{pages}{141}
  (\bibinfo{year}{1995}). 
  
  

  
  
  
     \bibitem[{\citenamefont{R.~Barbieri}(2000)}]{Barbieri}
\bibinfo{author}{\bibfnamefont{R.}~\bibnamefont{Barbieri}, \bibfnamefont{P.}~\bibnamefont{Creminelli}
 and \bibfnamefont{A.}~\bibnamefont{Strumia}, 
  \bibinfo{journal}{Nucl. Phys. B} \textbf{\bibinfo{volume}{585}}, \bibinfo{pages}{28}}
  (\bibinfo{year}{2000}). 


   \bibitem[{\citenamefont{R.~N.~Mohapatra}(1999)}]{Mohapatra}
\bibinfo{author}{\bibfnamefont{R.}~\bibfnamefont{N.}~\bibfnamefont{Mohapatra}, \bibfnamefont{S.}~\bibnamefont{Nandi}
 and \bibfnamefont{A.}~\bibfnamefont{Perez-Lorenzana}, 
  \bibinfo{journal}{Phys. Lett. B} \textbf{\bibinfo{volume}{466}}, \bibinfo{pages}{115}}
  (\bibinfo{year}{1999}); 
  \bibinfo{author}{\bibfnamefont{R.}~\bibfnamefont{N.}~\bibfnamefont{Mohapatra}
 and \bibfnamefont{A.}~\bibfnamefont{Perez-Lorenzana}, 
  \bibinfo{journal}{Nucl. Phys. B} \textbf{\bibinfo{volume}{576}}, \bibinfo{pages}{466}}
  (\bibinfo{year}{2000}); 
    \bibinfo{author}{\bibfnamefont{R.}~\bibfnamefont{N.}~\bibfnamefont{Mohapatra}
 and \bibfnamefont{A.}~\bibfnamefont{Perez-Lorenzana}, 
  \bibinfo{journal}{Nucl. Phys. B} \textbf{\bibinfo{volume}{593}}, \bibinfo{pages}{451}}
  (\bibinfo{year}{2001}). 

  
   \bibitem[{\citenamefont{H.~Davoudiasl}(2002)}]{Davoudiasl:2002fq}
\bibinfo{author}{\bibfnamefont{H.}~\bibnamefont{Davoudiasl}, \bibfnamefont{P.}~\bibnamefont{Langacker}
 and \bibfnamefont{M.}~\bibnamefont{Perelstein}, 
  \bibinfo{journal}{Phys. Rev. D} \textbf{\bibinfo{volume}{65}}, \bibinfo{pages}{105015}}
  (\bibinfo{year}{2002}). 


  
           \bibitem[{\citenamefont{BastoGonzalez:2012me}(2014)}]{BastoGonzalez:2012me}
\bibinfo{author}{\bibfnamefont{V.}~\bibfnamefont{S.}~\bibnamefont{Basto-Gonzalez},
\bibfnamefont{A.}~\bibnamefont{Esmaili},  \bibfnamefont{O.}~\bibfnamefont{L.}~\bibfnamefont{G.}~\bibnamefont{Peres}},
   \bibinfo{journal}{Phys. Lett. B} \textbf{\bibinfo{volume}{718}}, \bibinfo{pages}{1020}
  (\bibinfo{year}{2013}); 
  \bibinfo{author}{\bibfnamefont{W.}~\bibnamefont{Rodejohann} and
\bibfnamefont{H.}~\bibnamefont{Zhang}},
   \bibinfo{journal}{Phys. Lett. B} \textbf{\bibinfo{volume}{737}}, \bibinfo{pages}{81}
  (\bibinfo{year}{2014}). 
  
  
  

  \bibitem[{\citenamefont{K.~Abe}(2014)}]{Abe:2014ugx}
\bibinfo{author}{\bibfnamefont{K.}~\bibnamefont{Abe}} \bibnamefont{et~al.}
  (\bibinfo{collaboration}{T2K Collaboration}), \bibinfo{journal}{Phys.
  Rev. Lett.} \textbf{\bibinfo{volume}{112}}, \bibinfo{pages}{181801}
  (\bibinfo{year}{2014}). 
  
 
  
  
    \bibitem[{\citenamefont{F.~P.~An}(2014)}]{An:2013zwz}
\bibinfo{author}{\bibfnamefont{F.}~\bibfnamefont{P.}~\bibnamefont{An}} \bibnamefont{et~al.}
  (\bibinfo{collaboration}{Daya Bay Collaboration}), 
   \bibinfo{journal}{Phys. Rev. Lett.} \textbf{\bibinfo{volume}{112}}, \bibinfo{pages}{061801}
  (\bibinfo{year}{2014}). 
  
  
  
\bibitem[{\citenamefont{I.~Girardi}(2004)}]{Girardi:2014kca}
\bibinfo{author}{\bibfnamefont{I.}~\bibnamefont{Girardi}, \bibfnamefont{D.}~\bibnamefont{Meloni}
and \bibfnamefont{S.}~\bibfnamefont{T.}~\bibnamefont{Petcov}},
 \bibinfo{journal}{Nucl. Phys. B} \textbf{\bibinfo{volume}{886}}, \bibinfo{pages}{31} (\bibinfo{year}{2014}). 
 
 
          \bibitem[{\citenamefont{I.~Girardi}(2014)}]{Girardi:2014gna}
\bibinfo{author}{\bibfnamefont{I.}~\bibnamefont{Girardi} and \bibfnamefont{D.}~\bibnamefont{Meloni}},
   \bibinfo{journal}{Phys. Rev. D} \textbf{\bibinfo{volume}{90}}, \bibinfo{pages}{7,  073011}
  (\bibinfo{year}{2014}). 
 
 
  
     
  \bibitem[{\citenamefont{L.~Wolfenstein}(1978)}]{Wolf78}
\bibinfo{author}{\bibfnamefont{L.}~\bibnamefont{Wolfenstein},
  \bibinfo{journal}{Phys. Rev. D} \textbf{\bibinfo{volume}{17}}, \bibinfo{pages}{2369}}
  (\bibinfo{year}{1978});
\bibinfo{author}{\bibfnamefont{M.}~\bibfnamefont{M.}~\bibnamefont{Guzzo}, \bibfnamefont{A.}~\bibnamefont{Masiero} 
and \bibfnamefont{S.}~\bibfnamefont{T.}~\bibnamefont{Petcov}},
 \bibinfo{journal}{Phys. Lett. B} \textbf{\bibinfo{volume}{260}}, \bibinfo{pages}{154}
  (\bibinfo{year}{1991});
\bibinfo{author}{\bibfnamefont{E.}~\bibnamefont{Roulet}},
 \bibinfo{journal}{Phys. Rev. D} \textbf{\bibinfo{volume}{44}}, \bibinfo{pages}{935}
  (\bibinfo{year}{1991}).
  

  
  
 
     \bibitem[{\citenamefont{J.~Koppw}(2008)}]{Kopp:2007ne}
\bibinfo{author}{\bibfnamefont{J.}~\bibnamefont{Kopp}, \bibfnamefont{M.}~\bibnamefont{Lindner}, 
\bibfnamefont{T.}~\bibnamefont{Ota} and \bibfnamefont{J.}~\bibnamefont{Sato}, 
  \bibinfo{journal}{Phys. Rev. D} \textbf{\bibinfo{volume}{77}}, \bibinfo{pages}{013007}}
  (\bibinfo{year}{2008}). 
  
  
 
   \bibitem[{\citenamefont{N.~Kitazawa}(2006)}]{Kitazawa:2006iq}
\bibinfo{author}{\bibfnamefont{N.}~\bibnamefont{Kitazawa}, \bibfnamefont{H.}~\bibnamefont{Sugiyama}
 and \bibfnamefont{O.}~\bibnamefont{Yasuda}},
 \eprint{arXiv:0606013}.
 
 
 
     \bibitem[{\citenamefont{M.~Blennow}(2008)}]{Blennow:2007pu}
\bibinfo{author}{\bibfnamefont{M.}~\bibnamefont{Blennow}, \bibfnamefont{T.}~\bibnamefont{Ohlsson}
 and \bibfnamefont{J.}~\bibnamefont{Skrotzki}},
   \bibinfo{journal}{Phys. Lett. B} \textbf{\bibinfo{volume}{660}}, \bibinfo{pages}{522}
  (\bibinfo{year}{2008}). 

 
 
      \bibitem[{\citenamefont{M.~Blennow}(2008)}]{Blennow:2008ym}
\bibinfo{author}{\bibfnamefont{M.}~\bibnamefont{Blennow}, \bibfnamefont{D.}~\bibnamefont{Meloni},  
\bibfnamefont{T.}~\bibnamefont{Ohlsson}, \bibfnamefont{F.}~\bibnamefont{Terranova}
 and \bibfnamefont{M.}~\bibnamefont{Westerberg}},
   \bibinfo{journal}{Eur. Phys. J. C} \textbf{\bibinfo{volume}{56}}, \bibinfo{pages}{529}
  (\bibinfo{year}{2008}). 
  
  
  
        \bibitem[{\citenamefont{T.~Ota}(2002)}]{Ota:2002na}
\bibinfo{author}{\bibfnamefont{T.}~\bibnamefont{Ota}, 
 and \bibfnamefont{J.}~\bibnamefont{Sato}},
   \bibinfo{journal}{Phys. Lett. B} \textbf{\bibinfo{volume}{545}}, \bibinfo{pages}{367}
  (\bibinfo{year}{2002}). 
  
  
 
 
        \bibitem[{\citenamefont{J.~Kopp}(2010)}]{Kopp:2010qt}
\bibinfo{author}{\bibfnamefont{J.}~\bibnamefont{Kopp}, 
\bibfnamefont{P.}~\bibfnamefont{A.}~\bibfnamefont{N.}~\bibnamefont{Machado}
and \bibfnamefont{S.}~\bibfnamefont{J.}~\bibnamefont{Parke}},
   \bibinfo{journal}{Phys. Rev. D} \textbf{\bibinfo{volume}{82}}, \bibinfo{pages}{113002}
  (\bibinfo{year}{2010}). 


    \bibitem[{\citenamefont{T.~Ohlsson}(2009)}]{Ohlsson:2008gx}
\bibinfo{author}{\bibfnamefont{T.}~\bibnamefont{Ohlsson} and \bibfnamefont{H.}~\bibnamefont{Zhang}, 
  \bibinfo{journal}{Phys. Lett. B} \textbf{\bibinfo{volume}{671}}, \bibinfo{pages}{99}}
  (\bibinfo{year}{2009}). 


   \bibitem[{\citenamefont{R.~Leitner}(2011)}]{Leitner:2011aa}
\bibinfo{author}{\bibfnamefont{R.}~\bibnamefont{Leitner}, \bibfnamefont{M.}~\bibnamefont{Malinsky},
 \bibfnamefont{B.}~\bibnamefont{Roskovec} and \bibfnamefont{H.}~\bibnamefont{Zhang}, 
  \bibinfo{journal}{JHEP} \textbf{\bibinfo{volume}{1112}}, \bibinfo{pages}{001}}
  (\bibinfo{year}{2011}). 



          \bibitem[{\citenamefont{A.~N.~Khan}(2013)}]{Khan:2013hva}
\bibinfo{author}{\bibfnamefont{A.}~\bibfnamefont{N.}~\bibnamefont{Khan}, 
\bibfnamefont{D.}~\bibfnamefont{W.}~\bibnamefont{McKay}
and \bibfnamefont{F.}~\bibnamefont{Tahir}},
   \bibinfo{journal}{Phys. Rev. D} \textbf{\bibinfo{volume}{88}}, \bibinfo{pages}{113006}
  (\bibinfo{year}{2013}); 
  A.~N.~Khan, D.~W.~McKay and F.~Tahir,
  Phys.\ Rev.\ D {\bf 90} (2014) 053008;
\bibinfo{author}{\bibfnamefont{Y.}~\bibfnamefont{F.}~\bibnamefont{Li} and  \bibfnamefont{Y.}~\bibfnamefont{L.}~\bibfnamefont{Zhou.}},
   \bibinfo{journal}{Nucl. Phys. B} \textbf{\bibinfo{volume}{888}}, \bibinfo{pages}{137}
  (\bibinfo{year}{2014}). 


  
  
    \bibitem[{\citenamefont{T.~Ohlsson}(2014)}]{Ohlsson:2013nna}
\bibinfo{author}{\bibfnamefont{T.}~\bibnamefont{Ohlsson}, \bibfnamefont{H.}~\bibnamefont{Zhang}
 and \bibfnamefont{S.}~\bibnamefont{Zhou},
  \bibinfo{journal}{Phys. Lett. B} \textbf{\bibinfo{volume}{728}}, \bibinfo{pages}{148}}
  (\bibinfo{year}{2014}). 
  

    \bibitem[{\citenamefont{D.~Meloni}(2009)}]{Meloni:2009cg}
\bibinfo{author}{\bibfnamefont{D.}~\bibnamefont{Meloni}, \bibfnamefont{T.}~\bibnamefont{Ohlsson},
 \bibfnamefont{W.}~\bibnamefont{Winter} and \bibfnamefont{H.}~\bibnamefont{Zhang},
  \bibinfo{journal}{JHEP} \textbf{\bibinfo{volume}{1004}}, \bibinfo{pages}{041}}
  (\bibinfo{year}{2010}). 




      \bibitem[{\citenamefont{T.~Ohlsson}(2013)}]{Ohlsson:2012kf}
\bibinfo{author}{\bibfnamefont{T.}~\bibnamefont{Ohlsson}},
   \bibinfo{journal}{Rept. Prog. Phys.} \textbf{\bibinfo{volume}{76}}, \bibinfo{pages}{044201}
  (\bibinfo{year}{2013}). 
  


  
 \bibitem[{\citenamefont{C.~Biggio}(2009)}]{enrique}
\bibinfo{author}{\bibfnamefont{C.}~\bibnamefont{Biggio}, \bibfnamefont{M.}~\bibnamefont{Blennow}
and \bibfnamefont{E.}~\bibnamefont{Fernandez-Martinez}}, 
  \bibinfo{journal}{JHEP} \textbf{\bibinfo{volume}{0908}}, \bibinfo{pages}{090}
  (\bibinfo{year}{2009}). 
  
  

        \bibitem[{\citenamefont{P.~Coloma}(2011)}]{Coloma:2011rq}
\bibinfo{author}{\bibfnamefont{P.}~\bibnamefont{Coloma}, \bibfnamefont{A.}~\bibnamefont{Donini}, \bibfnamefont{J.}~\bibnamefont{Lopez-Pavon}
and \bibfnamefont{H.}~\bibnamefont{Minakata}},
   \bibinfo{journal}{JHEP} \textbf{\bibinfo{volume}{1108}}, \bibinfo{pages}{036}
  (\bibinfo{year}{2011}). 
  
  
 

 \bibitem[{\citenamefont{B.~Pontecorvo}(1967)}]{01-BPont67}
\bibinfo{author}{\bibfnamefont{B.}~\bibnamefont{Pontecorvo} \bibnamefont{``Neutrino experiments and the question of 
leptonic-charge  conservation''}}, \bibinfo{journal}{Zh. Eksp. Teor. Fiz.} \textbf{\bibinfo{volume}{53}}, \bibinfo{pages}{1717}
  (\bibinfo{year}{1967}); 
   \bibnamefont{``Mesonium and Antimesonium''}, \bibinfo{journal}{Zh. Eksp. Teor. Fiz.} \textbf{\bibinfo{volume}{33}}, \bibinfo{pages}{549}
  (\bibinfo{year}{1957}); 
   \bibnamefont{``Inverse Beta Processes and Nonconservation of Lepton Charge''}, \bibinfo{journal}{Zh. Eksp. Teor. Fiz.} \textbf{\bibinfo{volume}{34}}, \bibinfo{pages}{247}
  (\bibinfo{year}{1958}); 
  \bibinfo{author}{\bibfnamefont{Z.}~\bibnamefont{Maki}, \bibfnamefont{M.}~\bibnamefont{Nakagawa}, and \bibfnamefont{S.}~\bibnamefont{Sakata}} 
  \bibnamefont{``Remarks on the Unified Model of Elementary Particles''}, \bibinfo{journal}{Prog. Theor. Phys.} \textbf{\bibinfo{volume}{28}}, \bibinfo{pages}{870}
  (\bibinfo{year}{1962}).
  
  
  
 \bibitem[{\citenamefont{K.~A.~Olive}(2014)}]{Agashe:2014kda}
\bibinfo{author}{\bibfnamefont{K.}~\bibnamefont{Nakamura} and \bibfnamefont{S.}~\bibfnamefont{T.}~\bibnamefont{Petcov} in 
\bibfnamefont{K.}~\bibfnamefont{A.}~\bibnamefont{Olive}} \bibnamefont{et~al.}
  (\bibinfo{collaboration}{Particle Data Group}), \bibinfo{journal}{Chin.
  Phys. C.} \textbf{\bibinfo{volume}{38}}, \bibinfo{pages}{090001}
  (\bibinfo{year}{2014}). 
  

 
  
        \bibitem[{\citenamefont{A.~Donini}(2006)}]{Donini:2005db}
\bibinfo{author}{\bibfnamefont{A.}~\bibnamefont{Donini}, \bibfnamefont{E.}~\bibnamefont{Fernandez-Martinez},
\bibfnamefont{D.}~\bibnamefont{Meloni} and \bibfnamefont{S.}~\bibnamefont{Rigolin}},
   \bibinfo{journal}{Nucl. Phys. B} \textbf{\bibinfo{volume}{743}}, \bibinfo{pages}{41}
  (\bibinfo{year}{2006}). 
  
  
  
  \bibitem[{\citenamefont{N.~Arkani-Hamed}(1998)}]{ADD}
\bibinfo{author}{\bibfnamefont{N.}~\bibfnamefont{Arkani-Hamed}, \bibfnamefont{S.}~\bibnamefont{Dimopoulos}
 and \bibfnamefont{G.}~\bibnamefont{Dvali}, 
  \bibinfo{journal}{Phys. Lett. B} \textbf{\bibinfo{volume}{429}}, \bibinfo{pages}{263}}
  (\bibinfo{year}{1998}); 
  \bibinfo{author}{\bibfnamefont{I.}~\bibnamefont{Antoniadis}, \bibfnamefont{N.}~\bibfnamefont{Arkani-Hamed}, \bibfnamefont{S.}~\bibnamefont{Dimopoulos}
 and \bibfnamefont{G.}~\bibnamefont{Dvali}, 
     \bibinfo{journal}{Phys. Lett. B} \textbf{\bibinfo{volume}{436}}, \bibinfo{pages}{257}}
  (\bibinfo{year}{1998});
 \bibinfo{author}{\bibfnamefont{N.}~\bibfnamefont{Arkani-Hamed}, \bibfnamefont{S.}~\bibnamefont{Dimopoulos}
 and \bibfnamefont{G.}~\bibnamefont{Dvali}, 
     \bibinfo{journal}{Phys. Rev. D} \textbf{\bibinfo{volume}{59}}, \bibinfo{pages}{086004}}
  (\bibinfo{year}{1999}).

  
  
      \bibitem[{\citenamefont{E.~G.~Adelberger}(2009)}]{Adelberger:2009zz}
\bibinfo{author}{\bibfnamefont{E.}~\bibnamefont{G.}~\bibnamefont{Adelberger}} \bibnamefont{et~al.}, 
\bibinfo{journal}{Prog. Part. Nucl. Phys.} \textbf{\bibinfo{volume}{62}}, \bibinfo{pages}{102}
  (\bibinfo{year}{2009});  
  \bibinfo{author}{\bibfnamefont{J.}~\bibnamefont{Beringer}} \bibnamefont{et~al.}
(\bibinfo{collaboration}{Particle Data Group Collaboration}), \bibinfo{journal}{Phys.
  Rev. D} \textbf{\bibinfo{volume}{86}}, \bibinfo{pages}{010001}
  (\bibinfo{year}{2012}). 
  
  
   \bibitem[{\citenamefont{P.~A.~N.~Machado}(2013)}]{Machado:2011jt}
\bibinfo{author}{\bibfnamefont{P.}~\bibfnamefont{A.}~\bibfnamefont{N.}~\bibnamefont{Machado}, \bibfnamefont{H.}~\bibnamefont{Nunokawa} 
and \bibfnamefont{R.}~\bibfnamefont{Z.}~\bibnamefont{Funchal}},
  \bibinfo{journal}{Phys. Rev. D} \textbf{\bibinfo{volume}{84}}, \bibinfo{pages}{013003}
  (\bibinfo{year}{2011}). 
  
  
    \bibitem[{\citenamefont{T.~A.~Mueller}(2011)}]{Mueller:2011nm}
\bibinfo{author}{\bibfnamefont{T.}~\bibfnamefont{A.}~\bibnamefont{Mueller} \bibnamefont{et~al.},
  \bibinfo{journal}{Phys. Rev. C} \textbf{\bibinfo{volume}{83}}, \bibinfo{pages}{054615}}
  (\bibinfo{year}{2011}). 


  \bibitem[{\citenamefont{P.~Huber}(2011)}]{Huber:2011wv}
\bibinfo{author}{\bibfnamefont{P.}~\bibnamefont{Huber},
  \bibinfo{journal}{Phys. Rev. C} \textbf{\bibinfo{volume}{84}}, \bibinfo{pages}{024617}}
  (\bibinfo{year}{2011}); 
  \bibinfo{author}{Erratum-ibid.,
  \bibinfo{journal}{Phys. Rev. C} \textbf{\bibinfo{volume}{85}}, \bibinfo{pages}{029901}}
  (\bibinfo{year}{2012}).

  
  
  
    \bibitem[{\citenamefont{P.~Vogel}(1999)}]{Vogel:1999zy}
\bibinfo{author}{\bibfnamefont{P.}~\bibnamefont{Vogel} and \bibfnamefont{J.}~\bibfnamefont{F.}~\bibnamefont{Beacom},
  \bibinfo{journal}{Phys. Rev. D} \textbf{\bibinfo{volume}{60}}, \bibinfo{pages}{053003}}
  (\bibinfo{year}{1999}). 
  
  
   \bibitem[{\citenamefont{P.~Huber}(2005)}]{GLOB2}
\bibinfo{author}{\bibfnamefont{P.}~\bibnamefont{Huber}, 
\bibfnamefont{M.}~\bibnamefont{Lindner} and \bibfnamefont{W.}~\bibnamefont{Winter}, 
  \bibinfo{journal}{Comput. Phys. Commun.} \textbf{\bibinfo{volume}{167}}, \bibinfo{pages}{195}}
  (\bibinfo{year}{2005});
  \bibinfo{author}{\bibfnamefont{P.}~\bibnamefont{Huber}, \bibfnamefont{J.}~\bibnamefont{Kopp},
\bibfnamefont{M.}~\bibnamefont{Lindner}, \bibfnamefont{M.}~\bibnamefont{Rolinec} and \bibfnamefont{W.}~\bibnamefont{Winter}, 
  \bibinfo{journal}{Comput. Phys. Commun.} \textbf{\bibinfo{volume}{177}}, \bibinfo{pages}{432}}
  (\bibinfo{year}{2007}).



  
  \bibitem[{\citenamefont{K.~Abe}(2013)}]{Abe:2013jth}
\bibinfo{author}{\bibfnamefont{K.}~\bibnamefont{Abe}} \bibnamefont{et~al.}
  (\bibinfo{collaboration}{T2K Collaboration}), \bibinfo{journal}{Phys.
  Rev. D} \textbf{\bibinfo{volume}{87}}, \bibinfo{pages}{092003}
  (\bibinfo{year}{2013}). 
  
  
  
      \bibitem[{\citenamefont{D.~Meloni}(2012)}]{Meloni:2012fq}
Private communication of T2K collaboration to the authors of the article:
\bibinfo{author}{\bibfnamefont{D.}~\bibnamefont{Meloni} and \bibfnamefont{M.}~\bibnamefont{Martini}}, 
 \bibinfo{journal}{Phys. Lett. B} \textbf{\bibinfo{volume}{716}}, \bibinfo{pages}{186}
  (\bibinfo{year}{2012}). 
  
  
 
   
         
       \bibitem[{\citenamefont{P.~Huber}(2002)}]{Huber:2002mx}
\bibinfo{author}{\bibfnamefont{P.}~\bibnamefont{Huber}, \bibfnamefont{M.}~\bibnamefont{Lindner},
and \bibfnamefont{W.}~\bibnamefont{Winter}}, 
 \bibinfo{journal}{Nucl. Phys. B} \textbf{\bibinfo{volume}{645}}, \bibinfo{pages}{3}
  (\bibinfo{year}{2002}). 
  
  
  
      \bibitem[{\citenamefont{P.~Huber}(2003)}]{Huber:2003pm}
\bibinfo{author}{\bibfnamefont{P.}~\bibnamefont{Huber}, \bibfnamefont{M.}~\bibnamefont{Lindner},
 \bibfnamefont{T.}~\bibnamefont{Schwetz} and \bibfnamefont{W.}~\bibnamefont{Winter}}, 
 \bibinfo{journal}{Nucl. Phys. B} \textbf{\bibinfo{volume}{665}}, \bibinfo{pages}{487}
  (\bibinfo{year}{2003}). 
  
  
        \bibitem[{\citenamefont{P.~Coloma}(2012)}]{Coloma:2012ji}
\bibinfo{author}{\bibfnamefont{P.}~\bibnamefont{Coloma}, \bibfnamefont{P.}~\bibnamefont{Huber}, \bibfnamefont{J.}~\bibnamefont{Kopp}
and \bibfnamefont{W.}~\bibnamefont{Winter}}, 
 \bibinfo{journal}{Phys. Rev. D} \textbf{\bibinfo{volume}{87}} \bibinfo{pages}{033004}
  (\bibinfo{year}{2013}). 
  
  
 
  
   \bibitem[{\citenamefont{C.~Giganti}(2013)}]{talk:T2K}
\bibinfo{author}{\bibfnamefont{C.}~\bibnamefont{Giganti}, GDR Neutrino Meeting, 16-17 June 2014, LAL, Universit\'e de Paris XI, Orsay, France.}



  
 \bibitem[{\citenamefont{K.~Eitel}(2005)}]{Eitel:2005hg}
\bibinfo{author}{\bibfnamefont{K.}~\bibnamefont{Eitel},
  \bibinfo{journal}{Nucl. Phys. Proc. Suppl.} \textbf{\bibinfo{volume}{143}}, \bibinfo{pages}{197}}
  (\bibinfo{year}{2005}). 
  
  


           \bibitem[{\citenamefont{A.~Esmaili}(2014)}]{Esmaili:2014esa}
\bibinfo{author}{\bibfnamefont{A.}~\bibnamefont{Esmaili},  \bibfnamefont{O.}~\bibfnamefont{L.}~\bibfnamefont{G.}~\bibnamefont{Peres} and \bibfnamefont{Z.}~\bibnamefont{Tabrizi}},
   \eprint{arXiv:1409.3502}.
  


  
 
       
\end{thebibliography}
\end{document}